\documentclass[useAMS,usenatbib]{mnras}
\usepackage{epsfig,rotate,graphicx}
\usepackage[fleqn]{amsmath}
\usepackage{lscape}
\usepackage{bm}
\usepackage{tikz}
\usepackage{paralist}
\usetikzlibrary{shapes.geometric, arrows}

\newcommand{\p}{\partial}

\newcommand{\be}{\begin{equation}}
\newcommand{\ee}{\end{equation}}
\newcommand{\gtrsim}{\;\raisebox{-.8ex}{$\buildrel{\textstyle>}\over\sim$}\;}
\newcommand{\lesssim}{\; \raisebox{-.8ex}{$\buildrel{\textstyle<}\over\sim$}\;}

\newcommand{\avg}[1]{\langle #1 \rangle}

\newcommand{\ii}{\mathrm{i}}
\newcommand{\bv}{\bm{v}}

\newcommand{\teta}{\tilde{\eta}}
\newcommand{\xim}{\xi_-}
\newcommand{\xip}{\xi_+}

\newcommand{\qthree}{Q_\mathrm{3D}}
\newcommand{\reynolds}{\alpha_\mathrm{R}}
\newcommand{\gstress}{\alpha_\mathrm{G}}

\newcommand{\heff}{H_\mathrm{eff}}
\newcommand{\omegav}{\Omega_\mathrm{v}}

\newcommand{\ro}{\mathrm{Ro}} 
\newcommand{\tstop}{\mathrm{St}}
\defcitealias{lesur09}{LP09}

\DeclareMathOperator{\real}{Re}

\usepackage{array,booktabs,tabularx}
\newcolumntype{R}{>{\centering\arraybackslash}X} 

\title[Self-gravitating 3D vortices]{Vortex survival in 3D self-gravitating accretion discs} 
\author[Lin \& Pierens]{Min-Kai Lin$^{1}$ \thanks{mklin@asiaa.sinica.edu.tw}, Arnaud Pierens$^{2,3}$ \\ 
$^1$Institute of Astronomy and Astrophysics, Academia Sinica, Taipei 10617, Taiwan\\
$^2$Universit\'e de Bordeaux, Observatoire Aquitain des Sciences de l'Univers, BP89 33271 Floirac Cedex, France\\
$^3$CNRS, Laboratoire d'Astrophysique de Bordeaux, BP89 33271 Floirac Cedex, France
}

\begin{document}

\maketitle
\begin{abstract} 
  Large-scale, dust-trapping vortices may account for observations of 
  asymmetric protoplanetary discs. Disc vortices are also potential sites for 
  accelerated planetesimal formation by concentrating dust grains. However, in 3D discs  
  vortices are subject to destructive `elliptic instabilities', which  
  reduces their viability as dust traps.   
  The survival of vortices in 3D accretion discs is thus an
  important issue to address. In this work, we 
  perform shearing box simulations to  
  show that 
  { disc self-gravity enhances the survival of 3D vortices, even when self-gravity is weak in the classic sense (e.g. with a Toomre $Q\simeq 5$).} 
  We find 
  a 3D, self-gravitating vortex can { grow} on secular timescales in
  spite of the elliptic instability.  
  The vortex aspect-ratio decreases as it strengthens, which feeds the
  elliptic instability. { The result is a 3D vortex with a
    turbulent core that persists for $\sim 10^3$ orbits.}
  We find when gravitational and hydrodynamic stresses become comparable,
  the vortex may undergo episodic bursts, which we interpret as
  interaction between elliptic and gravitational
  instabilities. We estimate the 
  distribution of dust particles in self-gravitating, turbulent
  vortices.  
  Our results suggest large-scale vortices in protoplanetary discs are
  more easily observed at large radii.  
\end{abstract}


\section{Introduction}\label{intro}


Vortex dynamics is playing an increasingly important role in protoplanetary
discs from both observational and theoretical perspectives. Several    
observations of transition discs reveal lopsided asymmetries in dust  
\citep{marel13,marel16,isella13,casassus13,fukagawa13,perez14,hashimoto15,marino15,ohta16,kraus17}.      
Disc vortices provide a possible explanation because they are 
associated with localized pressure bumps, which can act as dust-traps
\citep{barge95,lyra13,zhu14}. 
This interpretation suggests that large-scale vortices may be common in real
protoplanetary discs.   


The origin of { protoplanetary} disc vortices is usually attributed to 
hydrodynamic instabilities.  
{ To date, these include}: the Rossby Wave Instability 
\citep[RWI,][]{lovelace99,li00,meheut10,lin12c,yellin16,ono16}; the
Sub-critical Baroclinic Instability 
\citep[SBI,][]{peterson07a,lesur10,lyra11,raettig13,barge16}; the Convective
Overstability \citep[ConO,][]{klahr14,lyra14,latter16}; the Vertical Shear
Instability \citep[VSI,][]{nelson13,stoll14,barker15,lin15,richard16}; and the
Zombie Vortex Instability
\citep[ZVI,][]{marcus15,umurhan16,lesur16}. A common outcome of
these instabilities is vortex formation and/or amplification. 

Which of these hydrodynamic instabilities operate depends on the disc
structure and thermal state, but taken together they pertain to a wide
range of protoplanetary disc conditions \citep{umurhan16b}. Since 
these mechanisms form or amplify vortices, we can expect vortices to be
natural features of protoplanetary discs. These vortices may assist in
planetesimal formation by concentrating dust particles 
\citep{inaba06,meheut12}.  

Given the relevance of protoplanetary disc vortices to planet
formation, it is essential to understand how they evolve under
realistic conditions. A well-known result is that in {
  three-dimensional (3D)} discs vortices can develop 
secondary instabilities that weaken or even destroy them   
\citep{lithwick09,lesur09,chang10,railton14}. This would strongly
threaten their survival in protoplanetary discs. 
It is therefore necessary to  { model disc vortices in 3D and}
account for such 
`elliptic instabilities' to obtain realistic vortex lifetimes. 


Another effect that is often neglected in modeling protoplanetary
disc vortices is their self-gravity. Disc self-gravity is 
is usually described by the Toomre parameter
\begin{align} 
Q = \frac{c_s\Omega}{\pi G \Sigma},
\end{align}
where $c_s$ is the sound-speed, $\Omega$ is the disc's rotation
frequency, $\Sigma$ is the surface density, and $G$ is the
gravitational constant. Traditionally, self-gravity is considered
important when $Q \lesssim 1$, which is the formal criterion for 
gravitational instability in laminar, axisymmetric thin discs  
\citep{toomre64}. 
This typically translates to rather large
disc-to-star mass ratios, $M_\mathrm{d}/M_*\gtrsim 0.1$.  
However, the condition for gravitational 
instability, and hence the importance of self-gravity, can be relaxed
when additional physics are considered \citep{lin16}.  

{ 
  The effect of self-gravity on the equilibrium structure of
  a disc vortex was first investigated by \cite{adams95} assuming circular
  vortices. However, the shear in protoplanetary discs likely   
  only permits elliptical vortices \citep{lesur09}. 
  Self-gravity has also been included in studies of
  vortex formation and evolution via the RWI 
  \citep{lyra09,lin11,yellin16,regaly17} and in gravito-turbulent
  discs \citep{mamat09}.  
}

{ 
 In the case of the RWI, recent}  studies show that 
 disc self-gravity affects vortex evolution 
when $Q \lesssim \pi/2h$ 
\citep{lovelace13,zhu16}, where $h$ is the 
disc aspect-ratio. 
For typical protoplanetary discs with $h\sim 0.05$,
this means that self-gravity should be accounted when $Q\lesssim 20$; 
significantly larger than the classic condition. { However, 
  these calculations have been limited to
  razor-thin, two-dimensional (2D) discs. 
}

The above discussion motivates us to study { the evolution of elliptic vortices in} 3D 
self-gravitating protoplanetary discs. This was first attempted by 
\cite{lin12b} in global 3D numerical simulations of vortex 
formation { via the RWI} at planet gaps in self-gravitating discs.  
However, these simulations lacked the numerical resolution needed 
to capture the elliptic instability. Thus the 
survival of 3D vortices in self-gravitating discs remain unclear.  

In this work, we study the evolution of 3D vortices in a local 
patch of a self-gravitating protoplanetary disc. We use the shearing 
box framework and insert a vortex to evolve as an initial 
value problem. While less realistic than { \cite{lin12b}} in terms
of vortex formation, this approach permits clean numerical experiments
and higher resolution.   

We find self-gravity { enhances the survival of 3D vortices against the elliptic instability.}
The elliptic instability develops regardless of the 
strength of self-gravity, but with sufficient self-gravity the vortex
subsequently undergoes a slow { growth}, with increasing levels of
internal hydrodynamic turbulence. 
The vortex then undergo bursts; alternating between coherent and turbulent cores.  
We interpret this as `gravito-elliptic' feedback: a { growing}
vortex spins up and develops hydrodynamic turbulence because 
of the elliptic instability, which contributes to 
further collapse by removing 
the large-scale internal rotation. The latter 
effect is similar to secular gravitational instabilities in viscous or
dusty accretion discs \citep[e.g.][]{youdin11,takahashi14,lin16}

This paper is organized as follows. We list the basic equations in
\S\ref{model}. In \S\ref{sgvort_approx} we give a qualitative
discussion on the expected effect of self-gravity on disc vortices. We 
describe our simulation setup and diagnostic measures in 
\S\ref{sims}. We present simulation results in \S\ref{results}. 
We discuss implications of our simulations 
in \S\ref{discussion} before summarizing in \S\ref{summary}. 


\section{Local disc model}\label{model} 
We consider an inviscid, 3D self-gravitating
accretion disc orbiting a central star. Cylindrical co-ordinates
$(R,\phi, z)$ are centred on the star. The equilibrium rotation
profile is described by $\Omega(R) \propto R^{-q}$. We are interested
in Keplerian discs with $q=3/2$ and vertical oscillation frequency
$\Omega_z = \Omega$. We adopt the shearing box framework to study a
small patch of the disc \citep{goldreich65}. Local Cartesian
co-ordinates $(x,y,z)$ are anchored on a fiducial point
$(R_0,\phi_0,0)$ in the global disc co-rotating with the background
flow, so $\phi_0 = \Omega_0t$ where $\Omega_0\equiv\Omega(R_0)$. 
For simplicity, we drop the subscript zero hereafter. 

The Cartesian shearing box fluid equations are  
\begin{align}
  &\frac{\p\rho}{\p t} + \nabla\cdot\left(\rho\bm{v}\right) =
  0,\label{cont_eq}\\
  &\frac{\p\bm{v}}{\p t} + \bm{v}\cdot\nabla\bv = -\frac{1}{\rho}\nabla
  p -\nabla \Phi  - 2 \Omega \hat{\bm{z}}\times \bm{v} +
  \Omega^2(2qx, 0, -z)\label{mom_eq},\\
  & \nabla^2\Phi = 4 \pi G \rho \label{poisson}, 
\end{align}
where $\rho$ is the mass density, $\bm{v}$ is the 
total fluid velocity in the box, $p$ is the pressure and $\Phi$ is the
gravitational potential of the gas.  
We adopt an isothermal equation of 
state so that   
\begin{align}
  p = c_s^2 \rho,
\end{align}
where $c_s=H\Omega$ is the constant isothermal sound-speed, with $H$
being the disc scale-height at the reference radius. 

A simple, steady state equilibrium solution to the above equations
describes a stratified, axisymmetric shear flow with $\bm{v}=-q\Omega
x \hat{\bm{y}}$ and $\rho=\rho(z)$. In dimensionless variables
$\hat{\rho}\equiv \rho/\rho_0$, where $\rho_0$ is the mid-plane
density, and $Z
\equiv z/H$,  the density field satisfies 
\begin{align}
  \frac{d^2\ln{\hat{\rho}}}{dZ^2} +
  \frac{\hat{\rho}}{\qthree} + 1 = 0,   
\end{align}  
where 
\begin{align}
\qthree \equiv \frac{\Omega^2}{4\pi G \rho_0} 
\end{align} 
is a self-gravity parameter for 3D discs { \citep{mamat10}}. The boundary
conditions are $\hat{\rho}(0) = 1$ and $d\hat{\rho}/dz = 0$ at
$z=0$. In the non-self-gravitating limit, $\qthree\to\infty$, we
obtain the usual Gaussian solution, 
$\hat{\rho} = \exp{\left(-Z^2/2\right)}$. 

However, for non-self-gravitating discs there exists another class
of equilibrium flow solutions corresponding to a vortex, which we
discuss below. 



\section{Vortex solutions and the 
  qualitative effect of self-gravity}\label{sgvort_approx} 

In this section we discuss the effect of disc 
self-gravity on equilibrium flows that correspond to a vortex. To to
this, it is useful to work in co-ordinates that reflect such a flow
pattern. We thus adopt the non-orthogonal `elliptico-polar'
co-ordinates $(s,\varphi,z)$ defined by 
\begin{align}
  &x = s\cos{\varphi},\\
  &y = \chi s \sin{\varphi}, 
\end{align} 
where $\chi$ is a constant, and $z$ remains unchanged \citep[see,
e.g.][ for further discussion of this co-ordinate
system]{kerswell94}. Here, we take $\chi>1$ so that the co-ordinate
$s\in[0,\infty)$ corresponds to the semi-minor axis of a series of
similar ellipses, and $\varphi\in[0,2\pi]$ corresponds to the
azimuthal position along a given ellipse. We also define the fluid
velocity components $v_s$, $v_\varphi$ such that  
\begin{align} 
  v_x \hat{\bm{x}} + v_y \hat{\bm{y}} \equiv v_s \bm{s} +
  v_\varphi\bm{\varphi}, 
\end{align}
where
\begin{align}
  &\bm{s} \equiv \cos{\varphi} \hat{\bm{x}} +
  \chi\sin{\varphi}\hat{\bm{y}},\\
  &\bm{\varphi} \equiv -\sin{\varphi} \hat{\bm{x}} +
  \chi\cos{\varphi}\hat{\bm{y}}. 
\end{align}
The shearing box equations in elliptico-polar co-ordinates are listed
in Appendix \ref{elliptico-eqns}.  

\subsection{The GNG vortex in the absence of
  self-gravity} \label{gng_vortex}
\cite{goodman87} present exact vortex solutions to
Eq. \ref{cont_eq}---\ref{mom_eq} in the absence of self-gravity (by
setting $G = \Phi = 0$). Here, we review this `GNG' vortex in
elliptico-polar co-ordinates.  

The GNG vortex corresponds to a steady axisymmetric 
flow in elliptico-polar co-ordinates in vertical hydrostatic equilibrium, so that $\p_t = \p_\varphi =
v_s = v_z = 0$ and $v_\varphi = v_\varphi(s)$. The momentum equations then read 
\begin{align}
  0 =& \frac{v_\varphi^2}{s}-\frac{1}{2}\left(\xi_-\cos{2\varphi} + \xi_+\right)
  \frac{\p \eta}{\p s} + \Omega v_\varphi \chi \left(\xi_-\cos{2\varphi} +
    \xi_+\right) \notag\\ &+ q \Omega^2 s \left(\cos{2\varphi} +
    1\right),\label{ellip_xeq}\\
  0 = & \frac{\xi_-}{2}\sin{2\varphi}\frac{\p \eta}{\p s}  - \Omega \chi
  \xi_- v_\varphi\sin{2\varphi} \notag  - q \Omega^2 s
  \sin{2\varphi}, \label{ellip_yeq}\\
  0 =& z\Omega^2 + \frac{\p \eta}{\p z},
\end{align}
where $\xi_\pm = 1 \pm 1/\chi^2$, and 
\begin{align}
  \eta \equiv c_s^2\ln{\rho}
\end{align}
is the gas enthalpy. 

Since $\p_\varphi=0$ by construction, the trigonometric coefficients
in Eq. \ref{ellip_xeq}---\ref{ellip_yeq} must vanish. Then we obtain the flow solution  
\begin{align}
  v_\varphi(s) = - \Omega_\mathrm{v} s \text{ with } \Omega_\mathrm{v} =
  \sqrt{\frac{2q}{\chi^2 - 1}}\Omega. \label{gng_sol}
\end{align}
We have have chosen the sign for $\omegav$ so that the flow is
anti-cyclonic { \citep{bodo07}}. The corresponding density distribution is given via 
\begin{align}\label{gng_density}
  \ln{\frac{\rho}{\rho_\mathrm{c}}} =
    -\frac{s^2}{2H_\mathrm{eff}^2} - \frac{z^2}{2H^2},
\end{align}
where $\rho_\mathrm{c} $ is the density at $s=z=0$ and
the characteristic horizontal size $H_\mathrm{eff}$ is 
\begin{align} 
  H_\mathrm{eff} \equiv \frac{H}{F},
\end{align}
with 
\begin{align}
  F^2 (\chi; q) \equiv \frac{2}{\xi_-}\left[ \frac{\sqrt{2q\left(\chi^2 -
        1\right)}}{\chi} - q \right]. 
\end{align}
Note that, in a Keplerian disc, the requirement $F^2>0$ imply 
non-self-gravitating vortices only exist with aspect-ratios
$\chi>2$. 

\subsection{Effect of self-gravity on the GNG vortex density field} 
The GNG solution is not compatible with self-gravity
because the trigonometric coefficients of the Poisson equation in
elliptico-polar co-ordinates do not vanish (see 
Eq. \ref{elliptico-poisson}). This is because the 
gravitational potential of an ellipsoidal density distribution does
not have the same symmetry, except for $s\to0$ \citep{chandra69}. 

In order to get a sense of the effect of self-gravity, let us consider
the region about $s=0$ (vortex centroid), and simplify further by
neglecting { vertical }stratification, so $\p_z=z=0$. 
For this cylindrical 
problem, we assume the self-gravitational potential has the form 
\begin{align}\label{phi_approx}
  \Phi =  \frac{1}{2} A s^2 + \mathrm{const.},
\end{align}
for small $s$, where $A$ is a constant. Eq. \ref{phi_approx} is consistent with the Poisson
equation near the origin provided that  
\begin{align}
  A = \frac{\Omega^2}{\qthree\xi_+}.
\end{align}

Including self-gravity in the momentum equations amounts to the
replacement $\eta \to \eta + \Phi$. Then 
\begin{align}
  \frac{\p\ln{\rho}}{\p s}  &=
  -\frac{s}{H_\mathrm{eff}^2} - \frac{1}{c_s^2}\frac{\p \Phi}{\p
    s} 
   = - \frac{s}{H_\mathrm{eff}^2}\left(1 +
  \frac{1}{\qthree\xi_+F^2}\right)\notag\\
  & \equiv - \frac{s}{H_\mathrm{eff,sg}^2}, 
\end{align} 
where we have inserted the above approximation for $\Phi$. 
Writing $H_\mathrm{eff,sg} = H/F_\mathrm{sg}$, we have 
\begin{align}
F^2_\mathrm{sg}(\chi,\qthree;q) \equiv \frac{H^2}{H^2_\mathrm{eff,sg}}
  = F^2 + \frac{1}{\qthree\xi_+}. \label{fsg}
\end{align}
Thus self-gravity reduces the characteristic vortex size by increasing
$F^2$ to $F^2+1/\qthree\xi_+$. 

\begin{figure}
  \includegraphics[width=\linewidth]{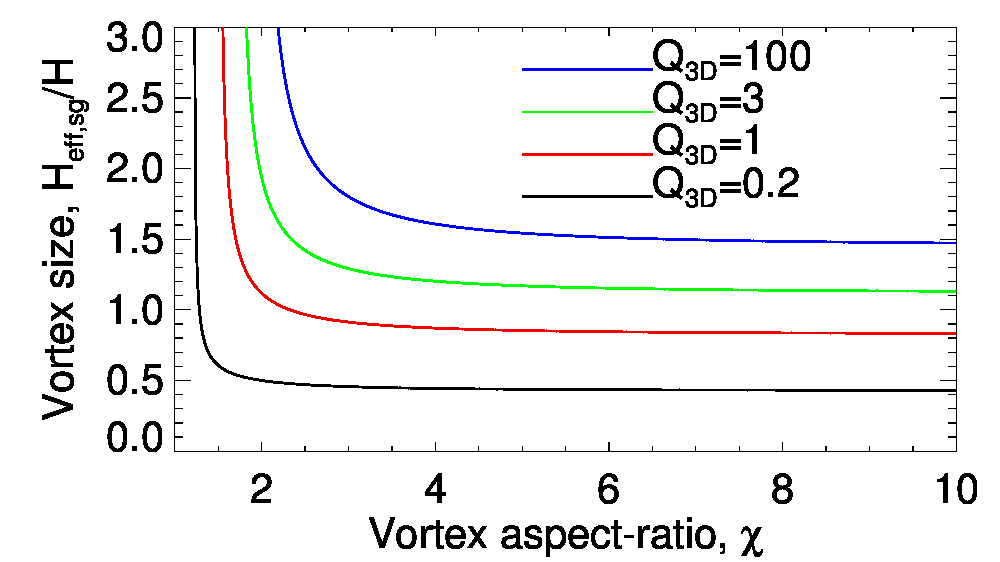}
  \caption{Characteristic horizontal vortex size, given by
    Eq. \ref{fsg}, as a function of the aspect-ratio, for a range of
    self-gravity parameters $\qthree$. \label{fsg_fig1} 
  }
\end{figure}

\begin{figure}
  \includegraphics[width=\linewidth]{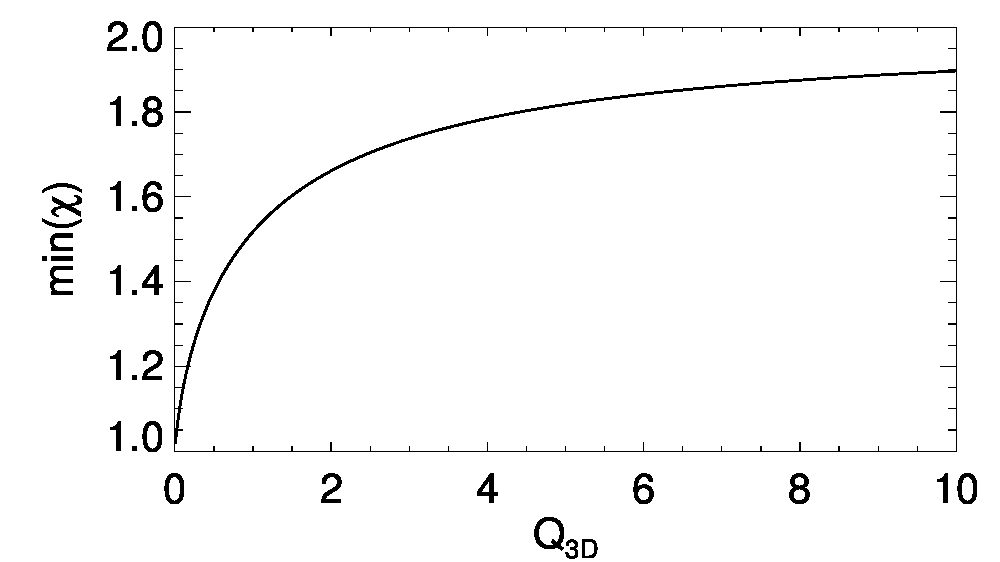}
  \caption{Minimum allowed aspect-ratio of the self-gravity-modified
    GNG vortex as a function of the 
    self-gravity parameter $\qthree$. \label{fsg_fig2} 
  }
\end{figure}

In Fig. \ref{fsg_fig1} we plot the characteristic vortex size
$H_\mathrm{eff,sg}$ as a function of the aspect-ratio and
self-gravity. The characteristic vortex size is insensitive to its 
aspect-ratio for $\chi\gtrsim4$. At fixed
$\chi$ the vortex size is reduced by self-gravity, but this effect is
small for elongated vortices. Notice the minimum allowed aspect-ratio 
[defined by $F_\mathrm{sg}^2(\chi)=0$] is reduced by
self-gravity. We show this in Fig. \ref{fsg_fig2}: as $\qthree\to 0$
the minimum aspect-ratio tends to unity.

{ Our estimate above differs from \cite{adams95}, who 
 found that self-gravity increases the size of a circular vortex. By contrast,
  Fig. \ref{fsg_fig1}---\ref{fsg_fig2} indicates self-gravity
  reduces the size of elliptic vortices; and 
that circular vortices are
  only allowed in strongly self-gravitating discs.}

\subsection{Effect of self-gravity on the GNG vortex vorticity
  field}\label{sg_on_vorticity} 
In Appendix \ref{sgeffect}, we give a more careful discussion of the
effect of self-gravity.  There, we also consider how self-gravity modifies
the distribution of vertical vorticity of the GNG vortex. This effect is
not obvious since potentials do not directly source vorticity. 

We find the vorticity perturbation due to self-gravity is
\begin{align}\label{delta_omega_sg}
  \Delta \omega \propto 
\left(4\omegav -
  \Omega\chi\xip\right)\left(\frac{s}{H_\mathrm{eff}}\right)^4 
\end{align}
near the vortex centre (see Eq. \ref{sgeffect_vort}). For the GNG
solution this implies that $\Delta\omega<0$ for $\chi\gtrsim 2.5$. As
we will be considering elongated vortices in practice, we expect 
that self-gravity enhances the GNG vortex by making its vorticity more
negative. 
{ 
  However, this is a small effect as the above estimate is restricted to $s\ll H_\mathrm{eff}$. \cite{adams95} reached a similar conclusion for circular vortices, where they find self-gravity does not modify the vortex velocity field. 
}
%

\section{Numerical simulations}\label{sims} 
We simulate the evolution of a single vortex in the 3D,
self-gravitating shearing box using the \textsc{Athena} code
\citep{stone08} to evolve Eq. \ref{cont_eq}---\ref{poisson}. 
We adopt standard numerical configurations: third order reconstruction,
a Roe solver, and Corner Transport Upwind (CTU) integrator. Disc 
self-gravity is solved via Fast Fourier Transform, but assuming a
vacuum beyond the vertical boundaries. We also use orbital
advection \citep{stone10}.  

\subsection{Grid setup and boundary conditions}   
The simulation domain is $x\in[-L_x,L_x]/2$, uniformly spaced with 
$N_x$ grid points; and similarly for the $y$ and $z$ directions.  
Our fiducial box size and resolution is $(L_x,L_y,L_z)=(16H,32H,6H)$ 
with $(N_x,N_y,N_z)=(512,512,192)$, or $32$ cells per $H$ in the $x,z$ 
directions and $16$ cells per $H$ in $y$. 
We use a lower resolution in the azimuth since we consider elongated
vortices. The box size and numerical resolution is 
a compromise between having a sufficiently large box to minimize
the influence of neighbouring vortices due to periodic boundary
conditions (see below), and having enough resolution to
capture the elliptic instability. We find this setup makes it feasible
to simulate up to 2000 orbits. For selected cases we
also double the $x$ resolution.



We apply periodic boundary conditions in the $y$ direction and
shear-periodic boundary conditions in $x$. In the $z$ direction
we apply reflective boundary conditions for the momenta, and 
extrapolate the density field assuming vertical hydrostatic
equilibrium. In addition, we apply damping for $|x|\in[0.8L_x,L_x]$ in
which the hydrodynamic variables are relaxed towards their initial
values on a timescale of $0.01$ orbits. This is done to
minimize spiral density waves launched by
neighbouring vortices.



\subsection{Units and self-gravity parameter}

We adopt computational units such that $H=c_s=1$, then $\Omega=1$ as
well. Time is quoted in units of the orbital period at the fiducial
radius, $P_0=2\pi/\Omega $. The initial midplane density is 
$\rho_0=1$. In \textsc{Athena}, the strength of self-gravity is
parameterized by the value of $ 4\pi G$. We specify it 
through $\qthree = 1/4\pi G$ in computational units. 

For { the disc models} considered in this work, the initial 
density distribution is approximately Gaussian, 
$\rho \simeq \rho_0\exp{\left(-z^2/2H^2\right)}$. Then  
our self-gravity parameter $\qthree$ is related to the classic
Toomre parameter $Q$ by  
\begin{align}
  \qthree \simeq \sqrt{\frac{\pi}{8}}Q
\end{align}
Thus using $Q$ or $\qthree$ is immaterial as
they only differ by a factor of order unity. We will label our
simulations with $\qthree$, however, since this is the natural choice 
for 3D discs.   


\subsection{Initial condition and vortex perturbation}\label{vort_pert}

The GNG vortex described above is useful for analytic discussion due
to its simple form, but is inconvenient to set as initial
conditions in shearing box simulations because it is unbounded in the 
horizontal plane, which does not respect the standard shearing box
boundary conditions. In a global disc, we expect the flow far from the 
vortex to return to orbital motion about the star.  

We thus initialise an axisymmetric, 3D shearing box as described in
\S\ref{model} and apply the GNG vortex flow as a perturbation:
\begin{align}
  &\Delta v_x = \frac{\omegav}{\chi}y\exp{\left(-m^2\right)},\\
  &\Delta v_y = \left(q\Omega x - \omegav\chi
    x\right)\exp{\left(-m^2\right)}, 
\end{align}
with
\begin{align}
  m^2(x,y) \equiv \frac{x^2}{b^2} + \frac{y^2}{a^2} =
  \frac{1}{b^2}\left(x^2 + \frac{y^2}{\chi^2}\right),
\end{align}
where $(a,b)$ is the semi-major and semi-minor axes of the vortex
perturbation, respectively. Near the origin, the total velocity field
is approximately that of the GNG vortex, while far from it one
recovers the Keplerian shear flow. 

\begin{figure}
  \includegraphics[scale=.39,clip=true,trim=0cm 0.cm 0cm 
    0cm]{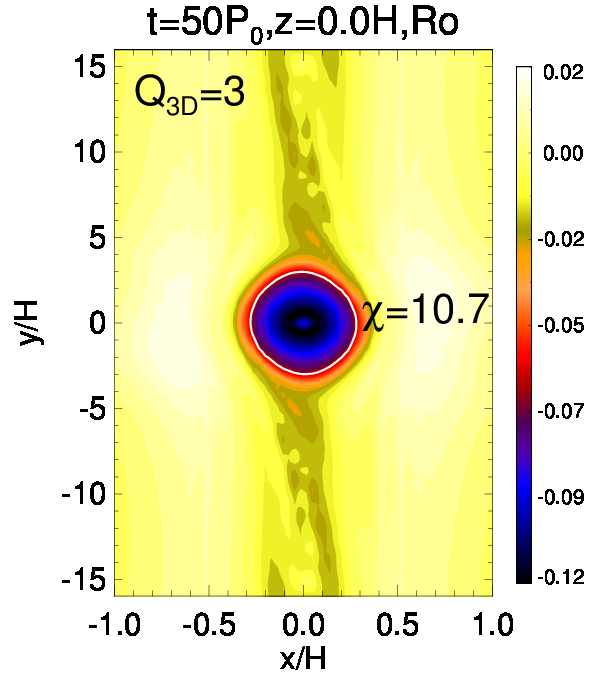}\includegraphics[scale=.39,clip=true,trim=2.3cm 0cm 0cm 
    0cm]{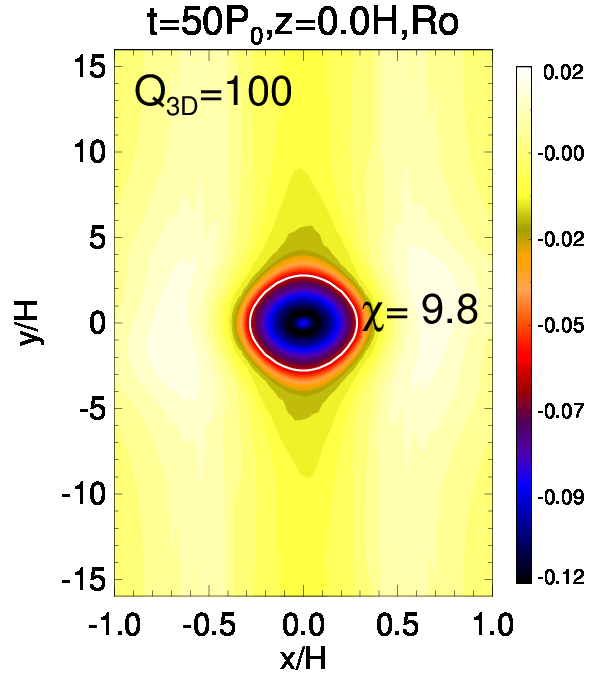}
  \caption{Vortex formed by perturbing the shearing box with the GNG
    vortex solution. Left: self-gravitating disc with
    $\qthree=3$. Right: effectively non-self-gravitating disc with
    $\qthree=100$. The colour scale corresponds to the Rossby number 
    $\ro = 
    \left(\nabla\times\bm{v}\right)/2\Omega + q/2$} { at the midplane.}
     Here, $\chi$ is the aspect-ratio of the loci 
    $\ro=0.5\mathrm{min}(\ro)$ marked by the white contour. \label{qss_vortex}
\end{figure}

\begin{figure}
  \includegraphics[scale=.66,clip=true,trim=0.5cm 1.7cm 0cm 
    0cm]{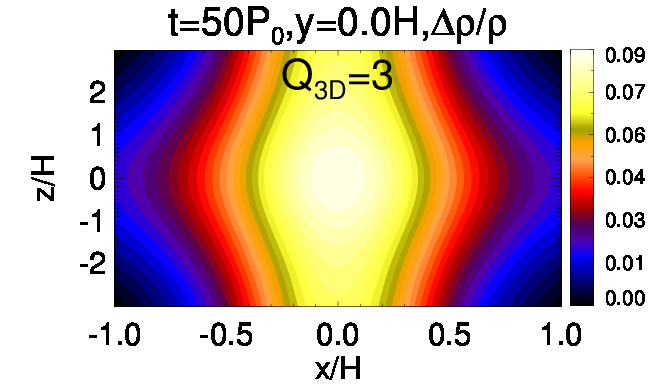}\\
  \includegraphics[scale=.66,clip=true,trim=0.5cm 0cm 0cm 
    0.95cm]{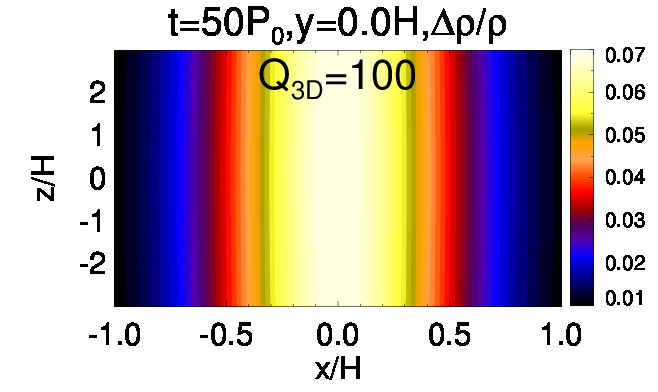}
  \caption{Vertical structure of vortices formed by perturbing the shearing box with the GNG 
    vortex solution. Upper: self-gravitating disc with
    $\qthree=3$. Lower: non-self-gravitating disc with
    $\qthree=100$. The colour scale is the density perturbation {
      relative to $t=0$(thus removing the background disc stratification)}. \label{qss_vortex_vdens} 
    }
\end{figure}

We introduce the above velocity perturbations through source terms in
the horizontal momentum equations over a timescale of $10P_0$.  
We specify the perturbation radial size $b$ and aspect ratio
$\chi$, and obtain the appropriate value of $\omegav$ from
Eq. \ref{gng_sol}. Fig. \ref{qss_vortex}---\ref{qss_vortex_vdens} shows that the initial 
vortex, formed by perturbing the disc with the GNG solution, is 
insensitive to $\qthree$. The self-gravitating vortex is {
  slightly} more
stratified \citep[as found by][]{lin12b}. 


\subsection{Diagnostics}\label{diagnostics} 
Our key diagnostics are:   
\begin{itemize}
\item The Rossby number 
\begin{align}
  \ro \equiv \frac{\hat{\bm{z}}\cdot\nabla\times\bm{v}}{2\Omega} +
  \frac{q}{2},  
\end{align}
as a dimensionless measure of the vertical vorticity relative to the
background shear flow. The Rossby number can be used to measure the
strength of the vortex, which can be defined as a coherent, extended
region of $\ro<0$ in the absence of turbulence (see Fig. \ref{qss_vortex}). 

\item The vertical Mach number
  \begin{align}
    M_z \equiv \frac{|v_z|}{c_s} 
  \end{align}
to characterize vertical motions and thus the elliptic instability,
which is expected to be  three-dimensional \citep{lesur09}. 

\item The Reynolds and gravitational stresses  
\begin{align}
  &\reynolds \equiv \frac{1}{c_s^2}v_x\left(v_y + q\Omega x\right),\\
  &\gstress \equiv \frac{1}{c_s^2}\frac{\p_x \Phi\p_y\Phi}{4\pi
   G\rho}, 
\end{align}
to characterize turbulent activity. The total stress is $\alpha \equiv
\reynolds + \gstress$.  

\item Vortex characteristics. We calculate the evolution of the vortex
  shape in post-processing, as follows. For this we use the vortex
  density field because it remains smooth throughout the 
  simulation. (The vorticity field can involve small-scale structures
  so it becomes difficult to define a large-scale coherent pattern.) 
We first locate the horizontal coordinates $(x_v,y_v)$ of 
$\mathrm{max}(\rho)$ near the origin. 
We define the vortex boundary as
the contour around $(x_v,y_v)$ on which $\rho =
0.5\mathrm{max}(\rho)$. This contour is usually close to an ellipse,
except at late evolution. We then define the semi-minor and major
vortex axes, $b_\rho$ and $a_\rho$, as half the length and width of
the contour, respectively. The corresponding aspect ratio is
$\chi_\rho = a_\rho/b_\rho$.  
\end{itemize}

We define the density-weighted average value of a quantity $f$ as   
\begin{align}
  \avg{f} \equiv \frac{\int_\mathcal{V} \rho f
    dV}{M_\mathcal{V}},
\end{align}
where 
\begin{align}
  M_\mathcal{V} \equiv \int_\mathcal{V} \rho dV, 
\end{align}
and the sample volume $\mathcal{V}$ is $x\in[-b,b]$. 
 

\section{Results}\label{results}

Table \ref{sim_summary} summarizes the simulations we have carried
out. Our fiducial set of runs {, which are highlighted in bold and discussed in detail below, are}
Q3, Q4, Q5 and  Q100. The initial vortex aspect-ratio is $\chi=10$
with { radial} size $b=H/2$. Note  that axisymmetric instability in 3D shearing
boxes require  $\qthree\lesssim 0.2$ { \citep{mamat10}}. Thus all cases
are gravitationally stable in the absence of the vortex. (They
correspond to Toomre parameters $Q\gtrsim 5$.) 


\begin{table*}
  \caption{Summary of simulations. In all runs an  
  initial phase ($t\lesssim 300P_0$) of elliptic instability
  weakens/lengthens the vortex. The subsequent evolution is described below.
  The fiducial runs are in bold and discussed in the text.}   
  \label{sim_summary}
  \begin{tabular}{lrrrrrr}
    \hline
    Run &$\qthree$ & $z_\mathrm{max}$ & $\chi$ &
    $N_x/L_x$  & Vortex evolution \\
    \hline 

    {\bf Q100} &$100$ & $3H$ &10 & $32/H$ &
    \begin{minipage}[t]{0.55\linewidth} 
      Vortex destroyed by initial EI and decays into an axisymmetric density bump. 
      \vspace{.1cm} 
    \end{minipage}\\ 
    
    Q100z2HR &$100$ & $2H$ &10 & $64/H$ &
    \begin{minipage}[t]{0.55\linewidth} 
     Vortex destroyed by the initial EI. 
      \vspace{.1cm} 
    \end{minipage}\\ 

    {\bf Q5} &$5$  & $3H$ &10 & $32/H$ &
     \begin{minipage}[t]{0.55\linewidth}
      Vortex survives until the end of the simulation ($t=2000P_0$) with a turbulent
      core with 3D motions. 
      \vspace{.1cm} 
    \end{minipage}\\

    {\bf Q4} & $4$  & $3H$ &10 & $32/H$&
    \begin{minipage}[t]{0.55\linewidth}  
      Similar to Q5, but the vortex undergoes  
       { an azimuthal} collapse from $t\sim1500P_0$ when
      $\reynolds\gtrsim\gstress$ in the vortex and oscillations in
      $|v_z|$ develops. For $t\gtrsim1700P_0$
      $\mathrm{max}\left|v_z\right|$ appears outside the vortex
      core. An over-dense vortex remains at the end of the simulation.  
      \vspace{.1cm}
    \end{minipage}\\

    Q4HR& $4$  & $3H$ &10 & $64/H$&
    \begin{minipage}[t]{0.55\linewidth}  
     Vortex destroyed by the initial EI. 
      \vspace{.1cm}
    \end{minipage}\\

    Q4z2HR& $4$  & $2H$ &10 & $64/H$&
    \begin{minipage}[t]{0.55\linewidth}  
      Vortex survives with a turbulent core. 
      \vspace{.1cm}
    \end{minipage}\\

    Q4c5 & $4$  & $3H$ & 5 & $32/H$ &
    \begin{minipage}[t]{0.55\linewidth}  
      Vortex destroyed by initial EI. 
      \vspace{.1cm}
    \end{minipage}\\

    Q4c20 &$4$  & $3H$ & 20 & $32/H$ &
    \begin{minipage}[t]{0.55\linewidth}  
      Similar to Q4, but { azimuthal} collapse occurs sooner at $t\sim 1000P_0$.  
      \vspace{.1cm}
    \end{minipage}\\

    {\bf Q3} & $3$ & $3H$ & 10 & $32/H$ &
    \begin{minipage}[t]{0.55\linewidth}
      Similar to $\qthree=4$, but { azimuthal} collapse begins earlier
      ($t\sim 1200P_0$). Vertical motions dominates outside the vortex
      core from $t\sim1400P_0$. Vortex decays after reaching a size of  
      $b\sim H$ at $t\sim 1500P_0$ and inducing spiral shock waves in
      the ambient disc. 
      \vspace{.1cm}
    \end{minipage}\\

    Q3HR& $3$ & $3H$ & 10 & $64/H$ &
    \begin{minipage}[t]{0.55\linewidth}
     Vortex survives with a turbulent core. 
      \vspace{.1cm}
    \end{minipage}\\

    Q3z2HR& $3$ & $2H$ & 10 & $64/H$ &
    \begin{minipage}[t]{0.55\linewidth}
     Vortex survives with a turbulent core. (Simulation { terminated} at $t=1200P_0$.)
      \vspace{.1cm}
    \end{minipage}\\

    \hline
  \end{tabular}
\end{table*}


\subsection{Overview}\label{overview}
\begin{figure}
  \includegraphics[width=\linewidth]{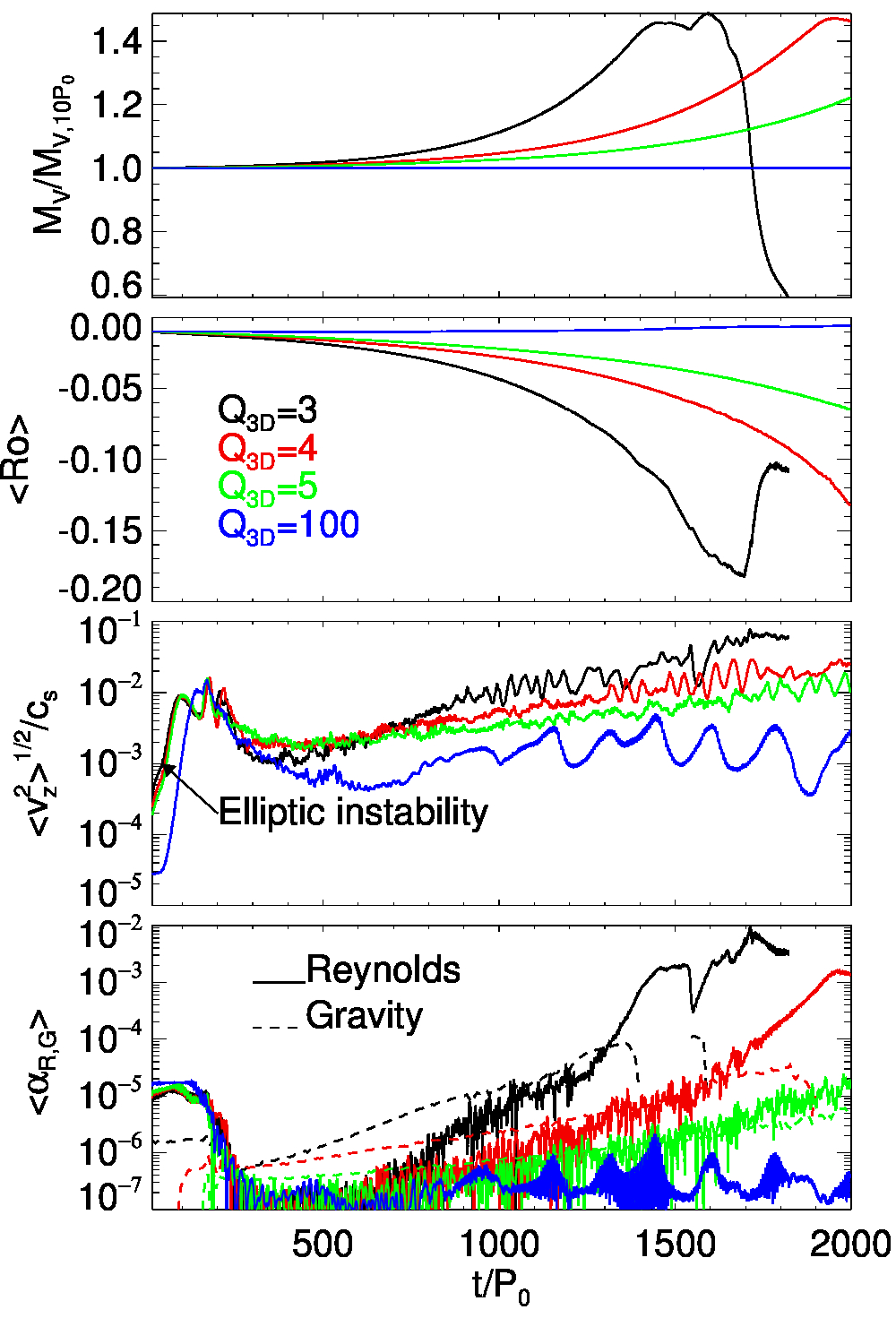}
  \caption{Top to bottom: total mass, Rossby number, vertical Mach
    number,  
    and stresses measured in the sample volume $|x|\leq H/2$ containing the
    vortex for 
    different strengths of self-gravity. These correspond to runs labeled Q3, Q4, Q5, and Q100 in Table \ref{sim_summary}.     
    \label{fiducial_comparison}}
\end{figure}

Fig. \ref{fiducial_comparison} shows the evolution of average mass, Rossby number, 
vertical Mach number, and  stresses  
in the sample volume $x\in[-b,b]$ and time-averaged over $5$
orbits. Without self-gravity ($\qthree = 100$) the vortex decays
by $t\sim 700P_0$, after which vertical motions remain small ($\left|v_z\right| \sim
10^{-3}c_s$) and the average Rossby number roughly constant, as is
$M_\mathcal{V}$. The box returns to a laminar state with $\alpha \sim
10^{-7}$.   

By contrast, the self-gravitating cases with $3\leq\qthree\leq5$
are characterized by an increase in $M_\mathcal{V}$ for $t\gtrsim
300P_0$ with a corresponding { increase in $\left|\ro\right|$, $\left|v_z\right|$, and $\alpha$}. These 
indicate sustained 3D turbulence developed from the elliptic instability as
the vortex spins up due to self-gravitational { growth}. The
self-gravitating vortices survive longer than the non-self-gravitating
case.  
However, the $\qthree=3$ vortex weakens again after becoming too
strong ($t\gtrsim 1500P_0$, see \S\ref{azi_collapse}). 


\begin{figure}
  \includegraphics[width=\linewidth, clip=true, trim=0cm 0cm 0cm 0cm]{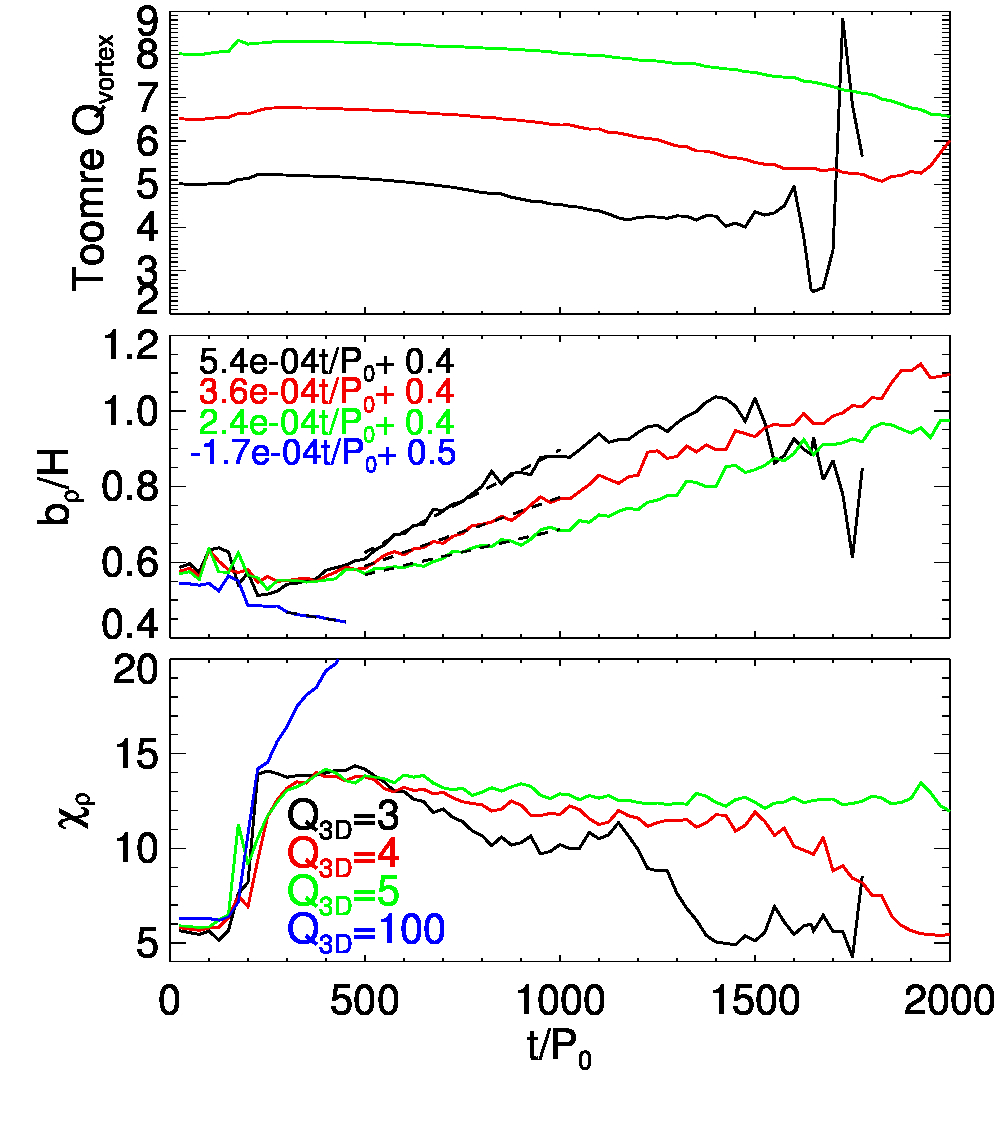}
  \caption{Time evolution of the { Toomre $Q$ parameter at the vortex centre (top)}, 
    { vortex semi-minor axis (middle)} and
    aspect-ratio (bottom). Measurements are based on the vortex density field. { For $\qthree=100$ the
    Toomre $Q \gtrsim 150$ and is therefore not shown.} 
    Dashed lines in the { middle} panel are linear fits to the
    growth in $b_\rho$. 
    \label{fiducial_comparison_dyn}}
\end{figure}

{ Fig. \ref{fiducial_comparison_dyn} shows the evolution of the
  Toomre $Q$ parameter measured at the vortex center, the}
density-based vortex semi-minor axis $b_\rho$ and aspect-ratio $\chi_\rho$. { The Toomre parameter 
  remains $Q>2$, which would suggest self-gravity to be unimportant in the classic sense. Evidently this is not the case: we observe vortex growth even with a Toomre $Q\sim 8$ initially.  
}

The {vortex} aspect-ratio first increases as the elliptic
instability weakens the vortex. Without self-gravity the
vortex eventually decays into a ring. For $\qthree=5$ the aspect-ratio 
is maintained at $\chi_\rho\sim 14$---15. For $\qthree=3,\,4$ the
aspect ratio decreases, first slowly then rapidly 
from $t\sim1200P_0,\, 1500P_0$ for $\qthree=3,\,4$,
respectively. { Notice} this drop coincides with when  
{ self-gravity becomes dominant with}  
$\gstress>\reynolds$. 

\begin{figure}
  \includegraphics[width=\linewidth]{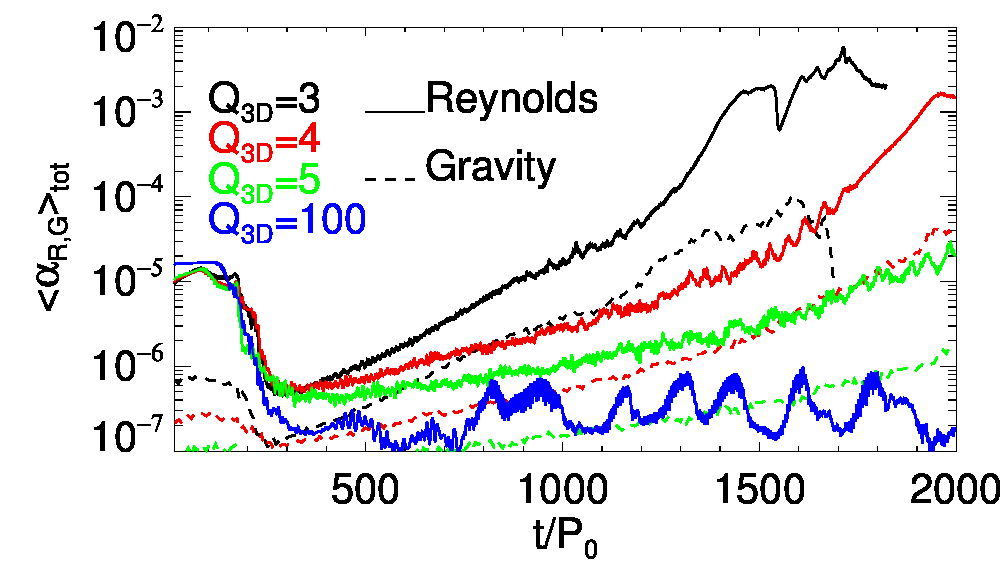}
  \caption{Stresses averaged over the simulation domain for the fiducial runs. 
    \label{alpha_tot}}
\end{figure}

Finally, in Fig. \ref{alpha_tot} we compare the stresses averaged over
the simulation domain excluding buffer zones. These are qualitatively
similar to stresses measured at the box centre, but 
the total stress is always dominated by the Reynolds
contribution. This reflects the fact that all of the discs under
consideration are weakly self-gravitating in the classic sense.  


\subsection{Elliptic instability} 
\begin{figure}
  \includegraphics[scale=.26,clip=true,trim=0cm 1.8cm 0cm 
    0cm]{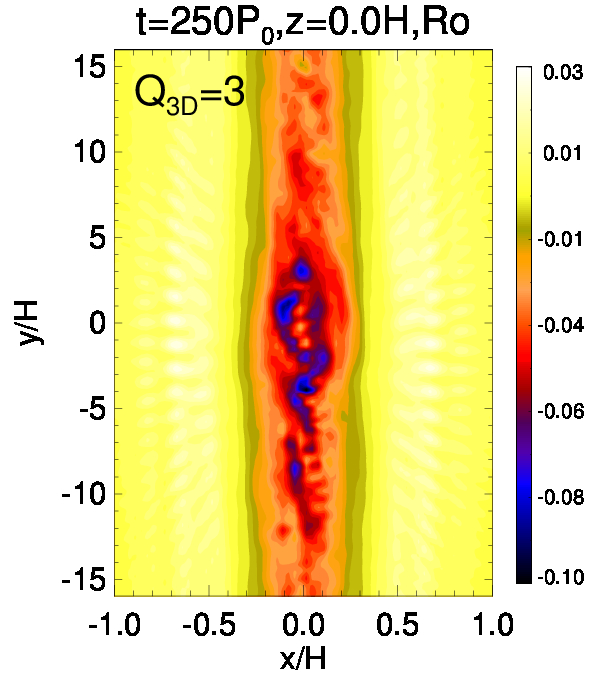}\includegraphics[scale=.26,clip=true,trim=2.3cm 1.8cm 0cm 
    0cm]{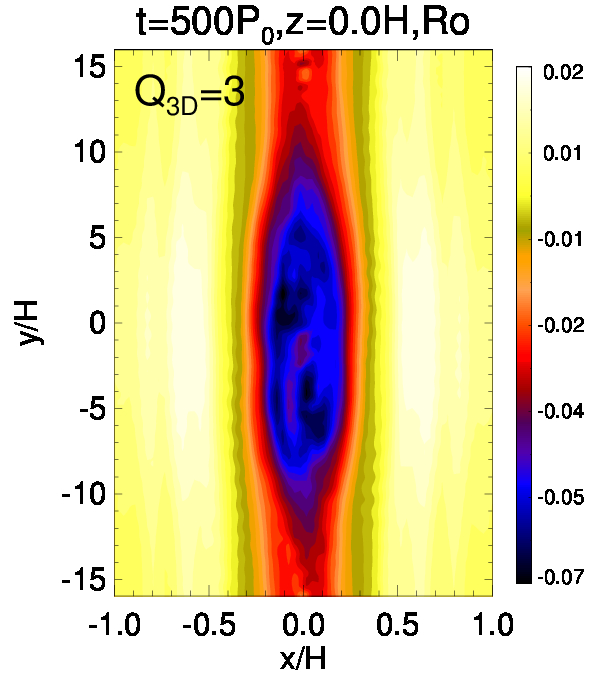}\includegraphics[scale=.26,clip=true,trim=2.3cm 1.8cm 0cm 
    0cm]{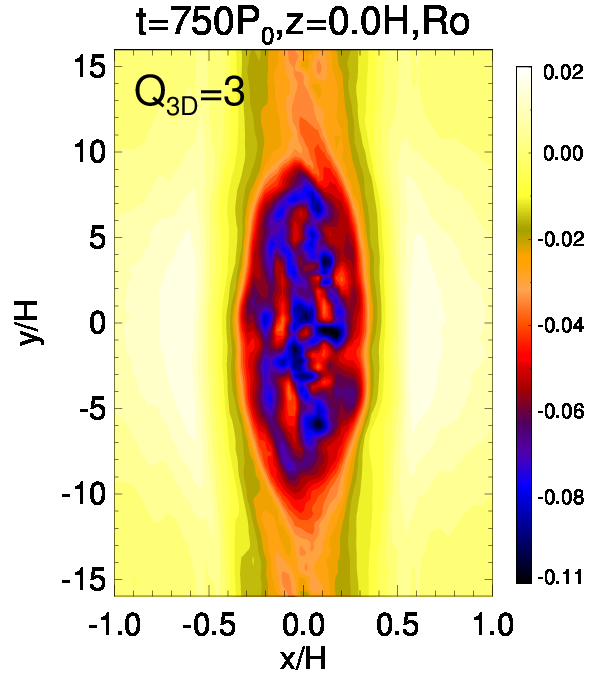}\\
  \includegraphics[scale=.26,clip=true,trim=0cm 0.cm 0cm 
    1cm]{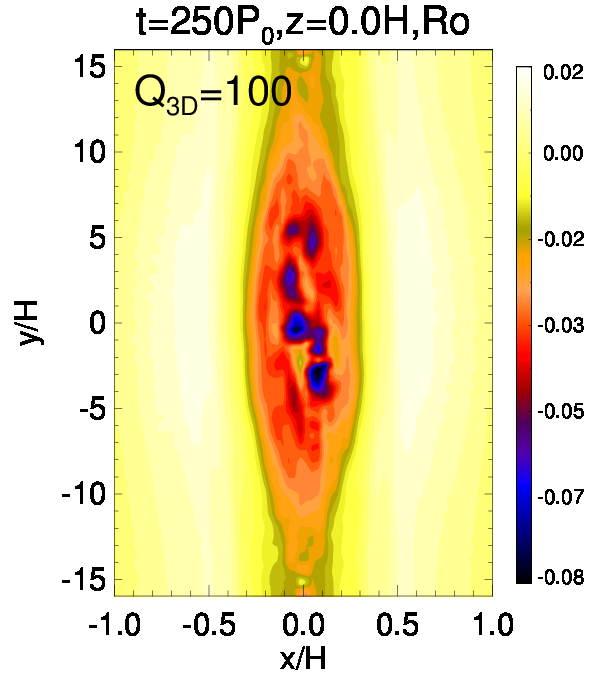}\includegraphics[scale=.26,clip=true,trim=2.3cm 0.cm 0cm 
    1cm]{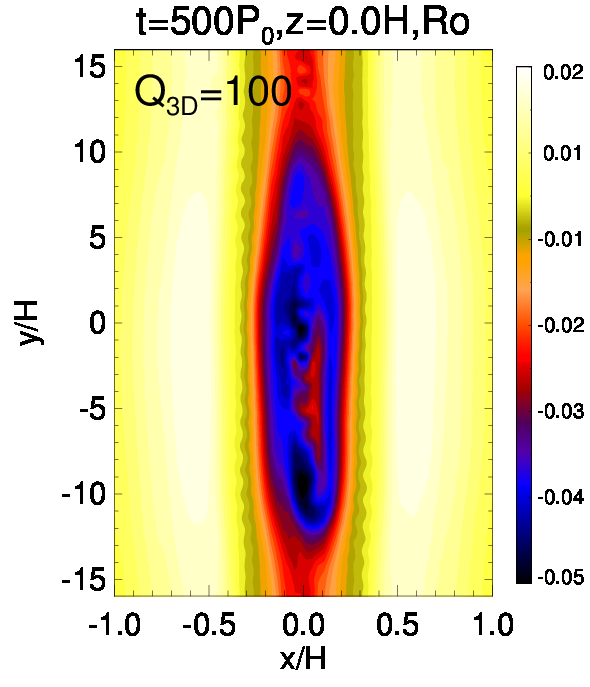}\includegraphics[scale=.26,clip=true,trim=2.3cm 0.cm 0cm 
    1cm]{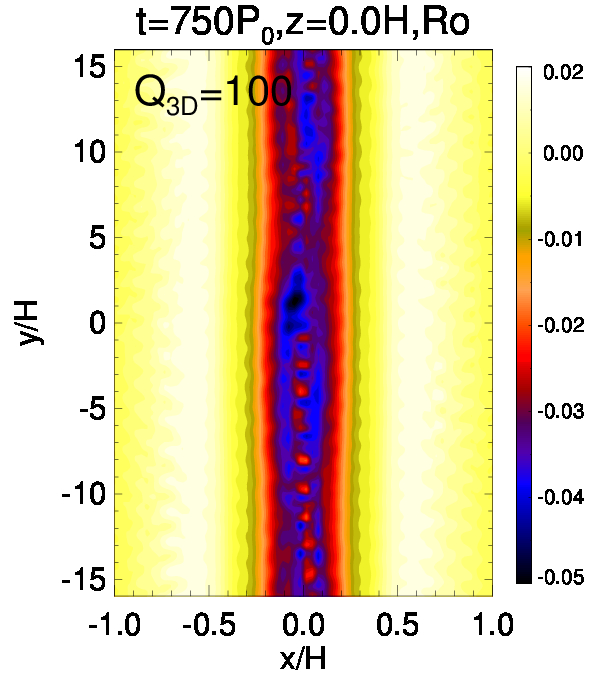}
  \caption{Hydrodynamic turbulence due to the elliptic instability of
    a 3D vortex in a self-gravitating disc (upper panels), and in a 
    non-self-gravitating disc (lower panels). Comparing the upper and
    lower panels show that hydrodynamic turbulence can be sustained
    (and grows) in the
    presence of self-gravity. Colors 
    show the { Rossby number} $\ro =
    \left(\nabla\times\bm{v}\right)/2\Omega + q/2$ at $z=0$. 
    \label{ei_vz}
    }
\end{figure}

We find early vortex evolution { and weakening} ($t\lesssim 300P_0$) is driven by 
elliptic instability (EI) and independent of
self-gravity\footnote{{ However, adjustment to the background Keplerian shear may also have minor contributions to the early vortex evolution.}}.
{ The EI} is signified by the exponential increase in $\sqrt{\avg{v_z^2}}$ in
$t\lesssim 100P_0$ with growth rate   
$\gamma\simeq 0.01\Omega$ and saturating at
$\sqrt{\avg{v_z^2}}\sim 0.01c_s$ in all cases. This is consistent with 
previous non-self-gravitating studies of the EI{, and the} 
weak dependence on self-gravity is expected since the EI is  
incompressible \citep{lesur09}. The left panels in Fig. \ref{ei_vz} show
the result of the EI is small-scale hydrodynamic turbulence. The
subsequent evolution then depends on self-gravity. 

\subsection{Secular 
  vortex growth and sustained turbulence}\label{stress-collapse}


{
  Fig. \ref{ei_vz} compares the secular evolution of the vortex after
  the initial EI. In both self-gravitating and non-self-gravitating
  cases, the initial EI results in an irregular vorticity field
  ($t=250P_0$). However, it then reorganises into a larger, coherent
  vortical structure by $t=500P_0$, which serve as initial conditions for secular evolution. 
  This initial vortex `revival' may be due
  to the Kelvin-Helmholtz instability of vorticity strips that result
  from the shearing of vorticity seeds by the background shear flow
  left by the initial EI \citep{lithwick07}. 
  This second-generation vortex then undergoes EI again. However, 
  in the non-self-gravitating case 
  the vortex simply decays: we do not observe the reemergence of a
  large-scale coherent vortex due to \citeauthor{lithwick07}'s mechanism.   
}

{ By contrast, we find that in self-gravitating discs the large-scale vortex persists with 
small-scale 3D hydrodynamic turbulence in its core (Fig. \ref{ei_vz}, upper panel). The vortex undergoes secular 
growth from} $t\sim 
300P_0$ to the end of the simulation for $\qthree=5$, but only to
$t\sim 1500P_0,\, 1200P_0$ for $\qthree=4,\,3$,
respectively. (In the latter cases the vortex { then undergoes an
  azimuthal} 
collapse, see \S\ref{azi_collapse}.) Stresses $\avg{\alpha}$ and vertical velocity
fluctuations $\sqrt{\avg{v_z^2}}$ grow exponentially, but with 
growth rate $\sim2$---$9\times10^{-4}\Omega$, significantly slower than the
initial EI because of the larger vortex aspect ratios { than those imposed initially} \citep{lesur09}. { However, notice 
$\sqrt{\avg{v_z^2}}$ grows faster with decreasing $\qthree$. This indicates stronger EI, which is due to smaller vortex aspect-ratios with increasing self-gravity (see Fig. \ref{fiducial_comparison_dyn})}.  

\begin{figure}
  \includegraphics[scale=.39,clip=true,trim=0cm 1.8cm 0cm
    0cm]{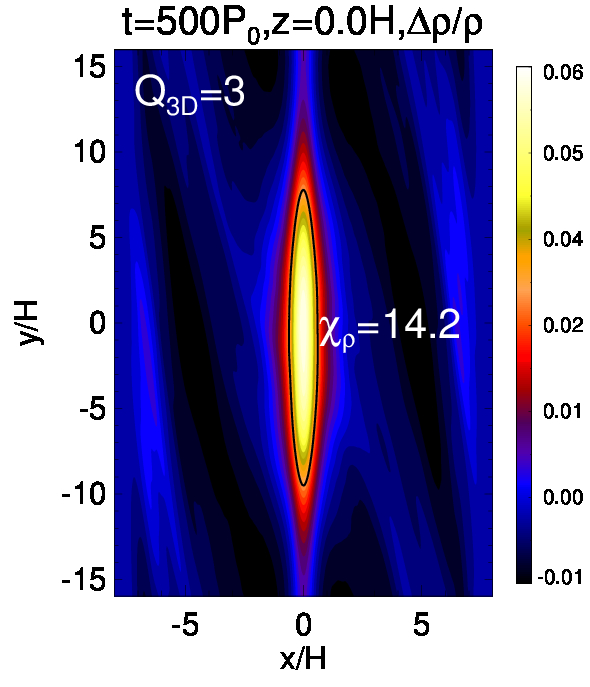}\includegraphics[scale=.39,clip=true,trim=2.3cm 1.8cm
    0cm 0cm]{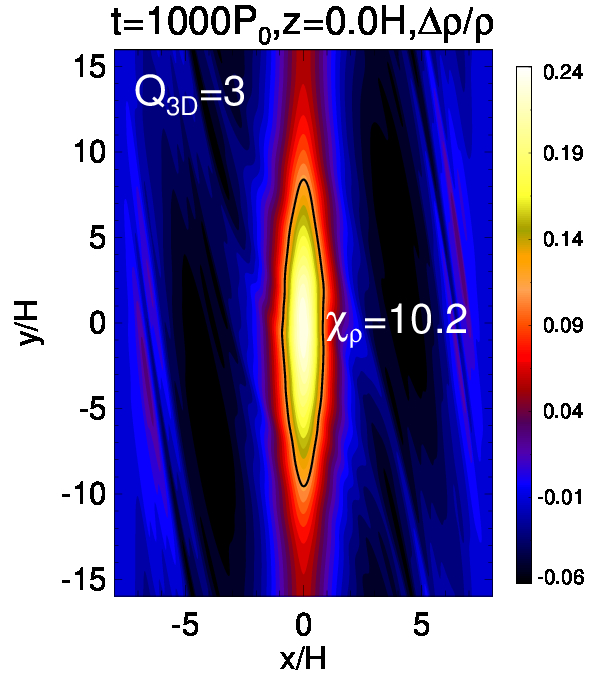}\\
  \includegraphics[scale=.39,clip=true,trim=0cm 0cm 0cm
    1cm]{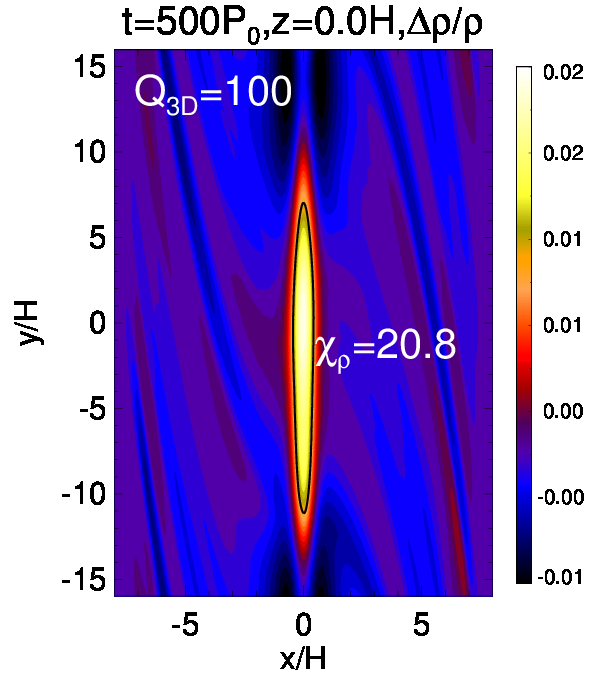}\includegraphics[scale=.39,clip=true,trim=2.3cm 0.0cm
    0cm 1cm]{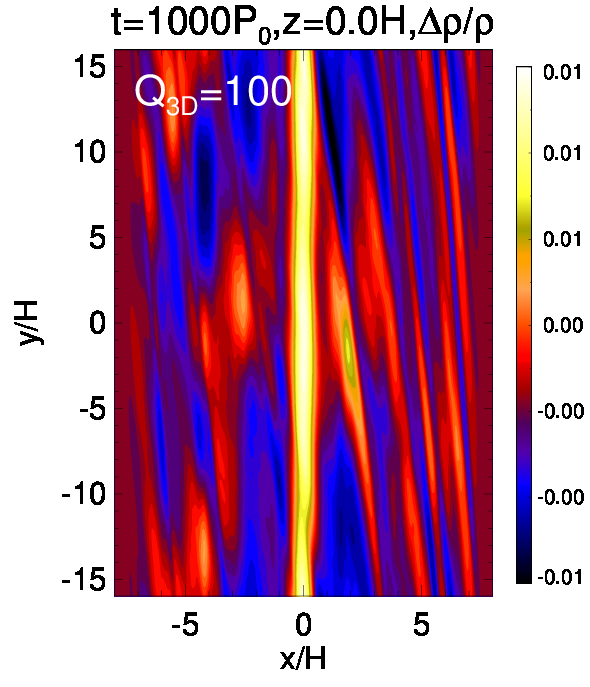}
  \caption{Secular vortex evolution, after the
    initial development of { elliptic instability} and corresponding hydrodynamic
    turbulence. Top: a self-gravitating disc. Bottom:
    non-self-gravitating disc. Colors show the relative density
    perturbation at the midplane. \label{secular_collapse}
    }
\end{figure}

Fig. \ref{secular_collapse} shows the evolution of the density field 
during secular { growth}. The self-gravitating vortex increases in
density and size, but with decreasing aspect-ratio. Notice the density field remains 
smooth even though the vorticity field is turbulent. This reflects the
incompressible nature of the EI. 
{ By contrast, the non-self-gravitating vortex is 
destroyed by the initial EI, leaving an axisymmetric density ring.}

{
  We note that during secular growth the monotonic increase in the vortex mass, rotation, and semi-minor axis 
  (Fig. \ref{fiducial_comparison}---\ref{fiducial_comparison_dyn}) in 
  the self-gravitating cases proceed  
  more rapidly than the decrease in its aspect-ratio. (For $\qthree=5$ the vortex aspect-ratio is nearly constant.)
  A possible reason 
   is that vertical stratification due to self-gravity (see
  Fig. \ref{qss_vortex_vdens}) makes the secular dynamics more similar
  to that of 2D vortices \citep{umurhan04,johnson05,shen06}. These
  studies show that 2D vortices 
  tend to merge into larger vortices \citep{godon99}. Though merging
  cannot occur for our single, isolated  vortex; secular evolution
  towards larger scales can be expected if the system behaves
  two-dimensionally due to the inverse cascade of energy
  \citep{kraichnan80}.  However, the additional stratification brought
  about by self-gravity is small; thus this effect is unlikely 
  important for the present $\qthree$ values. Here, the vortex growth in
  its radial size may be simply due to mass accretion.   
}

\subsection{Episodic bursts}\label{episodic_burst}

\begin{figure}
  \includegraphics[scale=.21,clip=true,trim=0.3cm 1.8cm
    0.15cm 0cm]{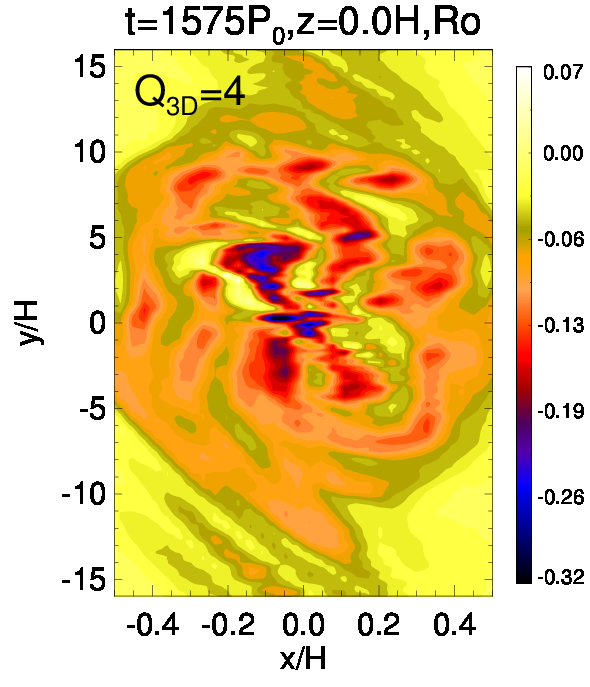}\includegraphics[scale=.21,clip=true,trim=2.3cm 1.8cm
    0.15cm 0cm]{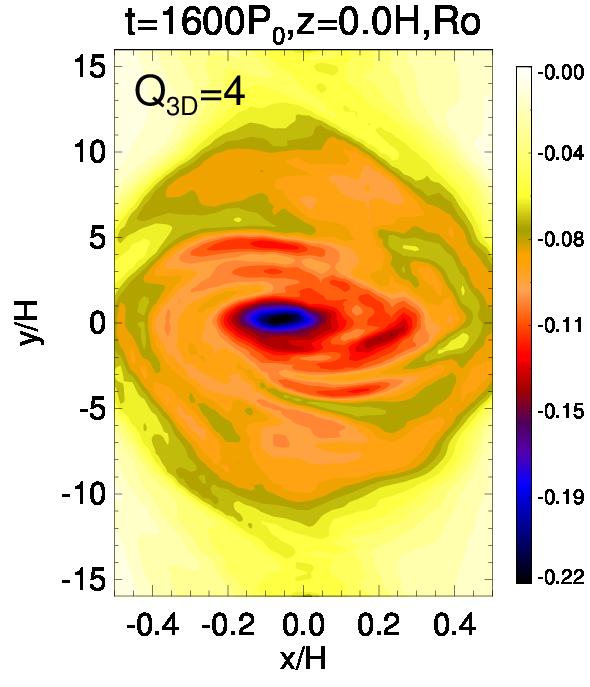}\includegraphics[scale=.21,clip=true,trim=2.3cm 1.8cm
    0.15cm 0cm]{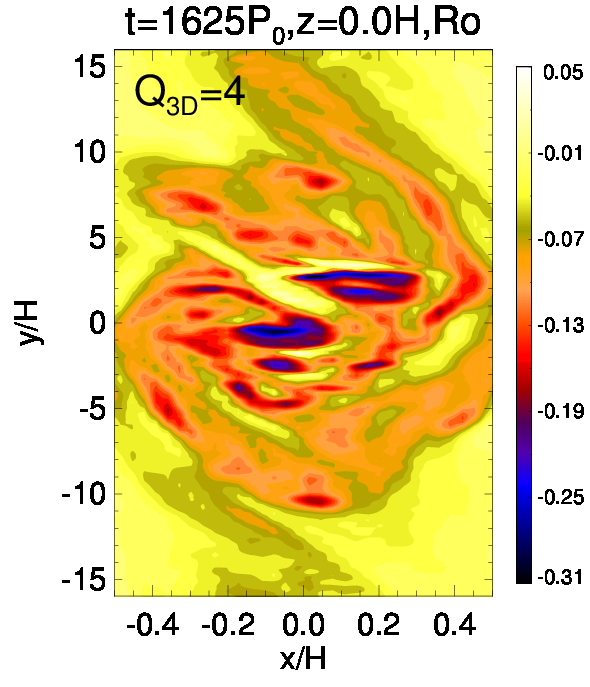}\includegraphics[scale=.21,clip=true,trim=2.3cm 1.8cm
    0.15cm 0cm]{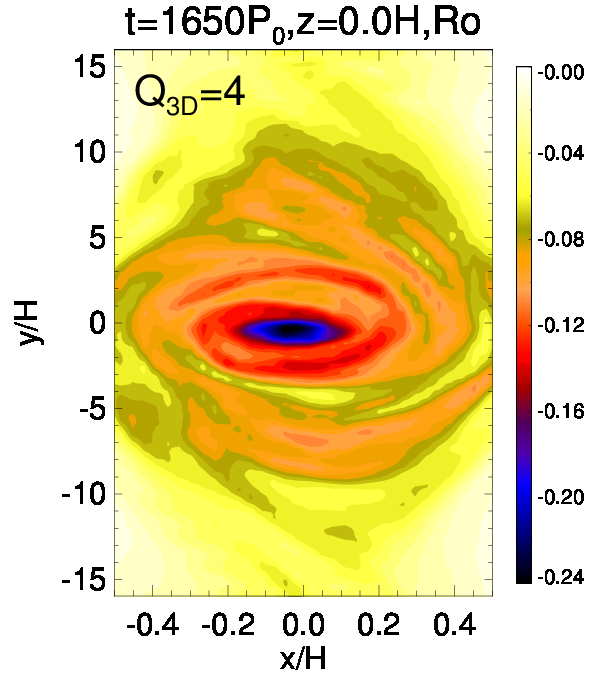}\\
     \includegraphics[scale=.21,clip=true,trim=0.3cm 0cm
     0.15cm 0cm]{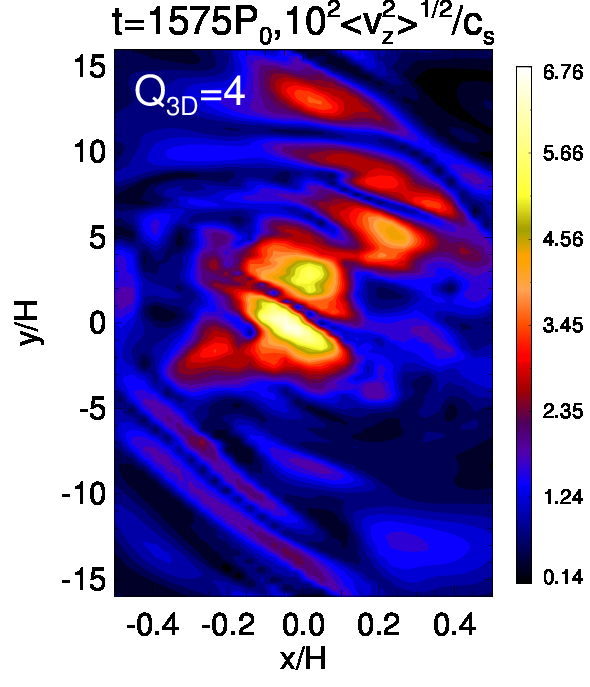}\includegraphics[scale=.21,clip=true,trim=2.3cm 0.cm
     0.15cm 0cm]{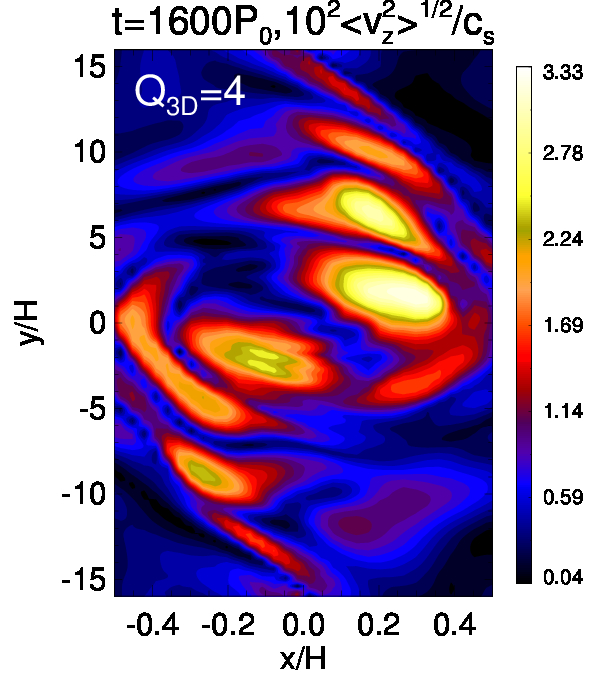}\includegraphics[scale=.21,clip=true,trim=2.3cm 0cm
     0.15cm 0cm]{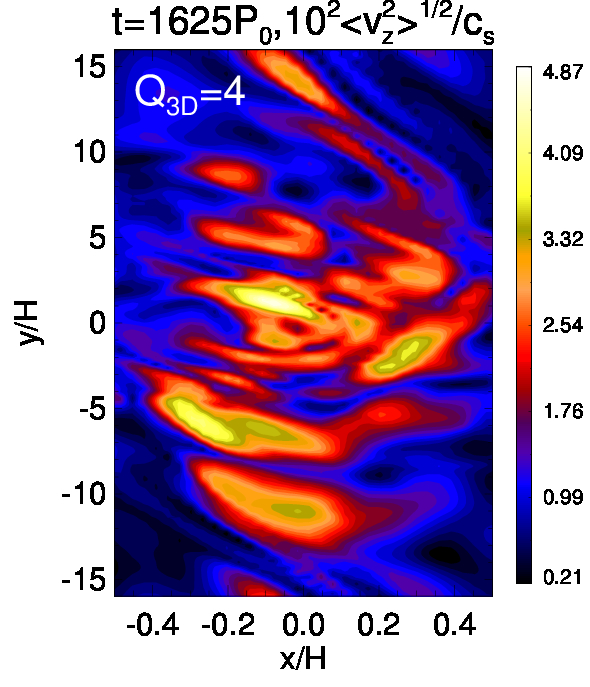}\includegraphics[scale=.21,clip=true,trim=2.3cm 0cm
     0.15cm 0cm]{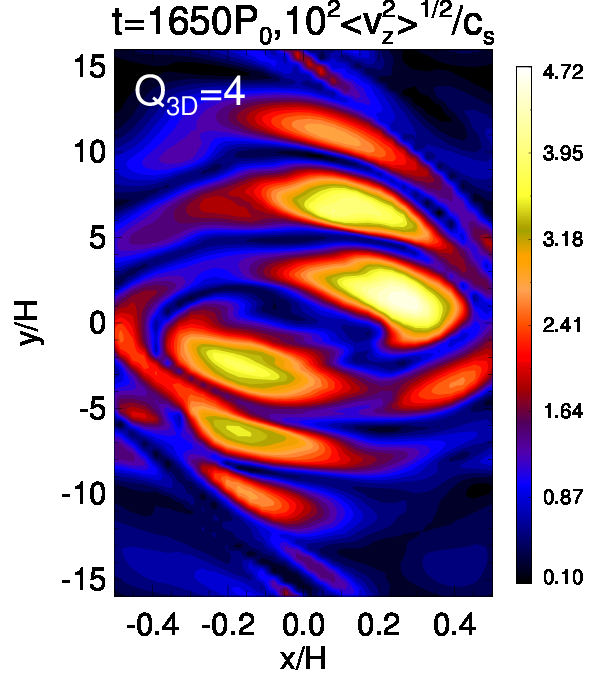}
  \caption{Example of episodic elliptic instability in a 
    self-gravitating 3D vortex. Top: Rossby number. Bottom: vertical
    Mach number. 
    \label{ei_burst} 
    }
\end{figure}

{ We find secular} vortex growth ends when $\reynolds\sim\gstress\sim10^{-5}$, at which 
point we observe \emph{episodic bursts} of EI in self-gravitating
cases. { This is reflected in Fig. \ref{fiducial_comparison} as
  oscillations in $\sqrt{\avg{v_z^2}}$, starting from} $t\sim 1000P_0, 1500P_0$ for $\qthree=3,4$, 
respectively; and { it} only just begins to develop at the end of the 
$\qthree=5$ simulation. 
Fig. \ref{ei_burst} show snapshots of this
episodic behaviour for $\qthree = 4$ { (where the simulation output
  frequency conveniently coincides with the burst.)} 
The vortex oscillates between 
having small-scale turbulence ($t=1575P_0,\,1625P_0$) and a coherent
patch of negative vorticity ($t=1600P_0,\,1650P_0$).  { Notice in
 that $\mathrm{max}\left|\ro\right|$ also
  oscillates. }  

{ We comment that the oscillations are not reflected by the Rossby number in
  Fig. \ref{fiducial_comparison} 
  because $\ro$ has been averaged over $x\in[-b,b]$, which 
  demonstrates that overall the vortex continues to strengthen. Nevertheless, we checked that $|\ro|$ does oscillate
  during the burst phase when the sampling volume is limited to near 
  the vortex centre. 
}   
 
{ This episodic behavior suggests competition between vortex growth due to self-gravity 
and vortex destruction by the EI}. 
{ As the vortex grows, it spins up ($|\ro|$ increases) and its aspect-ratio decreases.  
  This feeds the EI, which converts some of the vortex's large-scale,
  planar rotation into small-scale 3D turbulence (signified by an
  increasing $|v_z|$). The EI activity decays, but is
  re-triggered once the vortex reforms as it grows via
  self-gravity. This growth may be partly attributable to 
  the removal of the vortex's internal
  rotation by the EI and enabling secular gravitational instabilities (see \S\ref{gravito-elliptic}).
}

A similar bursty behaviour
has been reported in 3D simulations of vortex amplification by the
sub-critical baroclinic instability in competition with EI \citep[][
  their Fig. 16]{lesur10}. We discuss this analogy further in
\S\ref{sbi}.   


\subsection{Azimuthal collapse, gap opening, and self-destruction}\label{azi_collapse}
{ As Reynolds stresses ($\reynolds$) increases beyond the
  gravitational stresses ($\gstress$), the vortex in the 
  $\qthree=3,\, 4$ discs enter a second, faster collapse 
 phase where its aspect-ratio rapidly drops, although its radial size
 continues to grow, see Fig. \ref{fiducial_comparison}---\ref{fiducial_comparison_dyn}. 
This azimuthal collapse phase corresponds to $t \in [1200,1500]P_0$ and $t\in[1500,2000]P_0$ for $\qthree=3,\,4$,
respectively. The total stress also increases rapidly (see Fig. \ref{alpha_tot}).}  

\begin{figure}
  \includegraphics[scale=.27,clip=true,trim=0.3cm 1.8cm
    0cm 0cm]{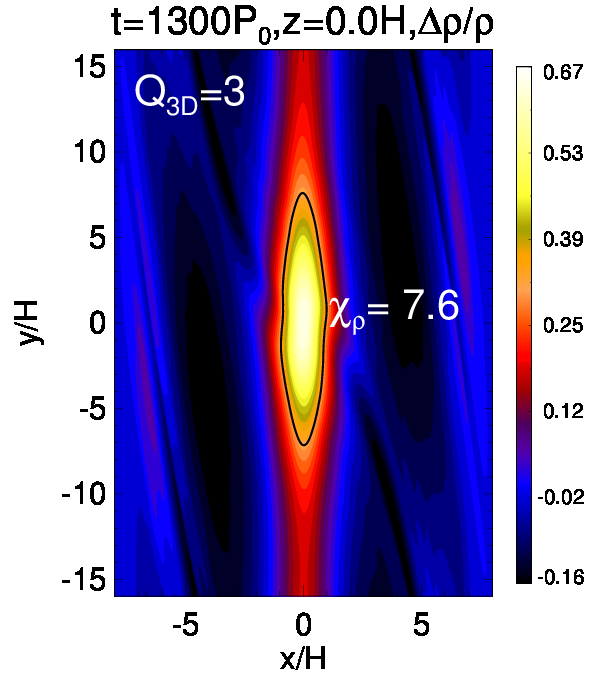}\includegraphics[scale=.27,clip=true,trim=2.3cm 1.8cm
    0cm 0cm]{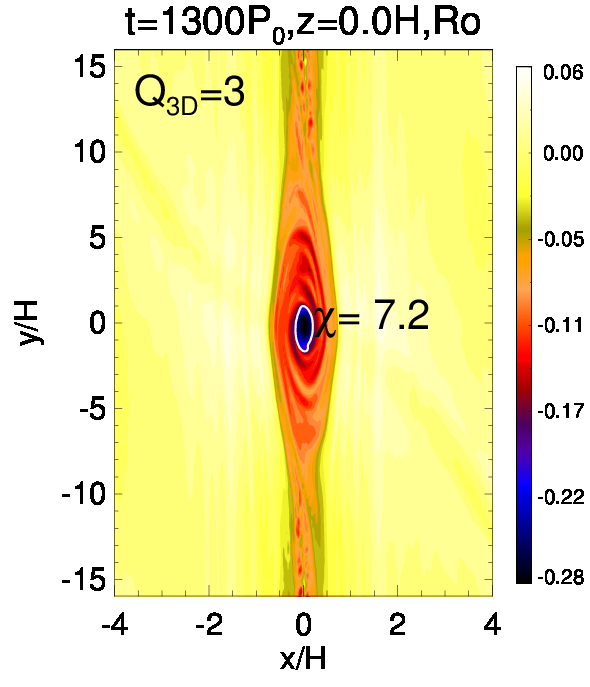}\includegraphics[scale=.27,clip=true,trim=2.3cm 1.8cm
    0cm 0cm]{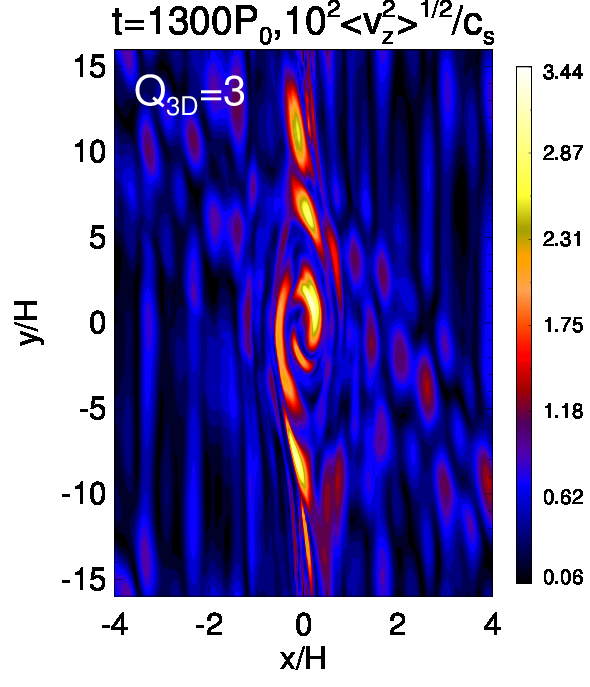}\\
   \includegraphics[scale=.27,clip=true,trim=0.3cm 0cm
    0cm 0cm]{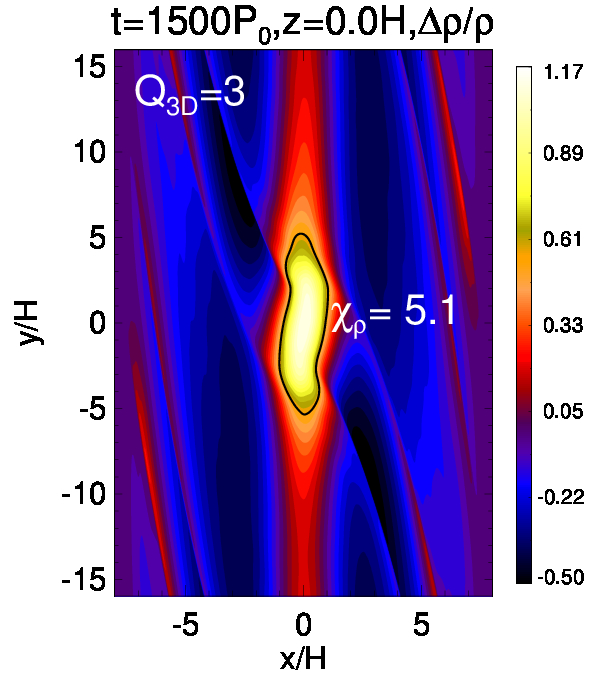}\includegraphics[scale=.27,clip=true,trim=2.3cm 0cm
    0cm 0cm]{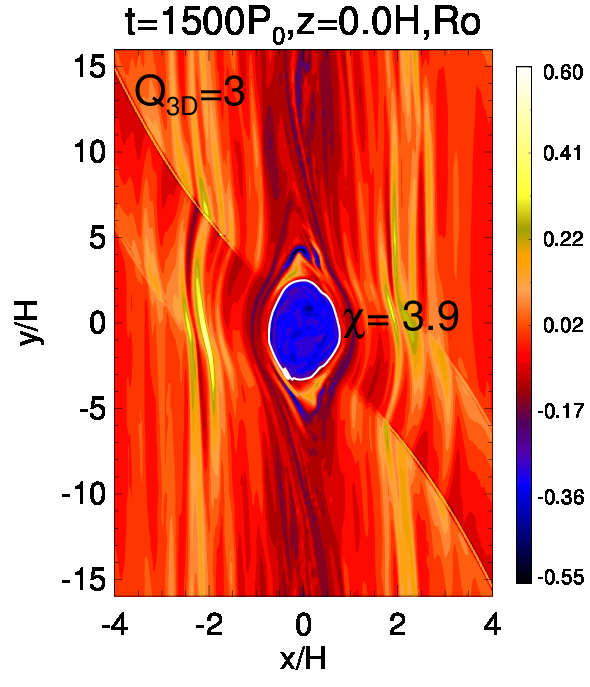}\includegraphics[scale=.27,clip=true,trim=2.3cm 0cm
    0cm 0cm]{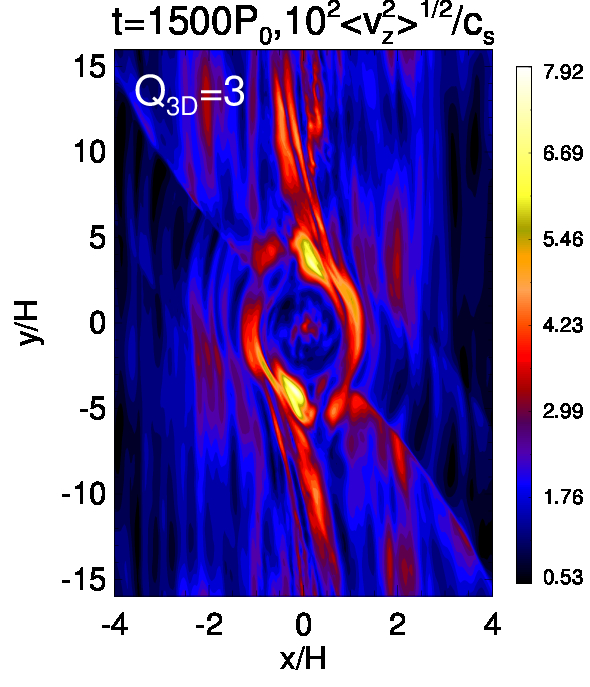}\\
  \caption{{ Evolution} of the $\qthree=3$ vortex during { 
      azimuthal collapse, when its aspect-ratio drops rapidly.}  Left: midplane density perturbation. 
    Middle: midplane Rossby number. Right: average vertical Mach 
    number. { Here, ellipses in $\ro$ are drawn by inspection, and correspond to the loci $\ro = 0.65\mathrm{min}(\ro)$ and $\ro=0.3\mathrm{min}(\ro)$ in the top and bottom snapshots, respectively. } 
    \label{second_collapse} 
    }
\end{figure}

Fig. \ref{second_collapse} shows { azimuthal} collapse for 
the { $\qthree=3$} simulation. The vortex exits the bursty phase with a
coherent vorticity patch of aspect-ratio $\chi\sim 7$ (upper middle
panel). {
 Notice $\mathrm{max}|v_z|$ moves outside the vortex core as its
 aspect-ratio decreases (lower right
 panel).  
}
This is in fact consistent with \cite{lesur09}, who find
the EI occurs outside the vortex core for { an intermediate range
  of aspect-ratios} 
in the absence of buoyancy, as is the case for isothermal gas
considered here. { In their specific case of Kida
  vortices \citep{kida81}, this range is $ 4\lesssim \chi\lesssim 6$,
 but the corresponding range may differ for the vortices in our simulations.}   

{
  We find azimuthal collapse is associated with vortex-driven spiral
  density waves. Spiral density waves are naturally excited by vortex
  modes in compressible gas \citep{bodo05,heinemann09,mamat07}. However, as the vortex
  half-width approaches $H$, these spiral density waves steepen into
shocks \citep{paardekooper10} and transfer angular momentum and energy
away from the vortex \citep{bodo07}. 
These spiral shocks can also be identified visually in Fig. \ref{second_collapse} at $t=1500P_0$ in both the
density and vorticity maps.

Azimuthal collapse is  terminated once the vortex
half-width exceeds $\sim H$.  This is seen in Fig. \ref{fiducial_comparison} as the 
rapid drop in $|\ro|$ towards the end of the simulation with
$\qthree=3$. The 
corresponding decrease in $M_\mathcal{V}$ indicate mass loss by the
vortex to the ambient disc.  
This final vortex weakening is likely related to the strong spiral shocks induced
by the vortex once it becomes too large in the radial direction.   
We observe a similar self-destruction behaviour as above in an extended run of the $\qthree=4$ case. 
However, a proper study of this final stage require global
simulations to minimize the influence of radial boundary conditions. We thus
limit our discussion here to when $b_\rho\lesssim H$. 
(See Appendix \ref{2d_sims} for examples of vortex evolution in 2D, global discs.)  

Nevertheless, our observation that vortex radial size being limited 
by the associated density waves is consistent with previous studies of
vortex evolution in 2D. The critical radial extent is expected to be $\sim H$ \citep{godon99}, beyond which the vortex flow becomes sonic \citep{paardekooper10}.
This is equivalent to the Jeans length $\lambda_\mathrm{J}\sim O(QH)$ in a  self-gravitating disc with $Q\sim 1$. Indeed, \cite{mamat09} suggest 2D vortices are limited to $\lambda_\mathrm{J}$ in their radial extent.   
They find in gravito-turbulent discs that upon reaching $\lambda_\mathrm{J}$, 2D vortices weaken and are sheared away by the background flow \citep[see also]{lyra09}.  
In our discs with $Q>1$, $\lambda_\mathrm{J}$ is somewhat larger than 
$H$. The vortex width limit in our 3D  simulations is thus similar to
that in corresponding 2D, non-self-gravitating discs.}

{
Notice in Fig. \ref{second_collapse} that two shallow gaps have
developed on either side of the vortex. Similar double-gap structures appear in low mass planet-disc 
simulations \citep[e.g.][]{zhu13, dong17}. This suggests that the vortex can interact with the ambient disc 
in a similar way to disc-planet interaction. 
} 
The strength of gravitational interaction between the vortex and the ambient disc
can be estimated via the total mass $M_\mathrm{pert}$ enclosed in the 
elliptical patch $(x^2 + y^2/\chi_\rho^2)/b_\rho^2\lesssim 1$: 
\begin{align}
  q_\mathrm{v}\equiv \frac{M_\mathrm{pert}}{M_*h^3} =
  \frac{\chi_\rho}{Q}
  \left(\frac{b_\rho}{H}\right)^2, 
\end{align} 
where $M_*$ is the mass of the central star and $h\equiv H/R$ is the
characteristic disc aspect-ratio at the fiducial radius { of the
  shearing box. Using Fig. \ref{fiducial_comparison_dyn} we find $q_\mathrm{v}\sim 1.3$ 
for the $\qthree=3$ simulation at $t=1500P_0$. 
}
This can be compared to the 
thermal condition for gap-opening by disc-planet interaction, 
$q_\mathrm{v}\gtrsim 1$ \citep{crida06}. Thus gap-opening by the 
vortex can be { indeed be} expected. 
{ However, we expect gap-opening by an elliptical, extended vortex
  to be weaker than by a point-mass planet of the same mass.}

The vortex-induced spiral shocks may also contributed to mass 
accretion onto the vortex, similar to shock-driven accretion in circumplanetary
discs \citep{zhu16b}. Both 
\begin{inparaenum}[1)]
\item
  the increasing vortex mass; and  
\item
  the decreasing vortex aspect-ratio 
\end{inparaenum}
are expected to enhance the vortex-driven spiral density waves. Thus
{ azimuthal} collapse proceeds more rapidly than the secular growth, when
the associated spiral waves are { weak}. Turbulence outside the vortex core may also 
contribute to angular momentum removal and hence mass accretion by 
the vortex.

\subsection{Vortex aspect-ratio, numerical resolution, and box height}\label{param_study}
We performed several variations of the fiducial runs
described above. These are summarized in Table  \ref{sim_summary} and
include different aspect-ratios for the initial vortex perturbation
($\chi$), the radial
resolution ($L_x/N_x$), and the vertical domain size ($z_\mathrm{max}$),

For the $\qthree=4$ { disc} we ran additional cases with $\chi=20$ and
$\chi=5$ for the initial vortex perturbation.
The former behaves similarly to our fiducial run with 
$\chi=10$. However, for $\chi=5$ the vortex is destroyed 
by the initial phase of EI, similar to the non-self-gravitating case 
$Q100$. This suggests that self-gravity cannot save vortices { born} with
small aspect-ratios.  

\begin{figure}
\includegraphics[width=\linewidth]{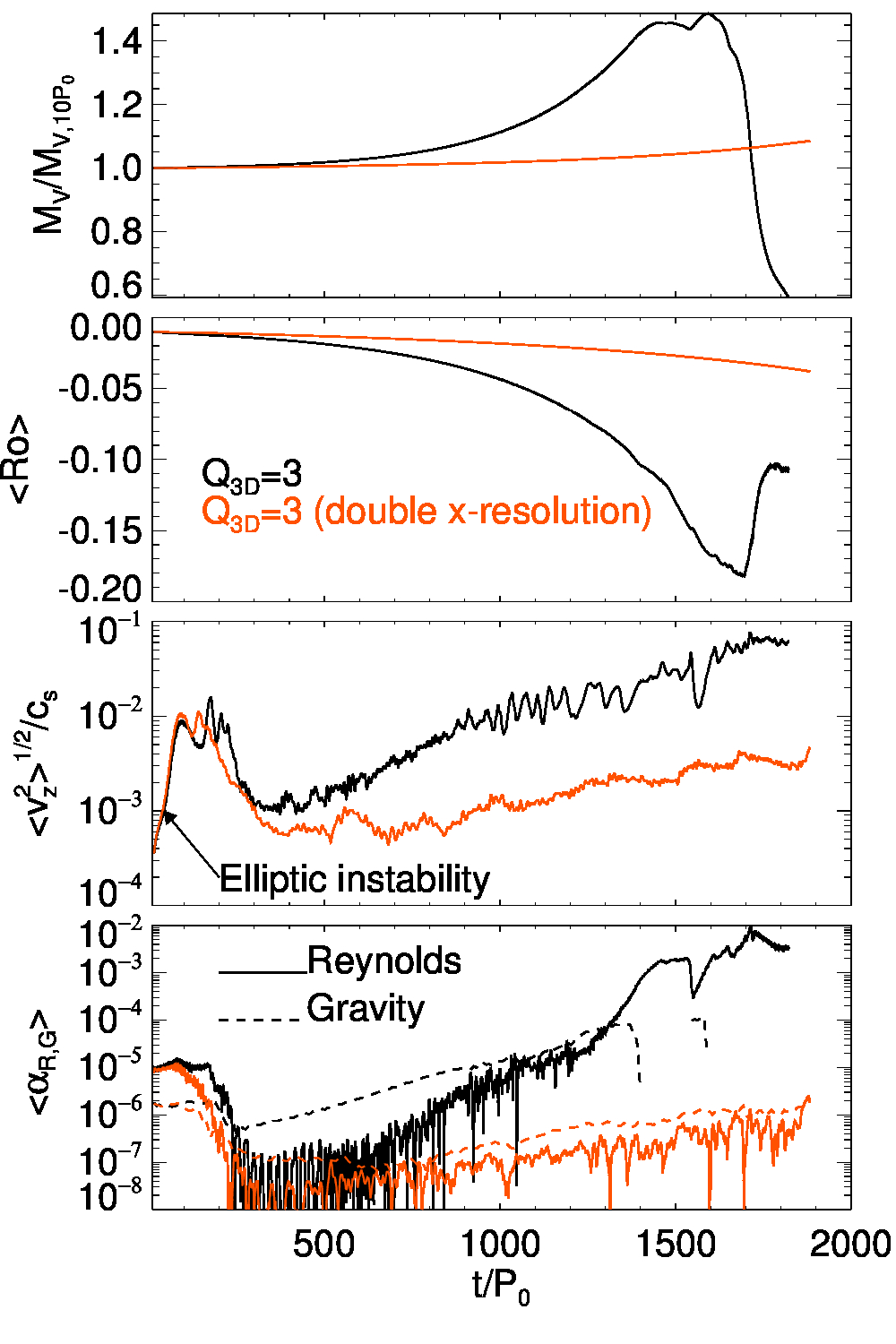}
\caption{{ Comparison between vortex evolution in a $\qthree=3$ disc, but different 
  radial resolutions: $N_x/L_x =  32/H$ (black) and $64/H$
  (orange). The vortex persists in both simulations.}
\label{athevol2_HR}}
\end{figure}

\begin{figure}
  \includegraphics[scale=.4,clip=true,trim=0.5cm 1.8cm 0cm 
    0cm]{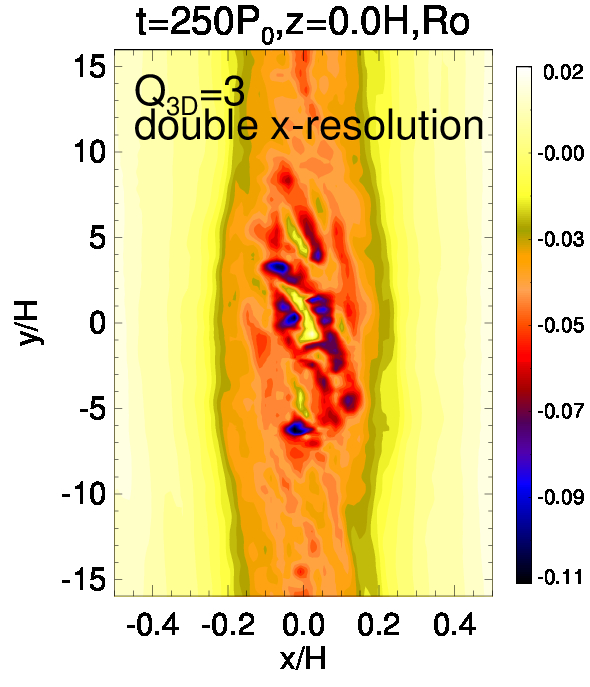}\includegraphics[scale=.4,clip=true,trim=2.3cm 1.8cm 0cm 
    0cm]{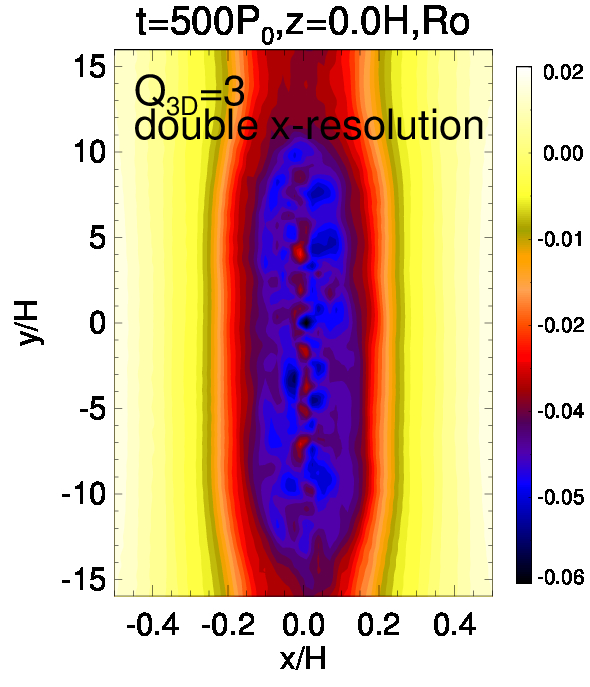}\\\includegraphics[scale=.4,clip=true,trim=.5cm 0.cm 0cm 
    0cm]{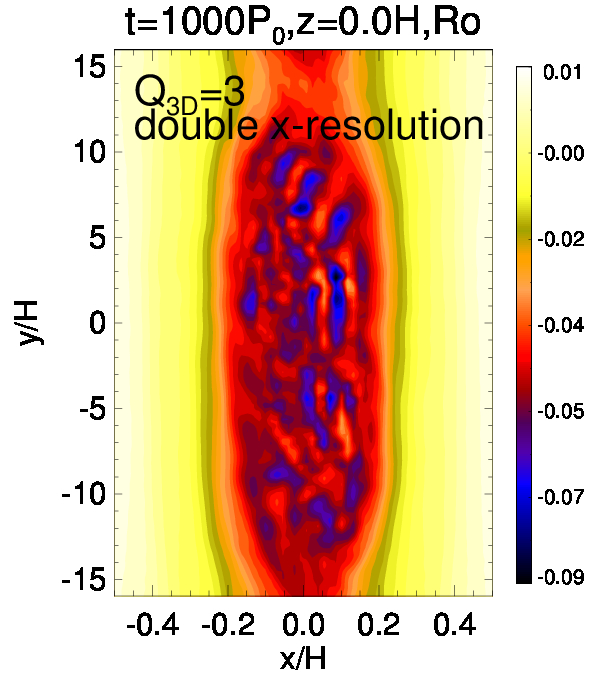}\includegraphics[scale=.4,clip=true,trim=2.3cm 0.cm 0cm 
    0cm]{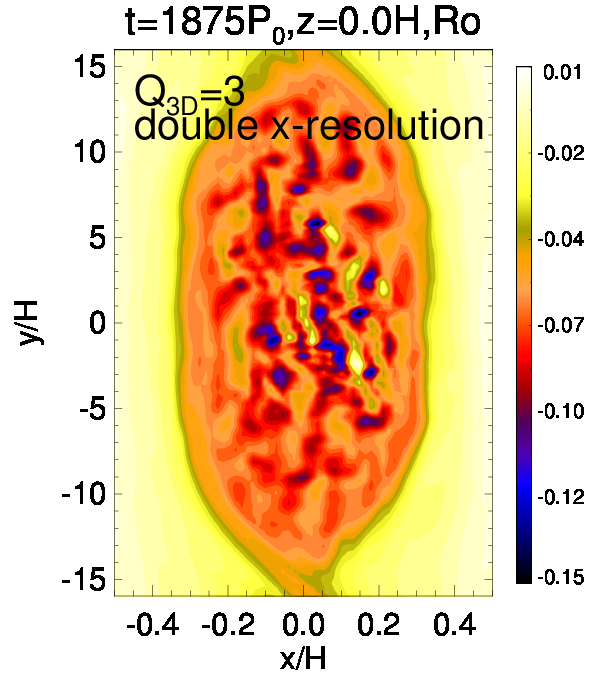}
\caption{{ Vortex evolution in the $\qthree=3$ disc but with
  double the radial resolution than the fiducial runs.}
\label{HR_sims_Q3}}
\end{figure}

{ We also repeated some runs with twice the radial
  resolution. These are compared in Fig. \ref{athevol2_HR} for the
  $\qthree=3$ disc. We find the initial EI growth rate is converged,
  but with better resolution the resulting turbulence leads to a
  weaker vortex to serve as initial conditions for secular growth. 
  This is shown in Fig. \ref{HR_sims_Q3} as the more elongated vortex at $t=500P_0$ compared to that in Fig. \ref{ei_vz}. 
  Since the EI turbulence is better resolved, it has a more
  destructive effect and the subsequent secular growth rate is slower and 
  a 3D vortex with a turbulent core persists until the end of the simulation.  
 }

\begin{figure}
\includegraphics[width=\linewidth]{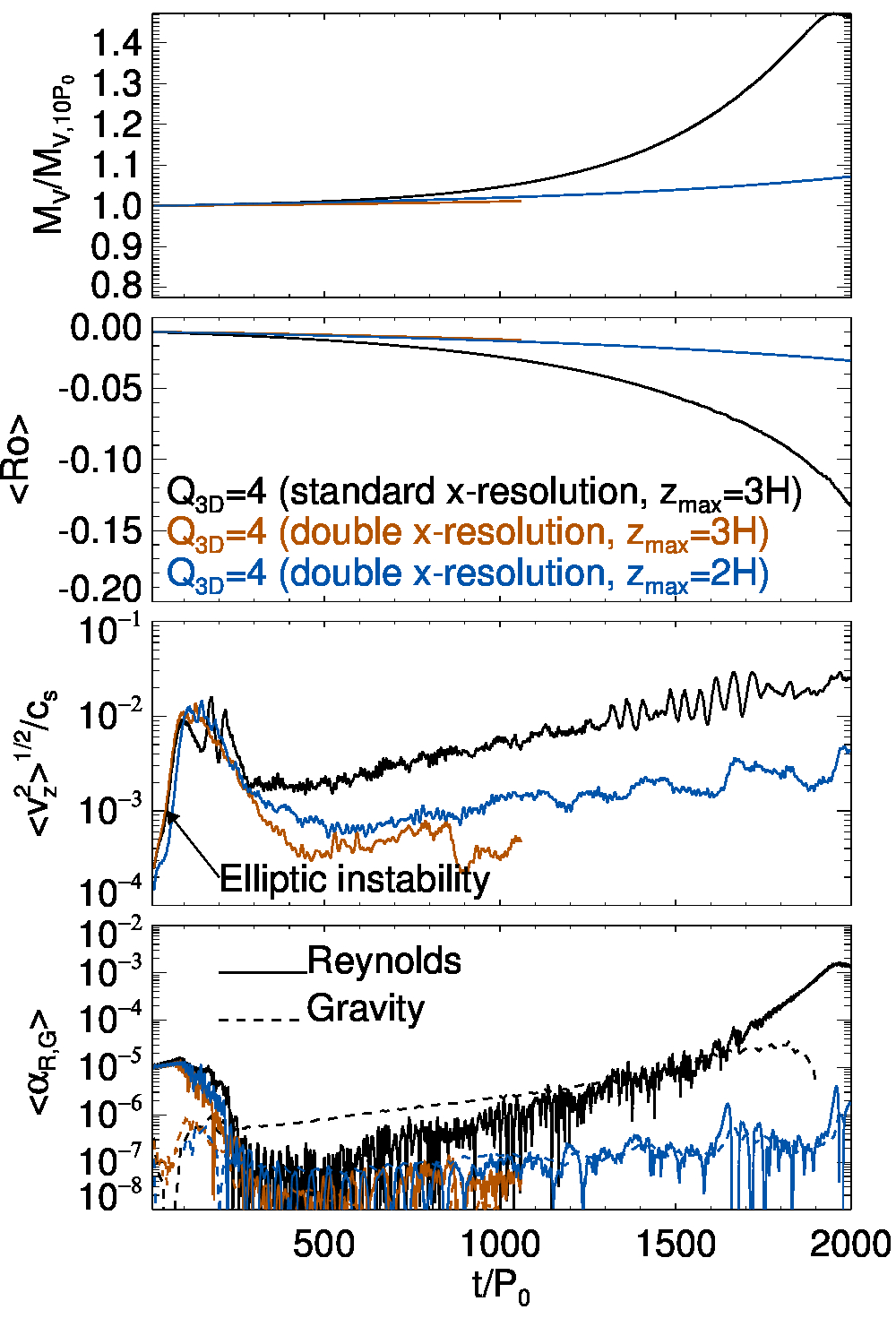}
\caption{Vortex evolution in discs with $\qthree=4$ but different radial resolutions and vertical domain sizes.    
\label{athevol2_z2HR}}
\end{figure}

\begin{figure}
  \includegraphics[scale=.26,clip=true,trim=0cm 1.8cm 0cm 
    0cm]{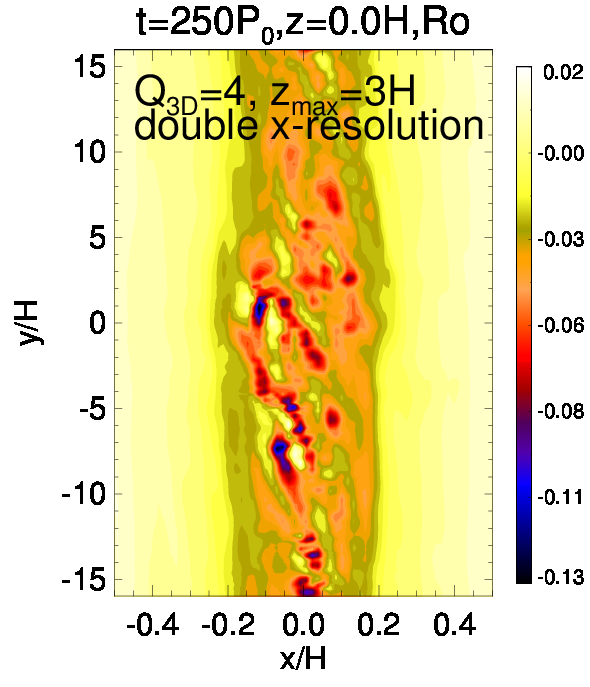}\includegraphics[scale=.26,clip=true,trim=2.3cm 1.8cm 0cm 
    0cm]{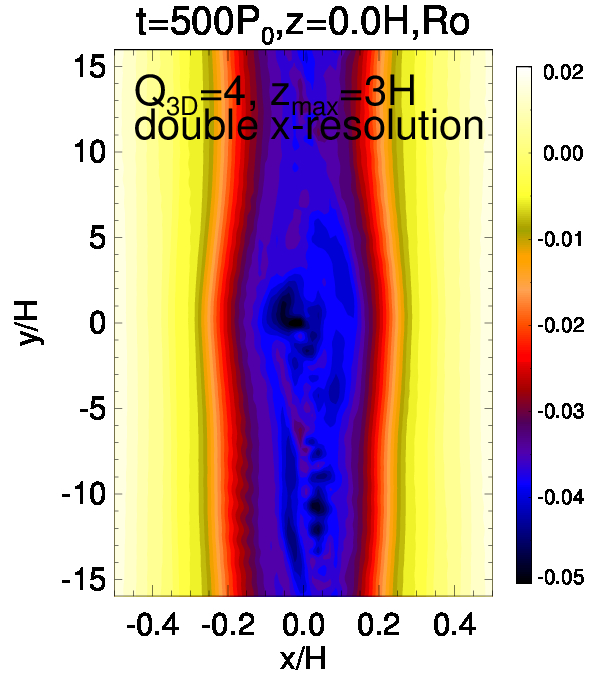}\includegraphics[scale=.26,clip=true,trim=2.3cm 1.8cm 0cm 
    0cm]{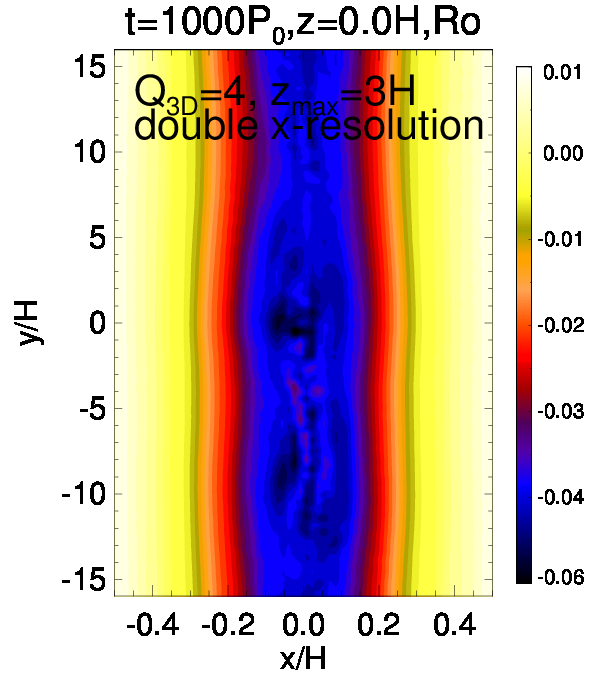}\\
  \includegraphics[scale=.26,clip=true,trim=0cm 0.cm 0cm 
    0cm]{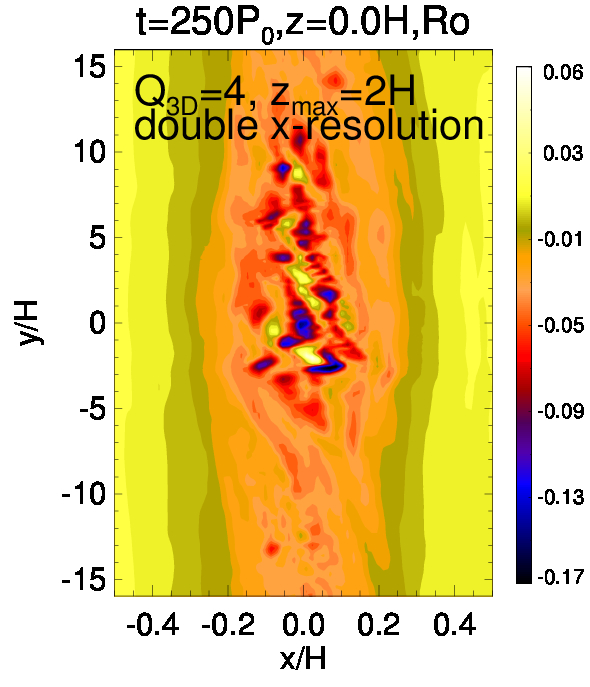}\includegraphics[scale=.26,clip=true,trim=2.3cm 0.cm 0cm 
    0cm]{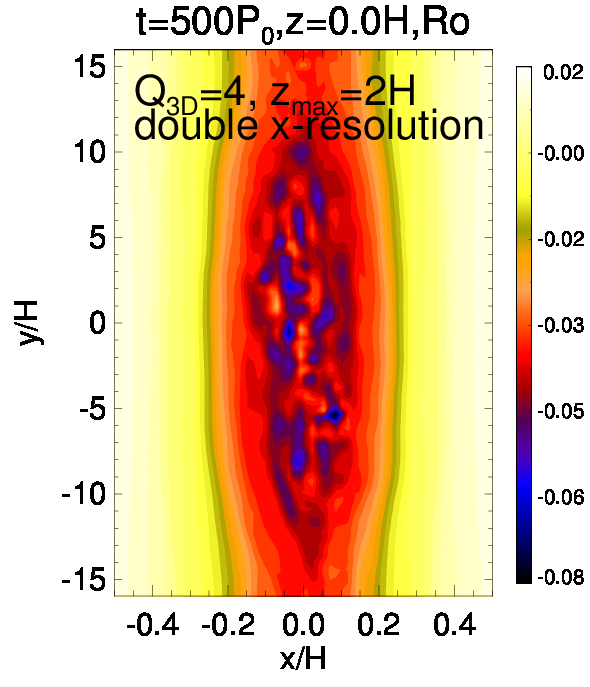}\includegraphics[scale=.26,clip=true,trim=2.3cm 0.cm 0cm 
    0cm]{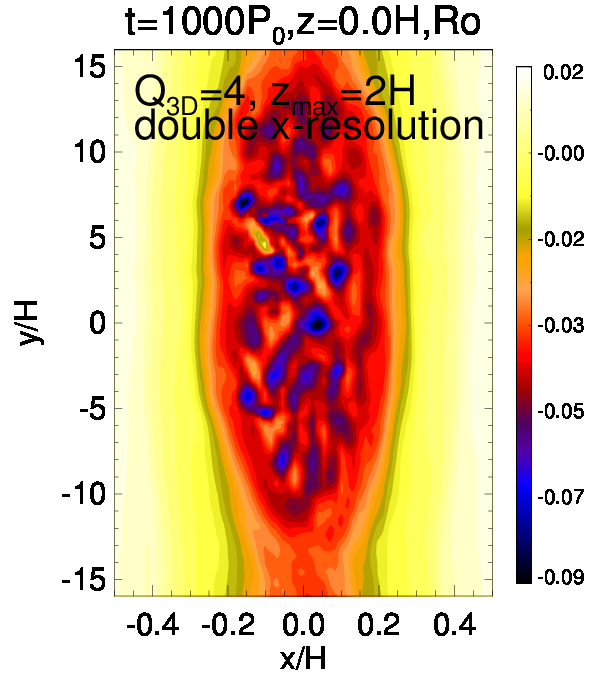}
  \caption{Effect of vertical domain size on vortex evolution. The disc has $\qthree=4$ and the simulations were performed with twice the fiducial radial resolution. The vortex does not survive with $z_\mathrm{max}=3H$ but does with $z_\mathrm{max}=2H$. 
\label{HR_sims_Q4}}
\end{figure}

{ We find that reducing the vertical domain also enhances vortex survival, which is as expected as the system is forced to behave more two-dimensionally \citep{lithwick09}. This is demonstrated in Fig. \ref{athevol2_z2HR} with simulations using $\qthree=4$. Starting from the fiducial run with $z_\mathrm{max}=3H$, we double the radial resolution and find the vortex is destroyed by the initial EI. (Comparing with  Fig. \ref{HR_sims_Q3} confirms that self-gravity enhances vortex survival.) 
  However, retaining the same resolution but setting
  $z_\mathrm{max}=2H$ results in a persistent 3D vortex with a
  turbulent core, also shown in Fig. \ref{HR_sims_Q4}. 
}

These additional runs indicate that, in order to undergo
self-gravitational { growth}, a 3D vortex needs to maintain its 
large-scale internal flow. Thus the initial EI cannot be too
strong. This can be limited by diffusion (numerical in our case)
and/or limited vertical domain.
{
 Notice in the simulations where a vortex persists that $\gstress >
 \reynolds$ during secular growth, as observed in the standard run. We
 suggest this condition required for long-term vortex survival. 
} 
If the initial EI-turbulence destroys
the large-scale vortex topology, then there is no vortex  for self-gravity to { maintain or amplify}. 

\section{Discussion}\label{discussion}

\subsection{Self-gravitational vortex growth and hydrodynamic turbulence} 

Our main finding is that { self-gravitating} 3D vortices can survive the elliptic instability. Self-gravity enables
vortex survival { and growth} even when the Toomre parameter 
$Q\sim 5$, which is significantly larger than that required for
axisymmetric instability in 3D shearing boxes 
{ \citep[$Q\lesssim 0.6$,][]{mamat10}}. Thus the presence of a vortex  
reduces the gravitational stability of the flow.  

We suggest that, provided the initial EI does not completely destroy 
the large-scale vortex flow topology (see \S\ref{param_study} and \S\ref{caveats}), the vortex
can then slowly { grow} under its self-gravity.  (Without self-gravity 
the vortex always decays.) 
The { growth} spins up the  
vortex and its aspect-ratio decreases. This is expected for 
vortices in shearing discs, e.g. the GNG vortex 
spin increases with decreasing $\chi$ (see Eq. \ref{gng_sol}). 
Smaller aspect-ratio vortices favour EI  
\citep{lesur09}, which generates further hydrodynamic
turbulence. 


The vortex { persists} during secular { growth} because the 
subsequent EI is weak, { as the vortex has been elongated by the  
  initial EI.} Our numerical experiments indicate 
a necessary condition for vortex survival is
having gravitational stresses exceed Reynolds stresses ($\gstress\gtrsim\reynolds$). 
It signifies the evolution 
being dominated by self-gravity with hydrodynamic turbulence being a 
byproduct.

To check that self-gravity can directly cause vortex growth, we ran  
several 2D simulations in Appendix \ref{2d_sims}. We find for 
sufficiently strong self-gravity, the vortex can indeed grow without
EI-induced turbulence. However, growth does not occur in razor-thin
discs if $\qthree\geq 4$, whereas we find growth in corresponding 3D
simulations with EI. This suggests that EI-induced hydrodynamic
turbulence may also contribute to gravitational { instability}, as discussed below.  

\subsection{Turbulence-enabled gravitational instability?}\label{gravito-elliptic}



The EI converts the vortex's internal large-scale, planar 
rotation into small-scale 3D turbulence. However, removing rotation
in a self-gravitating flow allows gravitational
collapse. This type of gravitational instability 
rely on dissipative processes to remove rotational stabilization \citep{lyndenbell74}.
Comparing our 3D simulations that have EI and show { vortex growth} and supplementary 2D
simulations in 
Appendix \ref{2d_sims} that shows no { EI or growth} suggest that EI-turbulence may provide a similar effect; { thus 
enabling vortex growth at large $Q$}. This picture is similar to secular gravitational instabilities in 
viscous or dusty discs, whereby friction 
allow gravitational instabilities to develop even when $Q > 1$
\citep[e.g.][]{schmit95,takahashi14}.


\tikzstyle{process} = [rectangle, minimum width=2.5cm, minimum height=1cm, text centered, draw=black, fill=orange!30]
\tikzstyle{decision} = [diamond, minimum width=2.5cm, minimum height=1cm, text centered, draw=black, fill=green!30]
\tikzstyle{line} = [draw, -triangle 45,thick]

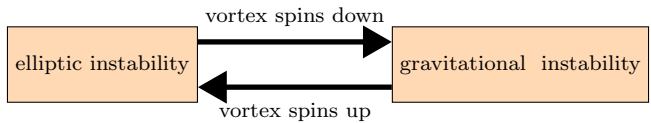
\begin{figure}
  \begin{center}
    \begin{tikzpicture}[node distance=1.8cm,
                    axis/.style={->, >=triangle 60, line width=2pt}
                    ]
                   \node[process](ellip){elliptic instability};
                   \node[process,right of=ellip, node distance =
                   5.5cm](gi){gravitational { instability}};               
                   \path (ellip) -- node [yshift=.63cm]{ vortex spins down} (gi);
                   \path (gi) -- node [yshift=-.63cm] { vortex spins up} (ellip);                 
                   \draw[axis] ([yshift=0.3cm]ellip.east) --  ([yshift=0.3cm]gi.west){};
                   \draw[axis] ([yshift=-.3cm]gi.west) -- ([yshift=-.3cm]ellip.east){};
                 \end{tikzpicture}
  \end{center}
\caption{The `gravito-elliptic' interpretation of the bursty behaviour
  observed towards the end of secular vortex growth.  A 
  vortex spins up and its aspect-ratio 
  decreases,  
  generating hydrodynamic turbulence via the EI. The associated
  turbulent viscosity favours { gravitational instability} by 
  removing rotational support against self-gravity. \label{gef}}
\end{figure}

  
Gravitational { instability} catalysed by EI-induced turbulence leads to the 
possibility of `gravito-elliptic' feedback, illustrated in 
Fig. \ref{gef}. Vortex { growth} leads to hydrodynamic turbulence
through the EI, which helps further collapse via secular gravitational instabilities, thus creating a cycle. 
The vortex attains 
a quasi-steady state if the destructive effect of EI is balanced by 
gravitational { instability}. This explains the 
`episodic bursts' seen in our fiducial simulations when $\reynolds\sim
\gstress$ (\S\ref{episodic_burst}).

\subsection{Analogy with SBI in 3D}\label{sbi}

{ The episodic EI bursts observed in \S\ref{episodic_burst} bears
some parallel to the Sub-critical Baroclinic Instability (SBI) in 3D. 
The SBI is a mechanism for  
amplifying vortices in non-isothermal discs and requires an unstable radial  
entropy gradient and an appropriate thermal diffusion or cooling
timescale \citep{lesur10}. 
Neither are present in our standard, isothermal shearing boxes, 
\citeauthor{lesur10} also found episodic EI in 3D
simulations of the SBI.}

In 3D, the SBI competes with the EI  
\citep{lesur10,lyra11,barge16}. These authors find the resulting
vortex maintains its large-scale flow, but also contain small-scale,
3D hydrodynamic turbulence in its core. This is similar to our
self-gravitating vortices (e.g. Fig. \ref{ei_vz}). { Self-gravity plays the role
of SBI} in our simulations by  amplifying the vortex.  

The important difference is that for SBI there exists an explicit
baroclinic vorticity source, which  
comes from the global radial entropy gradient. Self-gravity cannot
directly source vorticity. However, the vortex can spin up 
as it collapses due to self-gravity. The result in both cases is that
the vortex aspect-ratio decreases, feeding the EI,
which tries to remove the large-scale vortex { rotation}. 

\subsection{Application to Rossby vortices}   



We can relate our simulations to vortex formation by the Rossby Wave
Instability \citep[RWI,][]{lovelace99,li00,li01}. The RWI is a
dynamical instability associated with radially-structured discs. More
specifically, it requires a minimum in the disc's potential vorticity
profile ($=\omega/\Sigma$). Candidate sites 
include gap edges opened by planets \citep{lyra08,li09,lin12b} and dead zone boundaries 
\citep{lyra08b,lyra09}. These isolated radial structures often involve
length scales of $O(H)$, and are thus consistent our typical vortex { half-widths} ($\lesssim H$).   

The expected initial aspect ratio of Rossby vortices is  
$\chi = \pi/hm_\mathrm{max}$; here $m_\mathrm{max}\sim 5$---8 is the
most 
unstable azimuthal wavenumber of the RWI \citep{lin11a} and $h= H/R$
is the disc aspect ratio. 
For protoplanetary discs with $h=0.05$---0.1 we expect 
$\chi\sim 4$---12. Our nominal choice $\chi=10$ is consistent with
the lower bound on lower $m_\mathrm{max}$, which is 
applicable to thin and/or low mass discs  
\citep{lin11a} as considered here. This is also the limit where Rossby vortices can merge. Thus our model could 
also correspond to a single, post-merger Rossby vortex. 

\subsection{Dust-trapping and observational implications}

Our main result is that moderate disc self-gravity can help sustain  
3D vortices in spite of the elliptic instability. This means that vortices are more likely to be
observed 
in the outer parts of protoplanetary discs, where self-gravity becomes 
important. This could be relevant to vortex formation via the RWI due
to gap-opening planets at large radii \citep{hammer17} or the outer
dead zone boundary \citep{lyra15}.   

For definiteness, let us consider the class II circumstellar disc
models described in \citet[][their Fig. 2a]{kratter16}. For typical
global accretion rates $\dot{M}\in [10^{-8},10^{-7}] M_{\sun}
\text{yr}^{-1}$, the Toomre parameter ranges from $Q\simeq 10$ at
$30$AU to $Q\simeq 5$ at $100$AU. These are similar
to the $Q$ values in our simulations. Furthermore, at these radii, 
heating is dominated by stellar irradiation \citep{chiang97} and
thus disc behaves almost isothermally; also consistent with our equation of
state. Note also our simulation timescales of $O(10^3)$ orbits
corresponds $10^6$yrs at $100$AU, which is of order the disc
lifetime. 

Our fiducial simulation ($\qthree=3$ or $Q\simeq 5$) corresponds to
$100$AU in this physical disc model. Although $Q>2$ is traditionally
considered non-self-gravitating  \citep{kratter16}, neglecting 
self-gravity altogether would result in a weak vortex with vanishing  
over-density. However, properly including self-gravity gives a
concentrated vortex (Fig. \ref{secular_collapse}). This would suggest that 
the trapping of dust --- often the observable --- is favoured in self-gravitating discs.     

On the other hand, self-gravitating 3D vortices have \emph{turbulent}
cores, which tends to expel dust particles due to turbulent dust
diffusion \citep{youdin07}.  
To examine this effect we use the dust-trapping model described in
\citep{lyra13} { assuming} an underlying GNG vortex. They 
find the distribution of small, passive dust particles within the vortex
core is given by  
\begin{align}
  \rho_\mathrm{d}(s) &= \rho_\mathrm{d,max}
  \exp{\left(-\frac{s^2}{2H_\mathrm{v}^2}\right)}, \\
  H_\mathrm{v} & \equiv \frac{H}{f(\chi)}\sqrt{\frac{\delta}{\mathrm{St}
  + \delta}}\label{lldist}
\end{align}
where $\mathrm{St} = \Omega \tau_\mathrm{s}$ is the Stokes number and
$\tau_\mathrm{s}$ is the stopping time characterizing dust-gas
friction and $\delta$ is a dimensionless parameterization of
turbulent dust diffusion. The GNG vortex has 
$f(\chi)\sim 0.7$ for $\chi\gg 1$ \citep[see][for  
  details]{lyra13}. Recall $s$ is a co-ordinate that label 
elliptical streamlines in the vortex by their semi-minor axes (see \S\ref{sgvort_approx}). 

We evaluate Eq. \ref{lldist} using simulation 
results as input parameters. Assuming dust-gas interaction is
dominated by drag forces, we set $\delta = \reynolds$. We also use
$\chi_\rho$ in place of $\chi$, but this is
immaterial since the function $f$ is effectively constant for
elongated vortices. 

\begin{figure}
  \includegraphics[scale=0.42,clip=true,trim=0cm 1.8cm 0cm .5cm]{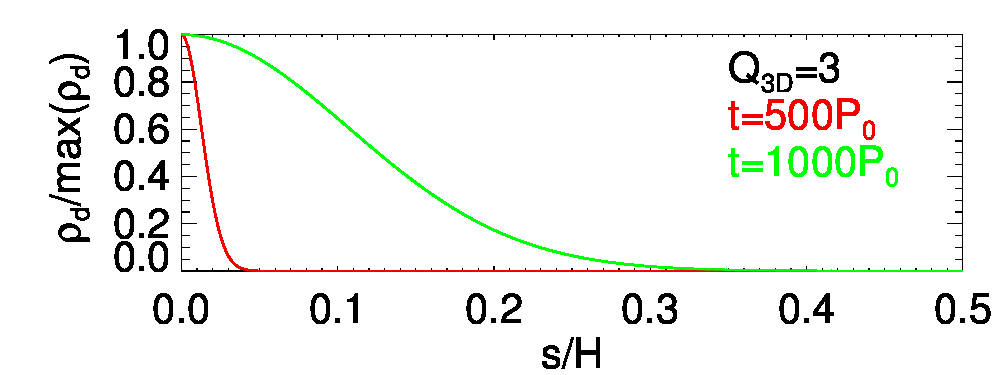}\\
  \includegraphics[scale=0.42,clip=true,trim=0cm 1.8cm 0cm .5cm]{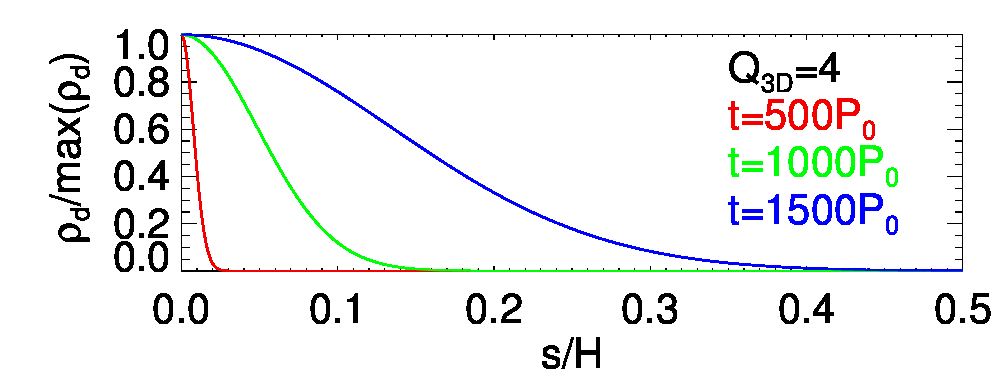}\\
  \includegraphics[scale=0.42,clip=true,trim=0cm 0cm 0cm .5cm]{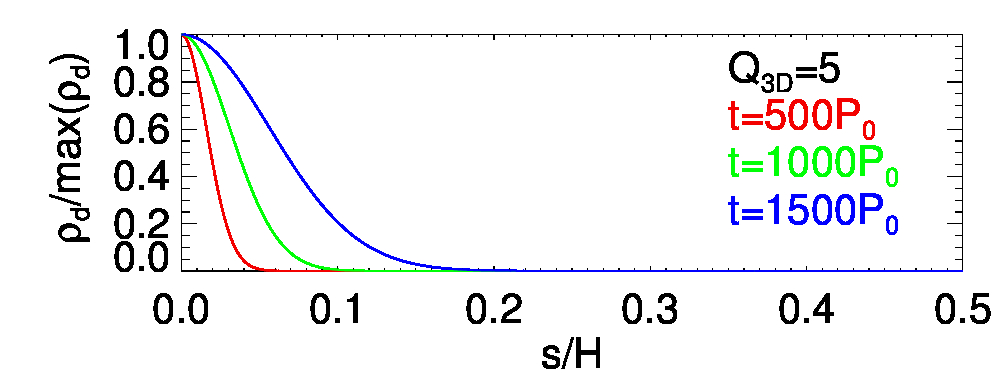}\\
  \caption{Normalised (mid-plane) dust density distribution obtained
    from the dust-trap model of \protect\cite{lyra13} with 
    vortex parameters
    taken from the fiducial simulations. We fix the Stokes number
    $\tstop=10^{-3}$.
    \label{LLdust}}
\end{figure}

In Fig. \ref{LLdust} we plot the expected distribution of dust
particles with $\mathrm{St}=10^{-3}$. We only consider the secular
growth stage with EI turbulence occurring inside the vortex 
core, where \citeauthor{lyra13}'s model is applicable. 
In all cases dust is initially concentrated towards the centre,
but over time it diffuses outward due to the increasing levels of
hydrodynamic turbulence, which in turn is more vigorous with
increasing self-gravity. Thus we can expect the dusty vortices to be
wider in self-gravitating discs. 

{
  We comment that the above result for a single vortex may not be
  applicable to strongly self-gravitating discs with
  gravtio-turbulence/fragmentation \citep{gammie01}. However, in that limit it has also  been shown that particle-trapping is enhanced by self-gravity \citep{gibbons12,gibbons14,gibbons15,shi16}. 
}

\subsection{Caveats and future directions}\label{caveats}

{
We have limited our study to classically stable discs, i.e. $Q>2$ (or
$\qthree > 1$). It would be interesting to study    
3D vortex evolution in 
strongly self-gravitating discs with $Q\lesssim 2$. However, in this
limit the vortex must be initialised with care. Simply perturbing a strongly 
self-gravitating disc with the non-self-gravitating GNG vortex
flow may result in transient effects. This problem may also be ill-posed
because it is unclear if  a single  
large-scale vortex can form in strongly self-gravitating discs in the
first place. For example, it has been shown that the vortex-forming RWI can be suppressed by
self-gravity \citep{lin11a,lovelace13}. Furthermore, classic 
gravitational instabilities can be expected in the ambient disc, which would complicate the
evolution of an isolated vortex.  We thus defer the strongly  
self-gravitating limit to a future study.  }

{ Returning to the $Q>5$ discs considered in our study, } the
additional runs described in \S\ref{param_study} indicate that 
3D vortices undergo secular self-gravitational { growth} only if the initial
EI-induced turbulence does not destroy the large-scale vortex 
topology. In our simulations the initial EI turbulence strength is limited by
a reduced vertical domain \citep{lithwick09} and/or 
numerical diffusion \citep{lesur09}. Since our models are inviscid, true 
numerical convergence may be  difficult to achieve.   

However, these effects may not be unrealistic. Numerical diffusion may
mimic other sources of low-level turbulent viscosity in a protoplanetary disc \citep{umurhan16b}.    
Vortices can { also} be limited in vertical extent if
the background disc has a layered structure \citep{bai16}. 
The presence of an { active} vortex-driving source may also help to retain the
large-scale vortex flow against the initial EI. 
Nevertheless, future works should incorporate
these physical effects explicitly. 

Local shearing box 
simulations do not permit a proper study of vortex 
gap-opening and migration \citep{paardekooper10}, which stem from global 
spiral density waves launched by the vortex. 
Clearly, one must eventually perform global disc simulations of  
self-gravitating 3D vortices. This is computationally challenging 
because of the need to simultaneously resolve the small-scale elliptic
instability. Though some recent non-self-gravitating, global
simulations have done so \citep{richard16,barge16}, including 
3D self-gravity can significantly increase the computational cost.  

However, our simulations show that even with EI-induced hydrodynamic
turbulence, the vortex maintains a smooth density field with moderate
amplitude. Furthermore, elongated vortices do not require high
resolution in azimuth. This means that an accurate global Poisson
solver (with sub-$H$ resolution) is not essential for this problem.

We have purposefully neglected vortex-forming mechanisms to isolate the
effect of self-gravity on vortex evolution. Nevertheless, future work
should account for hydrodynamic instabilities that drive 
vortex formation (RWI, ConO/SBI, ZVI, or VSI), which require improved
geometries and physics such as cooling.  

Finally, we performed pure gas simulations and only estimated the
extent of vortex dust-trapping in post-processing. It will be
desirable to conduct two-phase, gas-dust simulations to properly capture the 
evolution of dust particles and the back reaction of drag forces on 
the gas. The latter becomes important for high dust-to-gas ratios,
which can be expected in vortex centres
\citep{fu14,ivo15, surville16}. A simple starting point to model these
effects is the one-fluid, thermodynamic model of dusty gas 
recently developed by \cite{lin17}.
 
\section{Summary}\label{summary}

In this paper we study the evolution of 3D vortices in
self-gravitating protoplanetary discs via customised numerical
simulations in the shearing box framework. We find 
self-gravity can help 3D vortices survive against elliptic
instabilities that would otherwise destroy them in non-self-gravitating discs. 
With moderate self-gravity the result is a 3D vortex with 
a turbulent core. We emphasize that our disc models are 
gravitationally stable in the traditional sense that 
Toomre $Q$ values safely exceed unity; yet we observe vortex { growth},
albeit on timescales of $O(10^3)$ orbits. 
This indicates that the presence of a vortex { reduces the long-term stability of the flow}.   

In addition to hydrodynamic turbulence 
as a byproduct of { self-gravitational vortex growth} 
via elliptic instabilities, we
{ suggest}
secular gravitational instabilities { can be} catalysed by the hydrodynamic
turbulence { that result from the elliptic instability}. The elliptic instability removes
rotational support against self-gravity, which { favours
  gravitational instability}; but
the subsequent spin-up of the vortex feeds the elliptic instability. There is 
competition between the vortex destruction by the elliptic instability and 
vortex growth due to self-gravity (either directly or assisted by hydrodynamic turbulence); { thus a quasi-steady state is possible.}  
Our simulations suggest vortices survive more easily, and hence are more likely observed, in protoplanetary discs at large radii.

\section*{Acknowledgments}

{ We thank the anonymous referee for a detailed report that lead to an improved physical interpretation and presentation
  of our results.} 
This work is supported by the Theoretical Institute for Advanced
Research in Astrophysics (TIARA) based in Academia Sinica Institute of
Astronomy and Astrophysics (ASIAA). All { 3D} simulations were
performed on 
the TIARA High Performance Computing cluster. The early stages of this project was carried 
out at the University of Arizona under the support of the Steward
Theory Fellowship and made use of the El Gato and
Ocelote clusters. This work is also part of the 
NASA Astrophysics Theory Program NNX17AK59G. {
Computer time for the 2D runs presented in this paper was provided by HPC resources of Cines under allocation A0010406957 made by GENCI (Grand Equipement National de Calcul Intensif).}

\bibliographystyle{mnras}
\bibliography{ref}

\begin{thebibliography}{}
\makeatletter
\relax
\def\mn@urlcharsother{\let\do\@makeother \do\$\do\&\do\#\do\^\do\_\do\%\do\~}
\def\mn@doi{\begingroup\mn@urlcharsother \@ifnextchar [ {\mn@doi@}
  {\mn@doi@[]}}
\def\mn@doi@[#1]#2{\def\@tempa{#1}\ifx\@tempa\@empty \href
  {http://dx.doi.org/#2} {doi:#2}\else \href {http://dx.doi.org/#2} {#1}\fi
  \endgroup}
\def\mn@eprint#1#2{\mn@eprint@#1:#2::\@nil}
\def\mn@eprint@arXiv#1{\href {http://arxiv.org/abs/#1} {{\tt arXiv:#1}}}
\def\mn@eprint@dblp#1{\href {http://dblp.uni-trier.de/rec/bibtex/#1.xml}
  {dblp:#1}}
\def\mn@eprint@#1:#2:#3:#4\@nil{\def\@tempa {#1}\def\@tempb {#2}\def\@tempc
  {#3}\ifx \@tempc \@empty \let \@tempc \@tempb \let \@tempb \@tempa \fi \ifx
  \@tempb \@empty \def\@tempb {arXiv}\fi \@ifundefined
  {mn@eprint@\@tempb}{\@tempb:\@tempc}{\expandafter \expandafter \csname
  mn@eprint@\@tempb\endcsname \expandafter{\@tempc}}}

\bibitem[\protect\citeauthoryear{{Adams} \& {Watkins}}{{Adams} \&
  {Watkins}}{1995}]{adams95}
{Adams} F.~C.,  {Watkins} R.,  1995, \mn@doi [\apj] {10.1086/176221}, \href
  {http://adsabs.harvard.edu/abs/1995ApJ...451..314A} {451, 314}

\bibitem[\protect\citeauthoryear{{Bai}}{{Bai}}{2016}]{bai16}
{Bai} X.-N.,  2016, \mn@doi [\apj] {10.3847/0004-637X/821/2/80}, \href
  {http://adsabs.harvard.edu/abs/2016ApJ...821...80B} {821, 80}

\bibitem[\protect\citeauthoryear{{Barge} \& {Sommeria}}{{Barge} \&
  {Sommeria}}{1995}]{barge95}
{Barge} P.,  {Sommeria} J.,  1995, \aap, \href
  {http://adsabs.harvard.edu/abs/1995A%26A...295L...1B} {295, L1}

\bibitem[\protect\citeauthoryear{{Barge}, {Richard}  \& {Le Diz{\`e}s}}{{Barge}
  et~al.}{2016}]{barge16}
{Barge} P.,  {Richard} S.,   {Le Diz{\`e}s} S.,  2016, \mn@doi [\aap]
  {10.1051/0004-6361/201628381}, \href
  {http://adsabs.harvard.edu/abs/2016A%26A...592A.136B} {592, A136}

\bibitem[\protect\citeauthoryear{{Barker} \& {Latter}}{{Barker} \&
  {Latter}}{2015}]{barker15}
{Barker} A.~J.,  {Latter} H.~N.,  2015, \mn@doi [\mnras]
  {10.1093/mnras/stv640}, \href
  {http://adsabs.harvard.edu/abs/2015MNRAS.450...21B} {450, 21}

\bibitem[\protect\citeauthoryear{{Bodo}, {Chagelishvili}, {Murante},
  {Tevzadze}, {Rossi}  \& {Ferrari}}{{Bodo} et~al.}{2005}]{bodo05}
{Bodo} G.,  {Chagelishvili} G.,  {Murante} G.,  {Tevzadze} A.,  {Rossi} P.,
  {Ferrari} A.,  2005, \mn@doi [\aap] {10.1051/0004-6361:20041046}, \href
  {http://adsabs.harvard.edu/abs/2005A%26A...437....9B} {437, 9}

\bibitem[\protect\citeauthoryear{{Bodo}, {Tevzadze}, {Chagelishvili},
  {Mignone}, {Rossi}  \& {Ferrari}}{{Bodo} et~al.}{2007}]{bodo07}
{Bodo} G.,  {Tevzadze} A.,  {Chagelishvili} G.,  {Mignone} A.,  {Rossi} P.,
  {Ferrari} A.,  2007, \mn@doi [\aap] {10.1051/0004-6361:20077695}, \href
  {http://adsabs.harvard.edu/abs/2007A%26A...475...51B} {475, 51}

\bibitem[\protect\citeauthoryear{{Casassus} et~al.,}{{Casassus}
  et~al.}{2013}]{casassus13}
{Casassus} S.,  et~al., 2013, \mn@doi [\nat] {10.1038/nature11769}, \href
  {http://adsabs.harvard.edu/abs/2013Natur.493..191C} {493, 191}

\bibitem[\protect\citeauthoryear{{Chandrasekhar}}{{Chandrasekhar}}{1969}]{chandra69}
{Chandrasekhar} S.,  1969, {Ellipsoidal figures of equilibrium}

\bibitem[\protect\citeauthoryear{{Chang} \& {Oishi}}{{Chang} \&
  {Oishi}}{2010}]{chang10}
{Chang} P.,  {Oishi} J.~S.,  2010, \mn@doi [\apj]
  {10.1088/0004-637X/721/2/1593}, \href
  {http://adsabs.harvard.edu/abs/2010ApJ...721.1593C} {721, 1593}

\bibitem[\protect\citeauthoryear{{Chiang} \& {Goldreich}}{{Chiang} \&
  {Goldreich}}{1997}]{chiang97}
{Chiang} E.~I.,  {Goldreich} P.,  1997, \apj, \href
  {http://adsabs.harvard.edu/abs/1997ApJ...490..368C} {490, 368}

\bibitem[\protect\citeauthoryear{{Crida}, {Morbidelli}  \& {Masset}}{{Crida}
  et~al.}{2006}]{crida06}
{Crida} A.,  {Morbidelli} A.,   {Masset} F.,  2006, \mn@doi [Icarus]
  {10.1016/j.icarus.2005.10.007}, \href
  {http://adsabs.harvard.edu/abs/2006Icar..181..587C} {181, 587}

\bibitem[\protect\citeauthoryear{Crnkovic-Rubsamen, Zhu  \&
  Stone}{Crnkovic-Rubsamen et~al.}{2015}]{ivo15}
Crnkovic-Rubsamen I.,  Zhu Z.,   Stone J.~M.,  2015, \mn@doi [\mnras]
  {10.1093/mnras/stv828}, 450, 4285

\bibitem[\protect\citeauthoryear{{Dong}, {Li}, {Chiang}  \& {Li}}{{Dong}
  et~al.}{2017}]{dong17}
{Dong} R.,  {Li} S.,  {Chiang} E.,   {Li} H.,  2017, \mn@doi [\apj]
  {10.3847/1538-4357/aa72f2}, \href
  {http://adsabs.harvard.edu/abs/2017ApJ...843..127D} {843, 127}

\bibitem[\protect\citeauthoryear{{Fu}, {Li}, {Lubow}  \& {Li}}{{Fu}
  et~al.}{2014}]{fu14}
{Fu} W.,  {Li} H.,  {Lubow} S.,   {Li} S.,  2014, \mn@doi [\apjl]
  {10.1088/2041-8205/788/2/L41}, \href
  {http://adsabs.harvard.edu/abs/2014ApJ...788L..41F} {788, L41}

\bibitem[\protect\citeauthoryear{{Fukagawa} et~al.,}{{Fukagawa}
  et~al.}{2013}]{fukagawa13}
{Fukagawa} M.,  et~al., 2013, \mn@doi [\pasj] {10.1093/pasj/65.6.L14}, \href
  {http://adsabs.harvard.edu/abs/2013PASJ...65L..14F} {65, L14}

\bibitem[\protect\citeauthoryear{{Gammie}}{{Gammie}}{2001}]{gammie01}
{Gammie} C.~F.,  2001, \mn@doi [\apj] {10.1086/320631}, \href
  {http://adsabs.harvard.edu/abs/2001ApJ...553..174G} {553, 174}

\bibitem[\protect\citeauthoryear{{Gibbons}, {Rice}  \&
  {Mamatsashvili}}{{Gibbons} et~al.}{2012}]{gibbons12}
{Gibbons} P.~G.,  {Rice} W.~K.~M.,   {Mamatsashvili} G.~R.,  2012, \mn@doi
  [\mnras] {10.1111/j.1365-2966.2012.21731.x}, \href
  {http://adsabs.harvard.edu/abs/2012MNRAS.426.1444G} {426, 1444}

\bibitem[\protect\citeauthoryear{{Gibbons}, {Mamatsashvili}  \&
  {Rice}}{{Gibbons} et~al.}{2014}]{gibbons14}
{Gibbons} P.~G.,  {Mamatsashvili} G.~R.,   {Rice} W.~K.~M.,  2014, \mn@doi
  [\mnras] {10.1093/mnras/stu809}, \href
  {http://adsabs.harvard.edu/abs/2014MNRAS.442..361G} {442, 361}

\bibitem[\protect\citeauthoryear{{Gibbons}, {Mamatsashvili}  \&
  {Rice}}{{Gibbons} et~al.}{2015}]{gibbons15}
{Gibbons} P.~G.,  {Mamatsashvili} G.~R.,   {Rice} W.~K.~M.,  2015, \mn@doi
  [\mnras] {10.1093/mnras/stv1766}, \href
  {http://adsabs.harvard.edu/abs/2015MNRAS.453.4232G} {453, 4232}

\bibitem[\protect\citeauthoryear{{Godon} \& {Livio}}{{Godon} \&
  {Livio}}{1999}]{godon99}
{Godon} P.,  {Livio} M.,  1999, \mn@doi [\apj] {10.1086/307720}, \href
  {http://adsabs.harvard.edu/abs/1999ApJ...523..350G} {523, 350}

\bibitem[\protect\citeauthoryear{{Goldreich} \& {Lynden-Bell}}{{Goldreich} \&
  {Lynden-Bell}}{1965}]{goldreich65}
{Goldreich} P.,  {Lynden-Bell} D.,  1965, \mnras, \href
  {http://adsabs.harvard.edu/abs/1965MNRAS.130..125G} {130, 125}

\bibitem[\protect\citeauthoryear{{Goodman}, {Narayan}  \&
  {Goldreich}}{{Goodman} et~al.}{1987}]{goodman87}
{Goodman} J.,  {Narayan} R.,   {Goldreich} P.,  1987, \mnras, \href
  {http://adsabs.harvard.edu/abs/1987MNRAS.225..695G} {225, 695}

\bibitem[\protect\citeauthoryear{{Hammer}, {Kratter}  \& {Lin}}{{Hammer}
  et~al.}{2017}]{hammer17}
{Hammer} M.,  {Kratter} K.~M.,   {Lin} M.-K.,  2017, \mn@doi [\mnras]
  {10.1093/mnras/stw3000}, \href
  {http://adsabs.harvard.edu/abs/2017MNRAS.466.3533H} {466, 3533}

\bibitem[\protect\citeauthoryear{{Hashimoto} et~al.,}{{Hashimoto}
  et~al.}{2015}]{hashimoto15}
{Hashimoto} J.,  et~al., 2015, \mn@doi [\apj] {10.1088/0004-637X/799/1/43},
  \href {http://adsabs.harvard.edu/abs/2015ApJ...799...43H} {799, 43}

\bibitem[\protect\citeauthoryear{{Heinemann} \& {Papaloizou}}{{Heinemann} \&
  {Papaloizou}}{2009}]{heinemann09}
{Heinemann} T.,  {Papaloizou} J.~C.~B.,  2009, \mn@doi [\mnras]
  {10.1111/j.1365-2966.2009.14799.x}, \href
  {http://adsabs.harvard.edu/abs/2009MNRAS.397...52H} {397, 52}

\bibitem[\protect\citeauthoryear{{Inaba} \& {Barge}}{{Inaba} \&
  {Barge}}{2006}]{inaba06}
{Inaba} S.,  {Barge} P.,  2006, \mn@doi [\apj] {10.1086/506427}, \href
  {http://adsabs.harvard.edu/abs/2006ApJ...649..415I} {649, 415}

\bibitem[\protect\citeauthoryear{{Isella}, {P{\'e}rez}, {Carpenter}, {Ricci},
  {Andrews}  \& {Rosenfeld}}{{Isella} et~al.}{2013}]{isella13}
{Isella} A.,  {P{\'e}rez} L.~M.,  {Carpenter} J.~M.,  {Ricci} L.,  {Andrews}
  S.,   {Rosenfeld} K.,  2013, \mn@doi [\apj] {10.1088/0004-637X/775/1/30},
  \href {http://adsabs.harvard.edu/abs/2013ApJ...775...30I} {775, 30}

\bibitem[\protect\citeauthoryear{{Johnson} \& {Gammie}}{{Johnson} \&
  {Gammie}}{2005}]{johnson05}
{Johnson} B.~M.,  {Gammie} C.~F.,  2005, \mn@doi [\apj] {10.1086/497358}, \href
  {http://adsabs.harvard.edu/abs/2005ApJ...635..149J} {635, 149}

\bibitem[\protect\citeauthoryear{Kerswell}{Kerswell}{1994}]{kerswell94}
Kerswell R.~R.,  1994, \mn@doi [Journal of Fluid Mechanics]
  {10.1017/S0022112094002107}, 274, 219

\bibitem[\protect\citeauthoryear{{Kida}}{{Kida}}{1981}]{kida81}
{Kida} S.,  1981, Journal of the Physical Society of Japan, \href
  {http://adsabs.harvard.edu/abs/1981JPSJ...50.3517K} {50, 3517}

\bibitem[\protect\citeauthoryear{{Klahr} \& {Hubbard}}{{Klahr} \&
  {Hubbard}}{2014}]{klahr14}
{Klahr} H.,  {Hubbard} A.,  2014, \mn@doi [\apj] {10.1088/0004-637X/788/1/21},
  \href {http://adsabs.harvard.edu/abs/2014ApJ...788...21K} {788, 21}

\bibitem[\protect\citeauthoryear{{Kraichnan} \& {Montgomery}}{{Kraichnan} \&
  {Montgomery}}{1980}]{kraichnan80}
{Kraichnan} R.~H.,  {Montgomery} D.,  1980, \mn@doi [Reports on Progress in
  Physics] {10.1088/0034-4885/43/5/001}, \href
  {http://adsabs.harvard.edu/abs/1980RPPh...43..547K} {43, 547}

\bibitem[\protect\citeauthoryear{{Kratter} \& {Lodato}}{{Kratter} \&
  {Lodato}}{2016}]{kratter16}
{Kratter} K.,  {Lodato} G.,  2016, \mn@doi [\araa]
  {10.1146/annurev-astro-081915-023307}, \href
  {http://adsabs.harvard.edu/abs/2016ARA%26A..54..271K} {54, 271}

\bibitem[\protect\citeauthoryear{{Kraus} et~al.,}{{Kraus}
  et~al.}{2017}]{kraus17}
{Kraus} S.,  et~al., 2017, \mn@doi [\apjl] {10.3847/2041-8213/aa8edc}, \href
  {http://adsabs.harvard.edu/abs/2017ApJ...848L..11K} {848, L11}

\bibitem[\protect\citeauthoryear{{Latter}}{{Latter}}{2016}]{latter16}
{Latter} H.~N.,  2016, \mn@doi [\mnras] {10.1093/mnras/stv2449}, \href
  {http://adsabs.harvard.edu/abs/2016MNRAS.455.2608L} {455, 2608}

\bibitem[\protect\citeauthoryear{{Lesur} \& {Latter}}{{Lesur} \&
  {Latter}}{2016}]{lesur16}
{Lesur} G.~R.~J.,  {Latter} H.,  2016, \mn@doi [\mnras]
  {10.1093/mnras/stw2172}, \href
  {http://adsabs.harvard.edu/abs/2016MNRAS.462.4549L} {462, 4549}

\bibitem[\protect\citeauthoryear{{Lesur} \& {Papaloizou}}{{Lesur} \&
  {Papaloizou}}{2009}]{lesur09}
{Lesur} G.,  {Papaloizou} J.~C.~B.,  2009, \mn@doi [\aap]
  {10.1051/0004-6361/200811577}, \href
  {http://adsabs.harvard.edu/abs/2009A%26A...498....1L} {498, 1}

\bibitem[\protect\citeauthoryear{{Lesur} \& {Papaloizou}}{{Lesur} \&
  {Papaloizou}}{2010}]{lesur10}
{Lesur} G.,  {Papaloizou} J.~C.~B.,  2010, \mn@doi [\aap]
  {10.1051/0004-6361/200913594}, \href
  {http://adsabs.harvard.edu/abs/2010A%26A...513A..60L} {513, A60}

\bibitem[\protect\citeauthoryear{{Li}, {Finn}, {Lovelace}  \& {Colgate}}{{Li}
  et~al.}{2000}]{li00}
{Li} H.,  {Finn} J.~M.,  {Lovelace} R.~V.~E.,   {Colgate} S.~A.,  2000, \mn@doi
  [\apj] {10.1086/308693}, \href
  {http://adsabs.harvard.edu/abs/2000ApJ...533.1023L} {533, 1023}

\bibitem[\protect\citeauthoryear{{Li}, {Colgate}, {Wendroff}  \& {Liska}}{{Li}
  et~al.}{2001}]{li01}
{Li} H.,  {Colgate} S.~A.,  {Wendroff} B.,   {Liska} R.,  2001, \mn@doi [\apj]
  {10.1086/320241}, \href {http://adsabs.harvard.edu/abs/2001ApJ...551..874L}
  {551, 874}

\bibitem[\protect\citeauthoryear{{Li}, {Lubow}, {Li}  \& {Lin}}{{Li}
  et~al.}{2009}]{li09}
{Li} H.,  {Lubow} S.~H.,  {Li} S.,   {Lin} D.~N.~C.,  2009, \mn@doi [\apjl]
  {10.1088/0004-637X/690/1/L52}, \href
  {http://adsabs.harvard.edu/abs/2009ApJ...690L..52L} {690, L52}

\bibitem[\protect\citeauthoryear{{Lin}}{{Lin}}{2012a}]{lin12b}
{Lin} M.-K.,  2012a, \mn@doi [\mnras] {10.1111/j.1365-2966.2012.21955.x}, \href
  {http://adsabs.harvard.edu/abs/2012MNRAS.426.3211L} {426, 3211}

\bibitem[\protect\citeauthoryear{{Lin}}{{Lin}}{2012b}]{lin12c}
{Lin} M.-K.,  2012b, \mn@doi [\apj] {10.1088/0004-637X/754/1/21}, \href
  {http://adsabs.harvard.edu/abs/2012ApJ...754...21L} {754, 21}

\bibitem[\protect\citeauthoryear{{Lin} \& {Kratter}}{{Lin} \&
  {Kratter}}{2016}]{lin16}
{Lin} M.-K.,  {Kratter} K.~M.,  2016, \mn@doi [\apj]
  {10.3847/0004-637X/824/2/91}, \href
  {http://adsabs.harvard.edu/abs/2016ApJ...824...91L} {824, 91}

\bibitem[\protect\citeauthoryear{{Lin} \& {Papaloizou}}{{Lin} \&
  {Papaloizou}}{2011a}]{lin11}
{Lin} M.-K.,  {Papaloizou} J.~C.~B.,  2011a, \mn@doi [\mnras]
  {10.1111/j.1365-2966.2011.18797.x}, \href
  {http://adsabs.harvard.edu/abs/2011MNRAS.tmp..876L} {pp 876--+}

\bibitem[\protect\citeauthoryear{{Lin} \& {Papaloizou}}{{Lin} \&
  {Papaloizou}}{2011b}]{lin11a}
{Lin} M.-K.,  {Papaloizou} J.~C.~B.,  2011b, \mn@doi [\mnras]
  {10.1111/j.1365-2966.2011.18798.x}, \href
  {http://adsabs.harvard.edu/abs/2011MNRAS.415.1426L} {415, 1426}

\bibitem[\protect\citeauthoryear{{Lin} \& {Youdin}}{{Lin} \&
  {Youdin}}{2015}]{lin15}
{Lin} M.-K.,  {Youdin} A.~N.,  2015, \mn@doi [\apj]
  {10.1088/0004-637X/811/1/17}, \href
  {http://adsabs.harvard.edu/abs/2015ApJ...811...17L} {811, 17}

\bibitem[\protect\citeauthoryear{{Lin} \& {Youdin}}{{Lin} \&
  {Youdin}}{2017}]{lin17}
{Lin} M.-K.,  {Youdin} A.~N.,  2017, \apj, 849, 129

\bibitem[\protect\citeauthoryear{{Lithwick}}{{Lithwick}}{2007}]{lithwick07}
{Lithwick} Y.,  2007, \mn@doi [\apj] {10.1086/522074}, \href
  {http://adsabs.harvard.edu/abs/2007ApJ...670..789L} {670, 789}

\bibitem[\protect\citeauthoryear{{Lithwick}}{{Lithwick}}{2009}]{lithwick09}
{Lithwick} Y.,  2009, \mn@doi [\apj] {10.1088/0004-637X/693/1/85}, \href
  {http://adsabs.harvard.edu/abs/2009ApJ...693...85L} {693, 85}

\bibitem[\protect\citeauthoryear{{Lovelace} \& {Hohlfeld}}{{Lovelace} \&
  {Hohlfeld}}{2013}]{lovelace13}
{Lovelace} R.~V.~E.,  {Hohlfeld} R.~G.,  2013, \mn@doi [\mnras]
  {10.1093/mnras/sts361}, \href
  {http://adsabs.harvard.edu/abs/2013MNRAS.429..529L} {429, 529}

\bibitem[\protect\citeauthoryear{{Lovelace}, {Li}, {Colgate}  \&
  {Nelson}}{{Lovelace} et~al.}{1999}]{lovelace99}
{Lovelace} R.~V.~E.,  {Li} H.,  {Colgate} S.~A.,   {Nelson} A.~F.,  1999,
  \mn@doi [\apj] {10.1086/306900}, \href
  {http://adsabs.harvard.edu/abs/1999ApJ...513..805L} {513, 805}

\bibitem[\protect\citeauthoryear{{Lynden-Bell} \& {Pringle}}{{Lynden-Bell} \&
  {Pringle}}{1974}]{lyndenbell74}
{Lynden-Bell} D.,  {Pringle} J.~E.,  1974, \mnras, \href
  {http://adsabs.harvard.edu/abs/1974MNRAS.168..603L} {168, 603}

\bibitem[\protect\citeauthoryear{{Lyra}}{{Lyra}}{2014}]{lyra14}
{Lyra} W.,  2014, \mn@doi [\apj] {10.1088/0004-637X/789/1/77}, \href
  {http://adsabs.harvard.edu/abs/2014ApJ...789...77L} {789, 77}

\bibitem[\protect\citeauthoryear{{Lyra} \& {Klahr}}{{Lyra} \&
  {Klahr}}{2011}]{lyra11}
{Lyra} W.,  {Klahr} H.,  2011, \mn@doi [\aap] {10.1051/0004-6361/201015568},
  \href {http://adsabs.harvard.edu/abs/2011A%26A...527A.138L} {527, A138+}

\bibitem[\protect\citeauthoryear{{Lyra} \& {Lin}}{{Lyra} \&
  {Lin}}{2013}]{lyra13}
{Lyra} W.,  {Lin} M.-K.,  2013, \mn@doi [\apj] {10.1088/0004-637X/775/1/17},
  \href {http://adsabs.harvard.edu/abs/2013ApJ...775...17L} {775, 17}

\bibitem[\protect\citeauthoryear{{Lyra}, {Johansen}, {Klahr}  \&
  {Piskunov}}{{Lyra} et~al.}{2008}]{lyra08b}
{Lyra} W.,  {Johansen} A.,  {Klahr} H.,   {Piskunov} N.,  2008, \mn@doi [\aap]
  {10.1051/0004-6361:200810626}, \href
  {http://adsabs.harvard.edu/abs/2008A%26A...491L..41L} {491, L41}

\bibitem[\protect\citeauthoryear{{Lyra}, {Johansen}, {Klahr}  \&
  {Piskunov}}{{Lyra} et~al.}{2009a}]{lyra08}
{Lyra} W.,  {Johansen} A.,  {Klahr} H.,   {Piskunov} N.,  2009a, \mn@doi [\aap]
  {10.1051/0004-6361:200810797}, \href
  {http://adsabs.harvard.edu/abs/2009A%26A...493.1125L} {493, 1125}

\bibitem[\protect\citeauthoryear{{Lyra}, {Johansen}, {Zsom}, {Klahr}  \&
  {Piskunov}}{{Lyra} et~al.}{2009b}]{lyra09}
{Lyra} W.,  {Johansen} A.,  {Zsom} A.,  {Klahr} H.,   {Piskunov} N.,  2009b,
  \mn@doi [\aap] {10.1051/0004-6361/200811265}, \href
  {http://adsabs.harvard.edu/abs/2009A%26A...497..869L} {497, 869}

\bibitem[\protect\citeauthoryear{{Lyra}, {Turner}  \& {McNally}}{{Lyra}
  et~al.}{2015}]{lyra15}
{Lyra} W.,  {Turner} N.~J.,   {McNally} C.~P.,  2015, \mn@doi [\aap]
  {10.1051/0004-6361/201424919}, \href
  {http://adsabs.harvard.edu/abs/2015A%26A...574A..10L} {574, A10}

\bibitem[\protect\citeauthoryear{{Mamatsashvili} \&
  {Chagelishvili}}{{Mamatsashvili} \& {Chagelishvili}}{2007}]{mamat07}
{Mamatsashvili} G.~R.,  {Chagelishvili} G.~D.,  2007, \mn@doi [\mnras]
  {10.1111/j.1365-2966.2007.12274.x}, \href
  {http://adsabs.harvard.edu/abs/2007MNRAS.381..809M} {381, 809}

\bibitem[\protect\citeauthoryear{{Mamatsashvili} \& {Rice}}{{Mamatsashvili} \&
  {Rice}}{2009}]{mamat09}
{Mamatsashvili} G.~R.,  {Rice} W.~K.~M.,  2009, \mn@doi [\mnras]
  {10.1111/j.1365-2966.2009.14481.x}, \href
  {http://adsabs.harvard.edu/abs/2009MNRAS.394.2153M} {394, 2153}

\bibitem[\protect\citeauthoryear{{Mamatsashvili} \& {Rice}}{{Mamatsashvili} \&
  {Rice}}{2010}]{mamat10}
{Mamatsashvili} G.~R.,  {Rice} W.~K.~M.,  2010, \mn@doi [\mnras]
  {10.1111/j.1365-2966.2010.16825.x}, \href
  {http://adsabs.harvard.edu/abs/2010MNRAS.406.2050M} {406, 2050}

\bibitem[\protect\citeauthoryear{{Marcus}, {Pei}, {Jiang}, {Barranco},
  {Hassanzadeh}  \& {Lecoanet}}{{Marcus} et~al.}{2015}]{marcus15}
{Marcus} P.~S.,  {Pei} S.,  {Jiang} C.-H.,  {Barranco} J.~A.,  {Hassanzadeh}
  P.,   {Lecoanet} D.,  2015, \mn@doi [\apj] {10.1088/0004-637X/808/1/87},
  \href {http://adsabs.harvard.edu/abs/2015ApJ...808...87M} {808, 87}

\bibitem[\protect\citeauthoryear{{Marino}, {Casassus}, {Perez}, {Lyra},
  {Roman}, {Avenhaus}, {Wright}  \& {Maddison}}{{Marino}
  et~al.}{2015}]{marino15}
{Marino} S.,  {Casassus} S.,  {Perez} S.,  {Lyra} W.,  {Roman} P.~E.,
  {Avenhaus} H.,  {Wright} C.~M.,   {Maddison} S.~T.,  2015, \mn@doi [\apj]
  {10.1088/0004-637X/813/1/76}, \href
  {http://adsabs.harvard.edu/abs/2015ApJ...813...76M} {813, 76}

\bibitem[\protect\citeauthoryear{{Meheut}, {Casse}, {Varniere}  \&
  {Tagger}}{{Meheut} et~al.}{2010}]{meheut10}
{Meheut} H.,  {Casse} F.,  {Varniere} P.,   {Tagger} M.,  2010, \mn@doi [\aap]
  {10.1051/0004-6361/201014000}, \href
  {http://adsabs.harvard.edu/abs/2010A%26A...516A..31M} {516, A31+}

\bibitem[\protect\citeauthoryear{{Meheut}, {Meliani}, {Varniere}  \&
  {Benz}}{{Meheut} et~al.}{2012}]{meheut12}
{Meheut} H.,  {Meliani} Z.,  {Varniere} P.,   {Benz} W.,  2012, \mn@doi [\aap]
  {10.1051/0004-6361/201219794}, \href
  {http://adsabs.harvard.edu/abs/2012A%26A...545A.134M} {545, A134}

\bibitem[\protect\citeauthoryear{{Nelson}, {Gressel}  \& {Umurhan}}{{Nelson}
  et~al.}{2013}]{nelson13}
{Nelson} R.~P.,  {Gressel} O.,   {Umurhan} O.~M.,  2013, \mn@doi [\mnras]
  {10.1093/mnras/stt1475}, \href
  {http://adsabs.harvard.edu/abs/2013MNRAS.435.2610N} {435, 2610}

\bibitem[\protect\citeauthoryear{{Ohta} et~al.,}{{Ohta} et~al.}{2016}]{ohta16}
{Ohta} Y.,  et~al., 2016, \mn@doi [\pasj] {10.1093/pasj/psw051}, \href
  {http://adsabs.harvard.edu/abs/2016PASJ...68...53O} {68, 53}

\bibitem[\protect\citeauthoryear{{Ono}, {Muto}, {Takeuchi}  \& {Nomura}}{{Ono}
  et~al.}{2016}]{ono16}
{Ono} T.,  {Muto} T.,  {Takeuchi} T.,   {Nomura} H.,  2016, \mn@doi [\apj]
  {10.3847/0004-637X/823/2/84}, \href
  {http://adsabs.harvard.edu/abs/2016ApJ...823...84O} {823, 84}

\bibitem[\protect\citeauthoryear{{Paardekooper}, {Lesur}  \&
  {Papaloizou}}{{Paardekooper} et~al.}{2010}]{paardekooper10}
{Paardekooper} S.,  {Lesur} G.,   {Papaloizou} J.~C.~B.,  2010, \mn@doi [\apj]
  {10.1088/0004-637X/725/1/146}, \href
  {http://adsabs.harvard.edu/abs/2010ApJ...725..146P} {725, 146}

\bibitem[\protect\citeauthoryear{{P{\'e}rez}, {Isella}, {Carpenter}  \&
  {Chandler}}{{P{\'e}rez} et~al.}{2014}]{perez14}
{P{\'e}rez} L.~M.,  {Isella} A.,  {Carpenter} J.~M.,   {Chandler} C.~J.,  2014,
  \mn@doi [\apjl] {10.1088/2041-8205/783/1/L13}, \href
  {http://adsabs.harvard.edu/abs/2014ApJ...783L..13P} {783, L13}

\bibitem[\protect\citeauthoryear{{Petersen}, {Julien}  \& {Stewart}}{{Petersen}
  et~al.}{2007}]{peterson07a}
{Petersen} M.~R.,  {Julien} K.,   {Stewart} G.~R.,  2007, \mn@doi [\apj]
  {10.1086/511513}, \href {http://adsabs.harvard.edu/abs/2007ApJ...658.1236P}
  {658, 1236}

\bibitem[\protect\citeauthoryear{{Raettig}, {Lyra}  \& {Klahr}}{{Raettig}
  et~al.}{2013}]{raettig13}
{Raettig} N.,  {Lyra} W.,   {Klahr} H.,  2013, \mn@doi [\apj]
  {10.1088/0004-637X/765/2/115}, \href
  {http://adsabs.harvard.edu/abs/2013ApJ...765..115R} {765, 115}

\bibitem[\protect\citeauthoryear{{Railton} \& {Papaloizou}}{{Railton} \&
  {Papaloizou}}{2014}]{railton14}
{Railton} A.~D.,  {Papaloizou} J.~C.~B.,  2014, \mn@doi [\mnras]
  {10.1093/mnras/stu2060}, \href
  {http://adsabs.harvard.edu/abs/2014MNRAS.445.4409R} {445, 4409}

\bibitem[\protect\citeauthoryear{{Reg{\'a}ly} \& {Vorobyov}}{{Reg{\'a}ly} \&
  {Vorobyov}}{2017}]{regaly17}
{Reg{\'a}ly} Z.,  {Vorobyov} E.,  2017, \mn@doi [\mnras]
  {10.1093/mnras/stx1801}, \href
  {http://adsabs.harvard.edu/abs/2017MNRAS.471.2204R} {471, 2204}

\bibitem[\protect\citeauthoryear{{Richard}, {Nelson}  \& {Umurhan}}{{Richard}
  et~al.}{2016}]{richard16}
{Richard} S.,  {Nelson} R.~P.,   {Umurhan} O.~M.,  2016, \mn@doi [\mnras]
  {10.1093/mnras/stv2898}, \href
  {http://adsabs.harvard.edu/abs/2016MNRAS.456.3571R} {456, 3571}

\bibitem[\protect\citeauthoryear{{Schmit} \& {Tscharnuter}}{{Schmit} \&
  {Tscharnuter}}{1995}]{schmit95}
{Schmit} U.,  {Tscharnuter} W.~M.,  1995, \mn@doi [\icarus]
  {10.1006/icar.1995.1099}, \href
  {http://adsabs.harvard.edu/abs/1995Icar..115..304S} {115, 304}

\bibitem[\protect\citeauthoryear{{Shen}, {Stone}  \& {Gardiner}}{{Shen}
  et~al.}{2006}]{shen06}
{Shen} Y.,  {Stone} J.~M.,   {Gardiner} T.~A.,  2006, \mn@doi [\apj]
  {10.1086/508980}, \href {http://adsabs.harvard.edu/abs/2006ApJ...653..513S}
  {653, 513}

\bibitem[\protect\citeauthoryear{{Shi}, {Zhu}, {Stone}  \& {Chiang}}{{Shi}
  et~al.}{2016}]{shi16}
{Shi} J.-M.,  {Zhu} Z.,  {Stone} J.~M.,   {Chiang} E.,  2016, \mn@doi [\mnras]
  {10.1093/mnras/stw692}, \href
  {http://adsabs.harvard.edu/abs/2016MNRAS.459..982S} {459, 982}

\bibitem[\protect\citeauthoryear{{Stoll} \& {Kley}}{{Stoll} \&
  {Kley}}{2014}]{stoll14}
{Stoll} M.~H.~R.,  {Kley} W.,  2014, \mn@doi [\aap]
  {10.1051/0004-6361/201424114}, \href
  {http://adsabs.harvard.edu/abs/2014A%26A...572A..77S} {572, A77}

\bibitem[\protect\citeauthoryear{{Stone} \& {Gardiner}}{{Stone} \&
  {Gardiner}}{2010}]{stone10}
{Stone} J.~M.,  {Gardiner} T.~A.,  2010, \mn@doi [\apjs]
  {10.1088/0067-0049/189/1/142}, \href
  {http://adsabs.harvard.edu/abs/2010ApJS..189..142S} {189, 142}

\bibitem[\protect\citeauthoryear{{Stone}, {Gardiner}, {Teuben}, {Hawley}  \&
  {Simon}}{{Stone} et~al.}{2008}]{stone08}
{Stone} J.~M.,  {Gardiner} T.~A.,  {Teuben} P.,  {Hawley} J.~F.,   {Simon}
  J.~B.,  2008, \mn@doi [\apjs] {10.1086/588755}, \href
  {http://adsabs.harvard.edu/abs/2008ApJS..178..137S} {178, 137}

\bibitem[\protect\citeauthoryear{{Surville}, {Mayer}  \& {Lin}}{{Surville}
  et~al.}{2016}]{surville16}
{Surville} C.,  {Mayer} L.,   {Lin} D.~N.~C.,  2016, \mn@doi [\apj]
  {10.3847/0004-637X/831/1/82}, \href
  {http://adsabs.harvard.edu/abs/2016ApJ...831...82S} {831, 82}

\bibitem[\protect\citeauthoryear{{Takahashi} \& {Inutsuka}}{{Takahashi} \&
  {Inutsuka}}{2014}]{takahashi14}
{Takahashi} S.~Z.,  {Inutsuka} S.-i.,  2014, \mn@doi [\apj]
  {10.1088/0004-637X/794/1/55}, \href
  {http://adsabs.harvard.edu/abs/2014ApJ...794...55T} {794, 55}

\bibitem[\protect\citeauthoryear{{Toomre}}{{Toomre}}{1964}]{toomre64}
{Toomre} A.,  1964, \mn@doi [\apj] {10.1086/147861}, \href
  {http://adsabs.harvard.edu/abs/1964ApJ...139.1217T} {139, 1217}

\bibitem[\protect\citeauthoryear{{Umurhan} \& {Regev}}{{Umurhan} \&
  {Regev}}{2004}]{umurhan04}
{Umurhan} O.~M.,  {Regev} O.,  2004, \mn@doi [\aap]
  {10.1051/0004-6361:20040573}, \href
  {http://adsabs.harvard.edu/abs/2004A%26A...427..855U} {427, 855}

\bibitem[\protect\citeauthoryear{{Umurhan}, {Estrada}  \& {Cuzzi}}{{Umurhan}
  et~al.}{2016a}]{umurhan16b}
{Umurhan} O.~M.,  {Estrada} R.~R.,   {Cuzzi} J.~N.,  2016a, in Lunar and
  Planetary Science Conference. p.~2887

\bibitem[\protect\citeauthoryear{{Umurhan}, {Shariff}  \& {Cuzzi}}{{Umurhan}
  et~al.}{2016b}]{umurhan16}
{Umurhan} O.~M.,  {Shariff} K.,   {Cuzzi} J.~N.,  2016b, \mn@doi [\apj]
  {10.3847/0004-637X/830/2/95}, \href
  {http://adsabs.harvard.edu/abs/2016ApJ...830...95U} {830, 95}

\bibitem[\protect\citeauthoryear{{Yellin-Bergovoy}, {Heifetz}  \&
  {Umurhan}}{{Yellin-Bergovoy} et~al.}{2016}]{yellin16}
{Yellin-Bergovoy} R.,  {Heifetz} E.,   {Umurhan} O.~M.,  2016, \mn@doi
  [Geophysical and Astrophysical Fluid Dynamics]
  {10.1080/03091929.2016.1158816}, \href
  {http://adsabs.harvard.edu/abs/2016GApFD.110..274Y} {110, 274}

\bibitem[\protect\citeauthoryear{{Youdin}}{{Youdin}}{2011}]{youdin11}
{Youdin} A.~N.,  2011, \mn@doi [\apj] {10.1088/0004-637X/731/2/99}, \href
  {http://adsabs.harvard.edu/abs/2011ApJ...731...99Y} {731, 99}

\bibitem[\protect\citeauthoryear{{Youdin} \& {Lithwick}}{{Youdin} \&
  {Lithwick}}{2007}]{youdin07}
{Youdin} A.~N.,  {Lithwick} Y.,  2007, \mn@doi [\icarus]
  {10.1016/j.icarus.2007.07.012}, \href
  {http://adsabs.harvard.edu/abs/2007Icar..192..588Y} {192, 588}

\bibitem[\protect\citeauthoryear{{Zhu} \& {Baruteau}}{{Zhu} \&
  {Baruteau}}{2016}]{zhu16}
{Zhu} Z.,  {Baruteau} C.,  2016, \mn@doi [\mnras] {10.1093/mnras/stw202}, \href
  {http://adsabs.harvard.edu/abs/2016MNRAS.458.3918Z} {458, 3918}

\bibitem[\protect\citeauthoryear{{Zhu}, {Stone}  \& {Rafikov}}{{Zhu}
  et~al.}{2013}]{zhu13}
{Zhu} Z.,  {Stone} J.~M.,   {Rafikov} R.~R.,  2013, \mn@doi [\apj]
  {10.1088/0004-637X/768/2/143}, \href
  {http://adsabs.harvard.edu/abs/2013ApJ...768..143Z} {768, 143}

\bibitem[\protect\citeauthoryear{{Zhu}, {Stone}, {Rafikov}  \& {Bai}}{{Zhu}
  et~al.}{2014}]{zhu14}
{Zhu} Z.,  {Stone} J.~M.,  {Rafikov} R.~R.,   {Bai} X.-n.,  2014, \mn@doi
  [\apj] {10.1088/0004-637X/785/2/122}, \href
  {http://adsabs.harvard.edu/abs/2014ApJ...785..122Z} {785, 122}

\bibitem[\protect\citeauthoryear{{Zhu}, {Ju}  \& {Stone}}{{Zhu}
  et~al.}{2016}]{zhu16b}
{Zhu} Z.,  {Ju} W.,   {Stone} J.~M.,  2016, \mn@doi [\apj]
  {10.3847/0004-637X/832/2/193}, \href
  {http://adsabs.harvard.edu/abs/2016ApJ...832..193Z} {832, 193}

\bibitem[\protect\citeauthoryear{{van der Marel} et~al.,}{{van der Marel}
  et~al.}{2013}]{marel13}
{van der Marel} N.,  et~al., 2013, \mn@doi [Science] {10.1126/science.1236770},
  \href {http://adsabs.harvard.edu/abs/2013Sci...340.1199V} {340, 1199}

\bibitem[\protect\citeauthoryear{{van der Marel}, {Cazzoletti}, {Pinilla}  \&
  {Garufi}}{{van der Marel} et~al.}{2016}]{marel16}
{van der Marel} N.,  {Cazzoletti} P.,  {Pinilla} P.,   {Garufi} A.,  2016,
  \mn@doi [\apj] {10.3847/0004-637X/832/2/178}, \href
  {http://adsabs.harvard.edu/abs/2016ApJ...832..178V} {832, 178}

\makeatother
\end{thebibliography}

\appendix
\section{Shearing box equations in elliptico-polar
  co-ordinates}\label{elliptico-eqns} 
In elliptico-polar co-ordinates $(s,\varphi,z)$ the divergence and
advective derivative  
operators have identical form to that in the usual cylindrical
co-ordinates. That is,  
\begin{align}
&\nabla \cdot \bm{W} = \frac{1}{s}\frac{\p}{\p s}\left(sW_s\right) +
  \frac{1}{s}\frac{\p W_\varphi}{\p \varphi} + \frac{\p W_z}{\p z},\\ 
&\bm{W}\cdot\nabla   = W_s\frac{\p}{\p s} +
  \frac{W_\varphi}{s}\frac{\p}{\p\varphi} + W_z\frac{\p }{\p z}, 
\end{align} 
for any vector field $\bm{W}$. The continuity equation thus retain the
same form as in the Cartesian box, as does the vertical momentum
equation.  
The horizontal momentum equations, however, are modified to read 
\begin{align}
  & \frac{\p v_s}{\p t} + \bm{v}\cdot\nabla v_s - \frac{v_\varphi^2}{s} \notag\\
  &= \frac{1}{2}\left(\xi_+ + \xi_-\cos{2\varphi} \right)\left(2\Omega\chi
  v_\varphi-\frac{\p \eta}{\p s}\right) \notag\\
  &+\frac{\xi_-}{2}\sin{2\varphi}\left(2\Omega\chi v_s +
  \frac{1}{s}\frac{\p\eta}{\p\varphi}\right) 
  +q\Omega^2s\left(\cos{2\varphi}+1\right)\\
&\frac{\p v_\varphi}{\p t} + \bm{v}\cdot\nabla v_\varphi +
\frac{v_\varphi v_s}{s}\notag\\   
&=\frac{\xi_-}{2}\sin{2\varphi}\left(\frac{\p \eta}{\p s} -
2\Omega\chi v_\varphi\right)\notag\\
&-\frac{1}{2}\left(\xi_+ -
\xi_-\cos{2\varphi}\right)\left(\frac{1}{s}\frac{\p\eta}{\p
  \varphi} + 2\Omega\chi v_s\right)
- q\Omega^2 s \sin{2\varphi},
\end{align}
where 
\begin{align}
\xi_\pm = 1 \pm \frac{1}{\chi^2}. 
\end{align}
Finally, the Poisson equation becomes
\begin{align}\label{elliptico-poisson}
&\left[\frac{1}{2}\left(\xi_-\cos{2\varphi} + \xi_+\right)\frac{\p^2}{\p s^2} +\frac{1}{2s^2}\left( \xi_+ - \xi_-\cos{2\varphi}\right)\frac{\p^2}{\p\varphi^2}\right.\notag\\
&\left. - \frac{\xi_-}{s}\sin{2\varphi}\frac{\p^2}{\p\varphi\p s} + \frac{1}{2s}\left(\xi_+ - \xi_-\cos{2\varphi}\right)\frac{\p}{\p s} + \frac{\xi_-}{s^2}\sin{2\varphi}\frac{\p}{\p\varphi}\right. \notag \\
&\left. +\frac{\p^2}{\p z^2}
\right]\Phi = 4 \pi G \rho.
\end{align}

\section{Linear response of the GNG vortex to its gravitational
  potential}\label{sgeffect}
We explore the perturbative effect of introducing self-gravity on the
GNG vortex solution. We simplify the problem by considering an
unstratified disc with no vertical velocities. Then $v_z = \p_z = 0$.  
This should be an adequate approximation for regions close to the
vortex centre ($s,\,z\to0$). 
We set  
\begin{align}
  G \to 0 + \delta G. 
\end{align}
This induces perturbations
\begin{align}
  &\rho \to \rho(s) + \delta\rho,\\
  &v_s \to 0 + \delta v_s,\\
  &v_\varphi \to -\Omega_\mathrm{v}s + \delta v_\varphi,\\
  &\Phi \to 0 + \delta\Phi.
\end{align}
The background flow, $v_\varphi(s)$ and $\rho(s)$, is
given by the non-self-gravitating GNG vortex solution
(Eq. \ref{gng_sol}---\ref{gng_density}) with the vertical dimension
suppressed. Since the basic state is symmetric in $\varphi$, we
Fourier analyse in azimuth and write  
\begin{align}\label{fourier_expansion}
  \begin{bmatrix}
    \delta\rho \\
    \delta v_s  \\
    \delta v_\varphi\\ 
    \delta \Phi
  \end{bmatrix}
  =\real \sum_{l=0}^\infty
  \begin{bmatrix}
     \rho \mathcal{Q}_l/c_s^2 \\
    \ii U_l  \\
    V_l\\ 
    \Phi_l
   \end{bmatrix}
   \exp{\ii l \varphi}. 
\end{align}

We insert the above perturbations into the steady state continuity,
horizontal momentum and Poisson equations and linearize. We then
multiply by $\mathrm{e}^{-\ii n \varphi}$, and integrate in $\varphi$ to obtain 
\begin{align}\label{continutity_ellip}
  \frac{1}{s}\left(s\rho U_n\right)^\prime + \frac{n}{s}\rho V_n -
  n\omegav \frac{\rho \mathcal{Q}_n}{c_s^2} = 0 \\
\end{align}
from the continuity equation, where $^\prime$ denotes $d/ds$; and the
horizontal momentum equations give 
\begin{align}
  2\omegav V_n + n\omegav U_n = & 
  -\frac{1}{2}\left[\xi_+\tilde{\eta}_n^\prime +
    \frac{\xi_-}{2}\left(\tilde{\eta}_{n-2}^\prime +
    \tilde{\eta}_{n+2}^\prime \right)\right]    \notag\\
  & +
  \frac{\xi_-}{4s}\left[\left(n-2\right)\tilde{\eta}_{n-2}-\left(n+2\right)\tilde{\eta}_{n+2}\right] 
  \notag \\
  & + \frac{1}{2}\Omega\xi_-\chi \left(U_{n-2}-U_{n+2}\right)\notag\\
  & + \Omega\chi\left[\xi_+V_n +
    \frac{\xi_-}{2}\left(V_{n-2}+V_{n+2}\right)\right],\\
  2\omegav U_n +  n\omegav V_n = 
  & \frac{\xi_-}{4}\left(\tilde{\eta}_{n-2}^\prime-\tilde{\eta}_{n+2}^\prime\right)\notag\\ 
  & + \frac{1}{2s}\left\{ n \xi_+\tilde{\eta}_{n}
  \phantom{\frac{\xi_-}{2}}\right.\notag\\  
   &\phantom{\frac{1}{2s}n\xi_+}\left.-
  \frac{\xi_-}{2}\left[(n-2)\tilde{\eta}_{n-2} +
    (n+2)\tilde{\eta}_{n+2} 
    \right]
  \right\}\notag\\
  & - \frac{1}{2}\Omega\xi_-\chi\left(V_{n-2} - V_{n+2}\right)\notag\\
  & + \Omega\chi\left[\xi_+U_n - \frac{\xi_-}{2}\left(U_{n-2} +
    U_{n+2}\right)
    \right],
\end{align}
where 
\begin{align}\label{eta_def}
  \tilde{\eta}_n \equiv \mathcal{Q}_n + \Phi_n.
\end{align}

Adding and subtracting the horizontal momentum equations gives the
more compact form 
\begin{align}
  &\frac{\xip}{2}\left(\teta^\prime_n + \frac{n}{s}\teta_n\right) +
  \frac{\xim}{2}\left(\teta^\prime_{n-2} -
  \frac{n-2}{s}\teta_{n-2}\right)   \notag\\ 
  &= \left[\left(2-n\right)\omegav
    - \Omega\chi\xip\right]Y_n +  \Omega\chi\xim X_{n-2},\notag\\
  &\frac{\xip}{2}\left(\teta^\prime_n - \frac{n}{s}\teta_n\right) +
  \frac{\xim}{2}\left(\teta^\prime_{n+2} + 
  \frac{n+2}{s}\teta_{n+2}\right)   \notag\\ 
  &= \left[\Omega\chi\xip - \left(n+2\right)\omegav \right]X_n -
  \Omega\chi\xim Y_{n+2},
\end{align}
where
\begin{align}
  X_n &\equiv U_n + V_n,\\
  Y_n &\equiv U_n - V_n. 
\end{align}

The linearized Poisson equation is
\begin{align}
  &\frac{\xi_-}{4}\left[\frac{d^2}{ds^2} - \frac{(2n-3)}{s}\frac{d}{ds}
    + \frac{n(n-2)}{s^2}\right]\Phi_{n-2}\notag\\
  &+\frac{\xi_+}{2}\left(\frac{d^2}{ds^2} + \frac{1}{s}\frac{d}{ds} -
  \frac{n^2}{s^2}\right)\Phi_n\notag\\
  &+\frac{\xi_-}{4}\left[\frac{d^2}{ds^2} + \frac{(2n+3)}{s}\frac{d}{ds}
  + \frac{n(n+2)}{s^2}\right]\Phi_{n+2} \notag\\
  &=  4\pi\delta G \rho(s)
  \delta_{n0}. 
\end{align}

\subsection{The $n=2$ problem}
To make analytic progress we truncate the series
expansion (Eq. \ref{fourier_expansion}) at $n=2$. Then for $n=0$ we
have 
\begin{align}
  &\frac{\xip}{2}\teta^\prime_0 = \left(\Omega\chi\xip - 2 \omegav
  \right)V_0,\label{trunc_eq1}\\
  &\frac{\xip}{2}\teta^\prime_0 + \frac{\xim}{2}\left(\teta^\prime_2 +
  \frac{2}{s}\teta_2\right) =  \left(\Omega\chi\xip - 2 \omegav
  \right)V_0 - \Omega\chi\xim Y_2,\label{trunc_eq2}
\end{align}
where we have utilized the fact that $U_0=0$ as implied by the
continuity equation (Eq. \ref{continutity_ellip}) for $n=0$. 
Next, for $n=2$ we have
\begin{align}
  &\frac{\xip}{2}\left(\teta^\prime_2 +\frac{2}{s}\teta_2\right) +
  \frac{\xim}{2}\teta^\prime_0 = \Omega\chi\xim V_0 - \Omega\chi\xip
  Y_2,\label{trunc_eq3}\\
  &\frac{\xip}{2}\left(\teta^\prime_2 -\frac{2}{s}\teta_2\right) =
  \left(\Omega\chi\xip - 4\omegav\right)X_2.\label{trunc_eq4}
\end{align}

Eq. \ref{trunc_eq1}---\ref{trunc_eq4} imply
\begin{align}
  &  \teta^\prime_0 = V_0 = 0,\\
&  2\omegav V_2 = \frac{\xip}{s}\teta_2 + \left(\Omega\chi\xip -
  2\omegav\right)U_2,\label{v2_expression}
\end{align}
Hence, an ODE for $\teta_2$ is 
\begin{align}
  &\teta^\prime_2 + \frac{1}{s}\left(2-
  \frac{\Omega\chi\xip}{\omegav}\right)\teta_2 +
  \Omega\chi\left(4 - 
  \frac{\Omega\chi\xip}{\omegav}\right)U_2 = 0.
\end{align}
Next, the $n=2$ continuity equation, after using Eq. \ref{eta_def} and
\ref{v2_expression}, becomes  
\begin{align}\label{u2_ode}
  &U_2^\prime + \left[\frac{\rho^\prime}{\rho} +
    \frac{1}{s}\left(\frac{\Omega\chi\xip}{\omegav}-1\right)\right]U_2 +
  \left(\frac{\xip}{s^2\omegav} - \frac{2\omegav}{c_s^2}\right)\teta_2
  \notag\\ 
  &+ \frac{2\omegav}{c_s^2}\Phi_2 = 0.
\end{align}
Further eliminating $U_2$ gives a second order ODE for $\teta_2$:
\begin{align}\label{eta2_ode}
  &\teta_2^{\prime\prime} + \left(\frac{\rho^\prime}{\rho} +
  \frac{1}{s}\right)\teta^\prime_2 
  +\left[ \frac{\lambda\rho^\prime}{s\rho} + 
    2\Omega\chi(\lambda+2)\frac{\omegav}{c_s^2} - \frac{4}{s^2} 
    \right]\teta_2 \notag\\
  & =  2\Omega\chi(\lambda+2)\frac{\omegav}{c_s^2}\Phi_2,
\end{align}
where $\lambda\equiv 2 - \Omega\chi\xip/\omegav$. We require
$\teta^\prime_2(0) = \teta_2(0) = 0 $.

Eq. \ref{eta2_ode} describes the non-axisymmetric response of the GNG
vortex to the potential $\Phi_2$. 
We are interested in the case where the potential arises from the
vortex mass itself. Thus  
$\Phi_2$ is given by a solution to the linearized Poisson solution.   


\subsection{Approximate solution as $s\to 0$} 
Here we consider the approximate solution to Eq. \ref{eta2_ode} for
small $s$. We first obtain the potential solution $\Phi_2$. To do so 
in a tractable manner, we apply the `source term approximation' to 
solve the Poisson equation as follows. 


We first solve the $n=0$ cylindrical Poisson equation neglecting the
$n=2$ term, 
\begin{align}
  \Phi_0^{\prime\prime} + \frac{1}{s}\Phi_0^\prime =
  \frac{2\Omega^2}{\xip \qthree }\exp{\left(-s^2/2\heff^2\right)}. 
\end{align}
The solution is
\begin{align}
  \Phi_0^\prime = \frac{2\Omega^2\heff^2}{\xip \qthree }\frac{1}{s}\left[1 -
    \exp{\left(-s^2/2\heff^2\right)}\right]. 
\end{align} 

We then use the above solution for $\Phi_0^\prime$ as a source term in
the $n=2$ Poisson equation
\begin{align}
  \Phi_2^{\prime\prime} + \frac{1}{s}\Phi_2^\prime -
  \frac{4}{s^2}\Phi_2 &= -\frac{\xim}{2\xip}\left(\Phi_0^{\prime\prime}
    - \frac{1}{s}\Phi_0^\prime\right)\notag\\
  &\simeq\frac{\xim\Omega^2}{4\xip^2\qthree \heff^2}s^2\equiv \mathcal{A} s^2,
\end{align}
where for the approximation we have expanded the solution for
$\Phi_0^\prime$ to first order in $s^2/2\heff^2$. The approximate
solution for $\Phi_2$ is then
\begin{align}
  \Phi_2\simeq \frac{\mathcal{A}}{12}s^4. 
\end{align}

For small $s$ Eq. \ref{eta2_ode}, with the above solution
for $\Phi_2$, is approximately
\begin{align}
  \teta^{\prime\prime}_2 + \frac{1}{s}\teta^{\prime}_2 -
  \frac{4}{s^2}\teta_2 = \mathcal{B} s^4,
\end{align}
where
\begin{align}
  &\mathcal{B} \equiv \mathcal{A}\Omega\chi(\lambda+2)\omegav/6c_s^2.
\end{align}
The solution is
\begin{align}
  \teta_2 = \frac{\mathcal{B}s^6}{32}. 
\end{align}

The corresponding solution for $U_2$ and $V_2$ can be obtained from
Eq. \ref{u2_ode} and \ref{v2_expression}, respectively. We find
\begin{align}
  U_2 = -\frac{\mathcal{A}\omegav}{192c_s^2}(6+\lambda)s^5,
\end{align}
and 
\begin{align}\label{V2mU2}
  V_2 - U_2 = \frac{\mathcal{A}\omegav}{48c_s^2}(2+\lambda)s^5. 
\end{align}

\subsection{Vorticity perturbation}\label{lin_vort_pert}
The perturbation to the vertical component of vorticity,
$\delta\omega$, is defined in Cartesian co-ordinates as 
\begin{align}
  \delta\omega \equiv \frac{\p}{\p x}\delta v_y - \frac{\p}{\p y}\delta v_x.
\end{align} 
Expanding $\delta\omega$ similarly to Eq. \ref{fourier_expansion} and
written in terms of the perturbed elliptico-polar velocity components
$U_n,\,V_n$ we have
\begin{align}
  \omega_n =& \frac{\chi\xip}{2}V_n^\prime +
  \frac{\chi\xip}{2s}\left(V_n + nU_n\right)\notag\\
  & +\frac{\chi\xim}{4}\left(\frac{d}{ds} - \frac{n-1}{s} 
    \right)\left(U_{n-2} + V_{n-2}\right)\notag\\
  & -\frac{\chi\xim}{4}\left(\frac{d}{ds} + \frac{n+1}{s} 
    \right)\left(U_{n+2} - V_{n+2}\right).
\end{align}
Then
\begin{align}
  \omega_0 =
  \frac{\chi\xim}{4}\left(\frac{d}{ds}+\frac{1}{s}\right)\left(V_2-U_2\right). 
\end{align}

Inserting the above solution for small $s$, Eq. \ref{V2mU2}, we find  
\begin{align}\label{sgeffect_vort}
  \omega_0 = \frac{\xim^2\chi}{128\xip^2\qthree F^2}\left(4\omegav -
  \Omega\chi\xip\right)\left(\frac{s}{H_\mathrm{eff}}\right)^4. 
\end{align}
Because of the second term in brackets, the sign of the vorticity
perturbation depends on $\chi$. For $\chi\gtrsim 2.56$ the vorticity 
is made more negative by self-gravity and vice versa.

\section{Vortex evolution in 2D discs}\label{2d_sims}

We also simulated vortex evolution in 2D, razor-thin self-gravitating 
discs. These simulations were carried out in global cylindrical 
geometry $(R,\phi)$ using the \textsc{genesis} code in a setup similar
to Pierens \& Lin { (submitted)}. However, we adopt the same self-gravity 
parameter and vortex initialization as in our 3D simulations.  

\begin{figure}
\includegraphics[width=\linewidth,clip=true,trim=0cm 1cm 0cm 0cm]{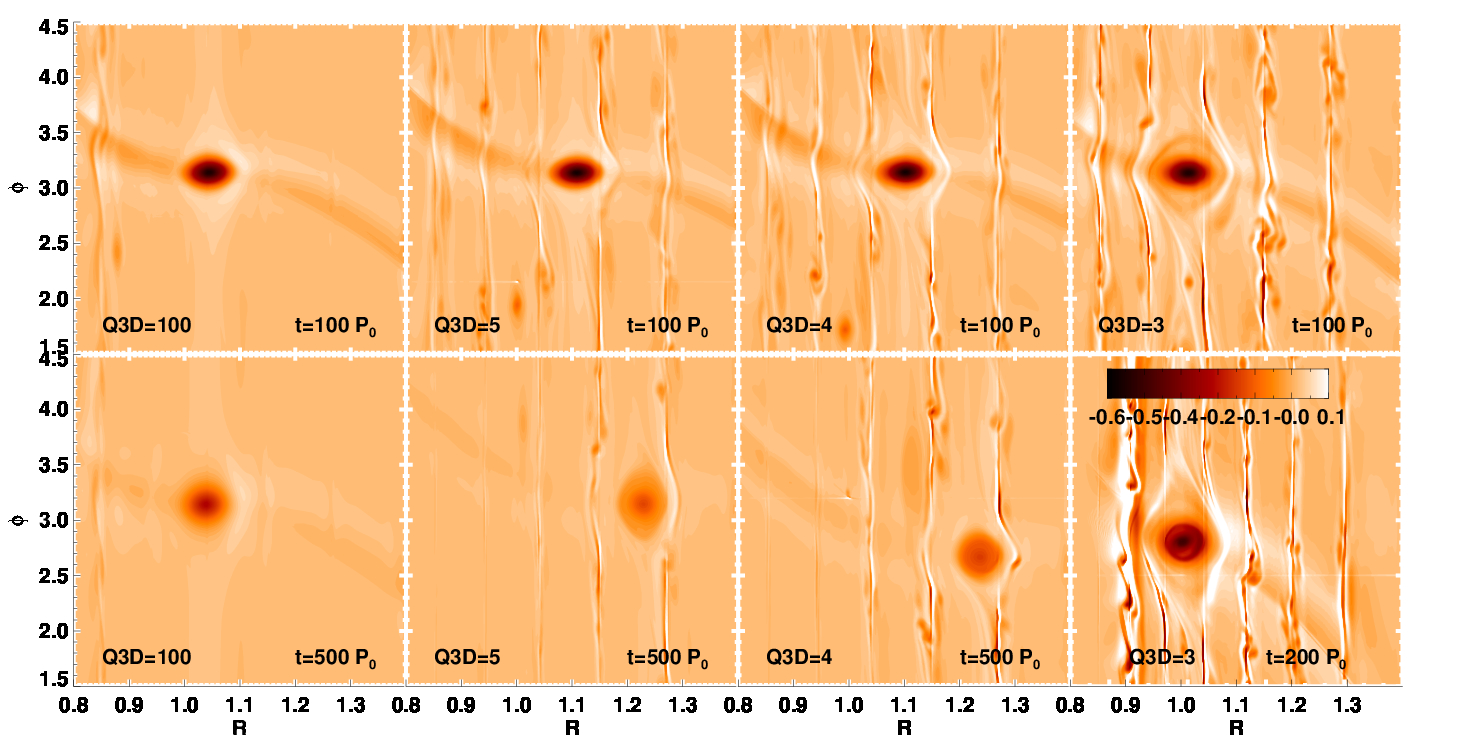}
\includegraphics[width=\linewidth]{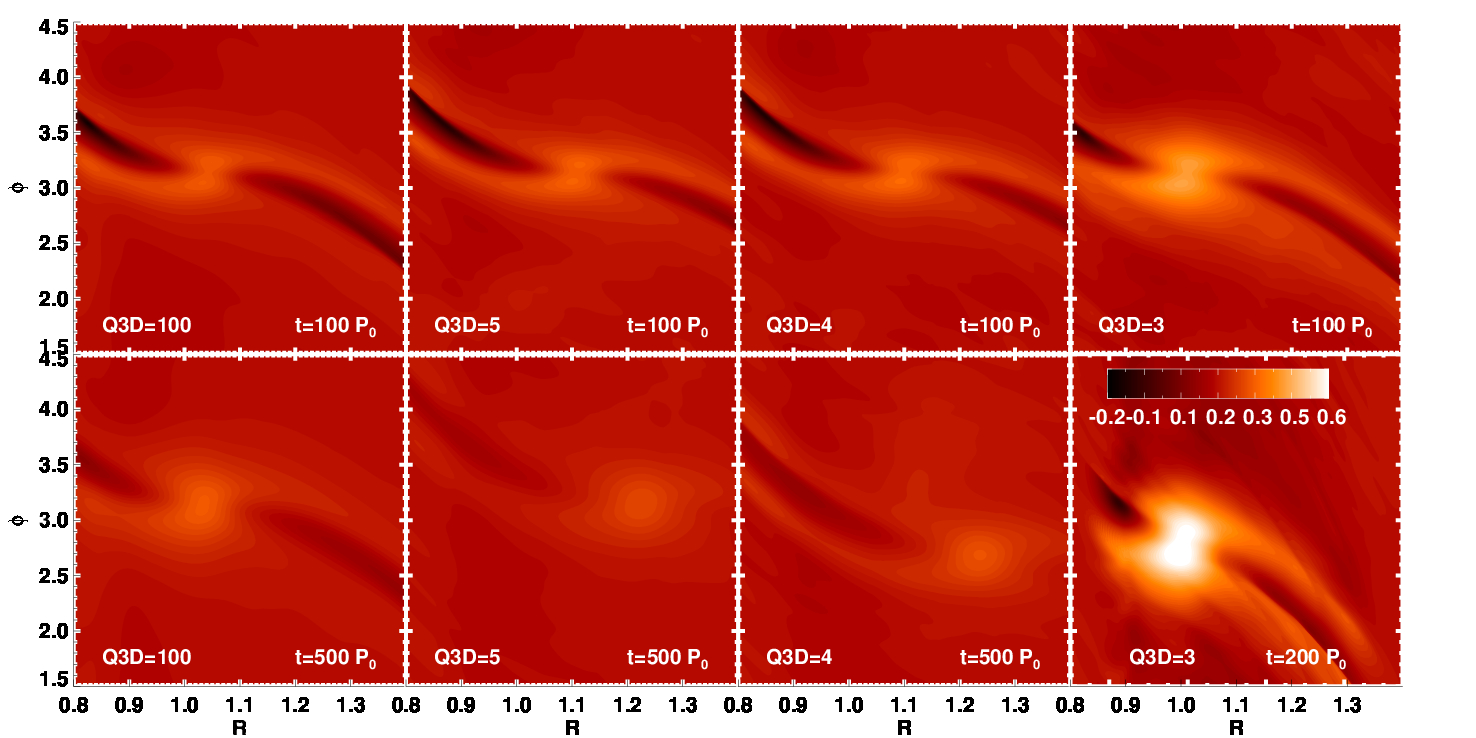}
\caption{
  Razor-thin, 2D global self-gravitating disc simulations of vortex
  evolution. Top: Rossby number; bottom: surface density. 
  \label{arnaud_sims}
  }
\end{figure}

Fig. \ref{arnaud_sims} compares the vortex evolution in discs with 
$\qthree=3,\,4,\,5$ and $100$. Being 2D, the { elliptic instability} is suppressed and thus
plays no role in these simulations. The $\qthree=100$ vortex 
decays due to numerical diffusion. This also occurs for the
$\qthree=4,\,5$, unlike in the corresponding 3D fiducial simulations where 
we find secular growth. { The $\qthree=3$ case does show vortex growth, similar to the corresponding 3D simulation.
  }

\begin{figure}
  \includegraphics[width=\linewidth]{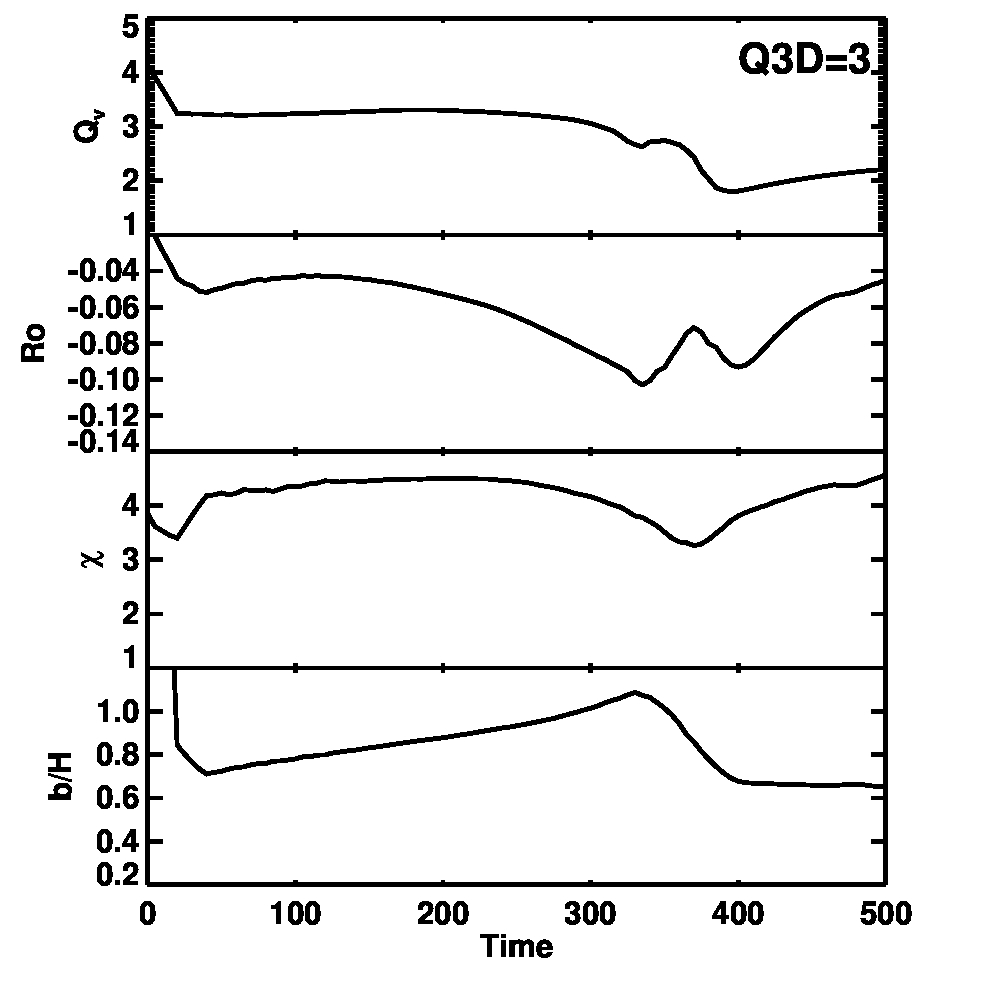}
  \caption{{ Vortex evolution in the razor-thin, 2D disc with
      $\qthree=3$ shown in Fig. \protect\ref{arnaud_sims}. Top to bottom: Toomre $Q$ at the vortex centre, Rossby number, aspect-ratio, and radial size. \label{arnaud_sims_q3} }}
\end{figure}

{
  In Fig. \ref{arnaud_sims_q3} we plot the time evolution of vortex
  properties in the $\qthree=3$, 2D disc. Similar to
  the 3D case, vortex growth is limited to $\lesssim H$ in its half-width. The global
  setup here allows us to simulate beyond this stage without
  interference from boundary conditions. We
  find the vortex induces strong spiral shocks and is weakened,
  signified by the drop in 
  $|\ro|$. The vortex is eventually
  destroyed by $t=500P_0$, i.e. it does not readjust into a new
  configuration \citep{bodo07}. This self-limited growth is
  qualitatively similar to \cite{mamat09}. However, \cite{mamat09}
  considered non-isothermal discs with $Q$ of order unity, which
  results in a gravito-turbulent state with multiple vortices, each
  lasting only two orbits; whereas our isolated vortex persists for $\sim 300$
  orbits.  
}

\begin{figure}
\includegraphics[width=\linewidth]{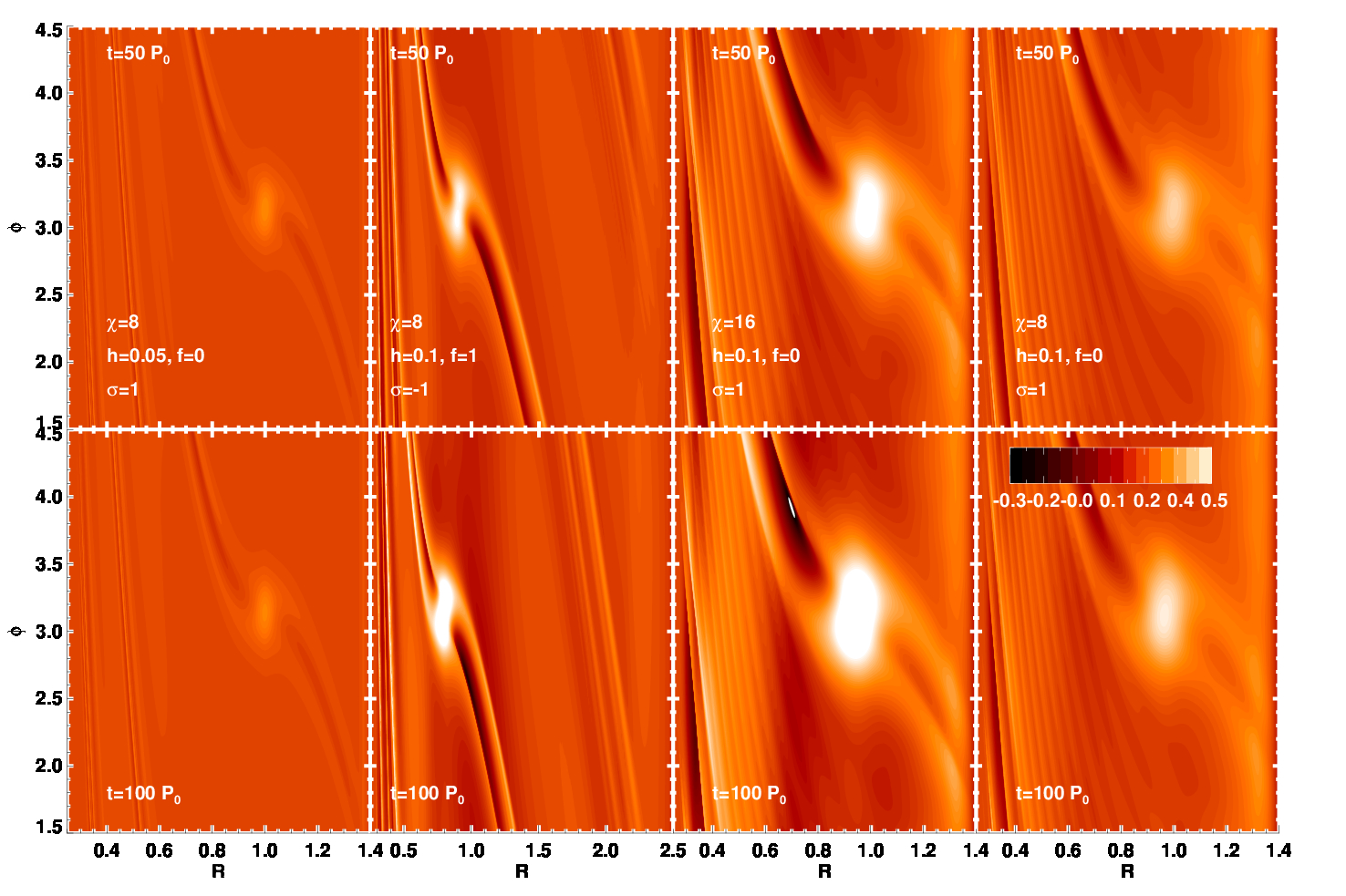}
\caption{
  Razor-thin, 2D global self-gravitating disc simulations of vortex
  evolution. We fix $\qthree=3$ but use different aspect-ratios $h$
  and flaring indices $f\equiv \p\ln{h}/\p\ln{R}$,  
  surface density profiles $\sigma \equiv - \p\ln{\Sigma}/\p\ln{R}$,
  and the initial vortex aspect-ratios $\chi$. 
  \label{arnaud_sims_survey}
  }
\end{figure}

For the $\qthree=3$, 2D disc we also experimented with different
parameters, including the disc aspect-ratio and flaring index, surface
density profile, and the initial vortex perturbation aspect-ratio. The
results are shown in Fig. \ref{arnaud_sims_survey}. We find vortex
growth is robust, unless the vortex { is} initially weak { (left panel)}. This supports the idea that a sufficiently strong 
vortex reduces the local gravitational stability of the 
disc. 

Comparing these 2D simulations to our 3D cases indicate that
for small $\qthree$ (e.g. $3$) EI-turbulence is not necessary for the vortex to undergo 
self-gravitational { growth}.  
However, at larger $\qthree$, self-gravity in 
itself is insufficient for { vortex growth}. Instead, we suggest for $\qthree\gtrsim 3$,   
EI-turbulence { can mediate vortex growth} via secular gravitational instabilities, since for 
$\qthree=4, 5$ we only observe growth in the 3D simulations with EI{, and not in the laminar, 2D discs.}

\end{document}